\documentclass[a4paper,11pt]{article}
\pdfoutput=1
\usepackage{jheppub}
\usepackage[T1]{fontenc}
\usepackage{tensor}
\usepackage{multirow, tabularx}
\usepackage[mathscr]{euscript}
\usepackage{float}
\usepackage{tikz}
\usepackage{soul}
\usepackage{mathtools}
\usepackage{dirtytalk}

\newcommand{\half}{\frac{\scriptstyle 1}{\scriptstyle 2}}
\newcommand{\C}{\mathbb{C}}

\newcommand{\CP}{\mathbb{CP}}

\newcommand{\R}{\mathbb{R}}
\renewcommand{\P}{\mathbb{P}}

\newcommand{\M}{\mathbb{M}}

\newcommand{\N}{\mathbb{N}}
\newcommand{\T}{\mathbb{T}}
\newcommand{\Z}{\mathbb{Z}}
\newcommand{\p}{\partial}

\renewcommand{\P}{\mathbb{P}}

\newcommand{\rd}{\, \mathrm{d}}

\newcommand{\be}{\begin{equation}\label}
\newcommand{\ee}{\end{equation}}
\newcommand{\bea}{\begin{eqnarray}\label}
\newcommand{\eea}{\end{eqnarray}}

\title{String theory in twistor space and minimal tension holography}

\author{N. M. McStay$^*$ \& R. A. Reid-Edwards$^{\dagger}$}

\affiliation{Department of Applied Mathematics and Theoretical Physics,\\University of Cambridge, CB3 0WA, United Kingdom}

\emailAdd{$^*$nm646@cam.ac.uk, $^{\dagger}$rar31@cam.ac.uk}

\abstract{Explicit examples of the AdS/CFT correspondence where both bulk and boundary theories are tractable are hard to come by, but the minimal tension string on $AdS_3 \times S^3 \times T^4$ is one notable example. In this paper, we discuss how one can construct sigma models on twistor space, with a particular focus on applying these techniques to the aforementioned string theory. We derive novel incidence relations, which allow us to understand to what extent the minimal tension string encodes information about the bulk. We identify vertex operators in terms of bulk twistor variables and a map from twistor space to spacetime is presented. We also demonstrate the presence of a partially broken global supersymmetry algebra in the minimal tension string and we argue that this implies that there exists an $\mathcal{N} = 2$ formulation of the theory. The implications of this are studied and we demonstrate the presence of an additional constraint on physical states.}

\begin{document} 
\maketitle
\flushbottom

\pagenumbering{arabic}

\section{Introduction}\label{sec:introduction}

The AdS/CFT correspondence is a conjecture that, if proven true, would be a clear point upon which the string theory programme, regardless of its ultimate status, will  have likely made major contributions to the discovery of a quantum theory of gravity. However, twenty-seven years after Maldacena's pioneering paper \cite{Maldacena:1997re}, we are still struggling to fully understand what deep truths the conjecture is telling us about quantum gravity. One of the major obstacles to progress has been the lack of a concrete detailed understanding of the string theory in tractable cases and away from the supergravity limit.

Recently, a proposal for the string theory describing deformations of $AdS_3\times S^3 \times \mathcal{M}_4$, where $\mathcal{M}_4$ is a compact four-dimensional hyperk\"{a}hler manifold, with pure NS-NS flux has been made \cite{Giribet:2018ada, Gaberdiel:2018rqv} (and further developed in \cite{Eberhardt:2018ouy,Eberhardt:2019ywk}). The focus of these papers is the case where $\mathcal{M}_4 = T^4$ and the flux takes its minimal possible value. The theory is based on a WZW model at level $k=1$ (corresponding to minimal tension, where stringy effects are important) and as such, we shall refer to this theory as the ``$k=1$ string''. Remarkably, it can be cast in terms of free fields and so the theory is tractable \cite{Dei:2020zui}. The expectation is that this string theory is equivalent to a two-dimensional symmetric orbifold CFT of $T^4$ and the proposal has passed many non-trivial checks, including the matching of the full non-protected spectra and correlation functions to all loops. The literature is extensive (see \cite{Gaberdiel:2018rqv,Eberhardt:2018ouy,Giribet:2018ada,Eberhardt:2019ywk,Eberhardt:2019qcl,Eberhardt:2020akk,Dei:2020zui,Gaberdiel:2020ycd,Knighton:2020kuh,Eberhardt:2020bgq,Eberhardt:2021jvj,Gaberdiel:2021kkp,Gaberdiel:2021njm,Gaberdiel:2022als,Gaberdiel:2022oeu,Naderi:2022bus,McStay:2023thk,Dei:2023ivl,Knighton:2023mhq,Knighton:2024ybs,Knighton:2024noc,Dei:2024sct} for a partial list), but despite its successes, there are still aspects that are not properly understood.

Whilst the $k=1$ string is described in terms of a concrete CFT, there is little understanding of to what extent we can view the theory as a sigma model. In conventional free field descriptions \cite{Dei:2020zui}, it has been suggested that the theory seems to naturally live, not in $AdS_3$ but in an associated twistor space \cite{McStay:2023thk}. Natural two-dimensional incidence relations, supported on the boundary, appear in the theory, but their geometrical origin, the precise twistor space that these fields live in and the role of the bulk twistor geometry has not been established. One can recast the twistor fields of \cite{Dei:2020zui} in terms of spacetime fields \cite{Dei:2023ivl}, but these are defined directly on the boundary $S^2$, rather than in the bulk of $AdS_3$. Hence the bulk physics remains poorly understood; indeed it is unclear to what extent the theory even probes the bulk. Moreover, at the level of the worldsheet theory, questions remain. The constraint algebra of the $k=1$ string is deduced from the hybrid formalism of \cite{Berkovits:1999im} and this includes two additional constraints, augmenting the usual Virasoro constraints. The first constraint generates a scaling symmetry, which has a natural geometric interpretation in the bulk twistor space. The origin of the second additional constraint, denoted $\mathcal{Q}$, is less clear from this twistorial perspective. String theories in twistor space have been considered before \cite{Witten:2003nn,Berkovits:2004hg,Berkovits:2004jj} but the physical principles underlying the $k=1$ string seem very different. In particular, no analogue of the ${\cal Q}$ constraint appears in the earlier twistor string constructions. This raises a wider question of what we can learn about constructing string theories in twistor space from the $k=1$ string.

The goal of this paper is to address some of these conceptual issues and to provide a more convincing rationale for this string theory as a sigma model. Putting this string on a firm conceptual basis is important if we are to take this construction as a prototype for other examples, such as $AdS_5\times S^5$ \cite{Gaberdiel:2021qbb,Gaberdiel:2021jrv}. Without a clear understanding of the physical interpretation of the $k=1$ sigma model in its own right, a generalisation to $AdS_5\times S^5$ appears to be difficult, since there is no currently known analogue of the hybrid formalism which can be applied to \cite{Gaberdiel:2021qbb,Gaberdiel:2021jrv}. We shall begin with a brief summary of the $k=1$ string before summarising the results in this paper. More details of the $k=1$ string may be found in \cite{Eberhardt:2018ouy,Eberhardt:2019ywk,Dei:2020zui}.\\

The sigma model is given by the hybrid formalism of \cite{Berkovits:1999im}, which consists of a WZW model of the superalgebra $\mathfrak{psu}(1,1|2)_1$, combined with a topologically twisted $T^4$, plus ghosts. The ghost sector contains the usual $(b,c)$ ghosts associated to reparameterisation invariance, as well as a scalar field $\rho$ --- a linear dilaton with central charge $c = +28$. This WZW model can be rewritten in terms of free fields, as we shall outline in \S\ref{sec:twistor_sigma_model}, leading to the action
\begin{equation*}
    S=\frac{1}{2\pi}\int_{\Sigma} \mathrm{d}^2 z \Big(\omega_{\alpha}\bar{D}\lambda^{\alpha} - \psi_{A}\bar{D}\chi^{A} - b\bar{\p}c - \frac{1}{2}\p\rho\bar{\p}\rho + \dots \Big).
\end{equation*}
The dots refer to the anti-holomorphic terms, as well as the (topologically twisted) $T^4$ fields. $(\omega_{\alpha},\lambda^{\alpha})$ are bosonic fields and $(\psi_A,\chi^A)$ are fermionic fields that we shall introduce in the main text. They are acted upon by a covariant derivative $\bar{D} = \bar{\p} + \tilde{A}$, which imposes a scaling constraint on the fields through the gauge field $\tilde{A}$. We will often bosonise the $(b,c)$ ghosts as $b = e^{-i\sigma}$ and $c = e^{i\sigma}$. The holomorphic BRST current is of the form
\begin{equation}\label{eq:hybrid_BRST}
    j_{BRST} = e^{i\sigma} \mathcal{T} + e^{-\rho}\mathcal{Q} + \dots
\end{equation}
where the dots refer to the compact terms and any ghost contributions for the scaling constraint. As one might expect, $\mathcal{T}$ is the stress tensor, whilst $\mathcal{Q} = (\chi)^2D\omega$, for a projective measure $D\omega$ --- see \eqref{eq:Q}. We aim to clarify the details of this sigma model from the perspective of twistor geometry in the main text.
In summary, our main motivation for this paper is to make progress in answering the following questions. Firstly, can we understand the $k=1$ string as the natural (and simplest) Type IIB sigma model embedding into the (mini-)twistor space of $AdS_3$, compactified on $S^3\times\mathcal{M}_4$? Secondly, using twistor methods, can we understand the extent to which the $k=1$ string depends on the bulk geometry of $AdS_3$? Finally, what can we learn about building string theories in twistor space from the hybrid formalism? In particular, what role does the ${\cal Q}$ constraint play?
The main results of this paper are as follows:
\begin{enumerate}
    \item We construct, from first principles, a sigma model embedding into the minitwistor space of $AdS_3$. We shall show that the correct framework for the sigma model involves two copies of the (mini-)twistor space of the bulk and the projective scaling constraint arises as the natural quadric condition that defines the ambitwistor space of the four-dimensional embedding space. We propose a novel, stringy,  incidence relation which, in contrast to other twistor string constructions \cite{Witten:2003nn,Berkovits:2004hg}, mixes left- and right-moving degrees of freedom. The physical states of the sigma model are found to live directly at the boundary of the $AdS_3$ spacetime, and the purely (anti-)holomorphic incidence relations of \cite{Dei:2020zui} are recovered in this near-boundary limit. We show that, inside the path integral, the entire worldsheet is localised to the boundary of $AdS_3$, clarifying the extent to which the $k=1$ string is constrained to live at the boundary of the spacetime. That the natural bulk incidence relations are not purely (anti-)holomorphic is reminiscent of the proposal in \cite{Bhat:2021dez} for $AdS_5$, where non-holomorphic incidence relations are also expected.
    \item We find evidence that there exists a ``doubly supersymmetric'' formulation of the $k=1$ string that gives a natural motivation for the presence of the $\rho$ ghost and the imposition of the constraint $\mathcal{Q}=0$. This leads to a critical sigma model on minitwistor space in a different manner compared to that of other twistor string constructions \cite{Berkovits:2004hg}.
    \item The vacuum vertex operators of the $k=1$ string have been studied extensively and are naturally defined in terms of the boundary data $\{x_i,w_i\}$, where $x_i$ is the location of a puncture and $w_i$ is the spectral flow. These vertex operators are recast in terms of the twistors and an integral transform is applied to recover the bulk-boundary propagators using our novel bulk incidence relation. This demonstrates how the $k=1$ string encodes information about the bulk spacetime geometry, despite its physical states being localised to the boundary of the spacetime.
    \item It is shown that vertex operators of the $k=1$ string with zero spectral flow decouple from the theory.
\end{enumerate}
Taken together, these results explain much of the structure of the $k=1$ string as natural and bring us close to a derivation of the theory as a sigma model on minitwistor space. This is in contrast to the lengthy field redefinitions that take us from the RNS string to the hybrid string. We note that there is not an RNS description of the $k=1$ string and so the results presented here provide an additional, independent, motivation for applying the hybrid formalism at $k=1$. This builds on work that was initiated in \cite{McStay:2023thk}. Since our interest is in the intersection of string theory and twistor theory, we shall only focus on the $k=1$ case, where supertwistors seem to be the ideal variables for describing the shortened representations that appear there. There is no evidence, as yet, that twistors will provide useful descriptions for $k>1$, where the physics seems to be qualitatively different.\\

Broadly, the paper is split into three parts. Firstly, we construct twistor sigma models of $AdS_3$ in \S\ref{sec:twistor_sigma_model} and \S\ref{sec:euclidean_AdS3}. We then move on to motivating the details of the hybrid formalism of \cite{Berkovits:1999im} at $k=1$ in \S\ref{sec:PBGS} and \S\ref{sec:Twistor_string} from the perspective of twistor theory. Lastly, we turn our attention to studying the vertex operators of the $k=1$ string in \S\ref{sec:bulk_VO} and \S\ref{sec:unflowed}, in light of our earlier discussions. A more detailed overview is as follows.

The starting point in  \S\ref{sec:twistor_sigma_model} is an alternative formulation for the $SL(2;\C)$ WZW theory. We propose the incidence relations
$$
\omega_{\alpha}=g_{\alpha\dot{\alpha}}\pi^{\dot{\alpha}},	\qquad		\mu_{\dot{\alpha}}=-\lambda^{\alpha}g_{\alpha\dot{\alpha}},
$$
where $g\in SL(2;\C)$ and $(\alpha,\dot{\alpha})$ are indices that label two-component spinors transforming under conjugate representations. We show that the requirement that the incidence relations are preserved by the natural left- and right-actions of $SL(2;\C)$ give right- and left-invariant currents, that generate the left- and right-actions, respectively. These are 
$(J_L)_{\alpha}{}^{\beta}\sim \omega_{\alpha}\lambda^{\beta}$ and 		$(J_R)_{\dot{\alpha}}{}^{\dot{\beta}}\sim\mu_{\dot{\alpha}}\pi^{\dot{\beta}}.$
Comparison with the WZW model suggests $\omega_{\alpha}$ and $\lambda^{\alpha}$ are holomorphic on-shell, whilst $\mu_{\dot{\alpha}}$ and $\pi^{\dot{\alpha}}$ should be anti-holomorphic on-shell. This is reproduced by the action
$$
    S=\frac{1}{2\pi}\int_{\Sigma} \mathrm{d}^2 z \Big(\omega_{\alpha}\bar{\partial}\lambda^{\alpha} + \pi^{\dot{\alpha}}  \partial\mu_{\dot{\alpha}}\Big).
$$
This is the $AdS_3$ content of the $k=1$ theory. In \S\ref{sec:euclidean_AdS3}, we then specialise to the case of Euclidean $AdS_3$ (or $H_3^+$) by introducing a notion of Hermitian conjugation and making the identifications $\bar{\omega}_{\dot{\alpha}} = \mu_{\dot{\alpha}}$ and $\bar{\pi}^{\alpha} = -\lambda^{\alpha}$. In this signature, the condition that the currents $J_L$ and $J_R$ are independently traceless is equivalent to the theory being constrained to the boundary; however, the signature condition only requires the weaker condition that the sum of the traces vanish. In this way we show how, in the Euclidean case, the physical states are constrained to live at the boundary. In the Euclidean $AdS_3$ setting, we work in terms of an explicit parameterisation of $SL(2;\C)/SU(2)$. By taking a near-boundary limit, we recover the boundary incidence relations,
$$
\omega_+-x\omega_-=0,
$$
that are the purely (anti-)holomorphic incidence relations that have appeared widely in the $k=1$ string literature \cite{Dei:2020zui,Knighton:2020kuh,Knighton:2023mhq,Bhat:2021dez}.

The constructions in \S\ref{sec:twistor_sigma_model} and \S\ref{sec:euclidean_AdS3} give the matter sector of the $k=1$ string and some, but not all, of the constraints that define the full theory. In \S\ref{sec:PBGS}, we begin to address the issue of understanding where the additional constraints come from. The constraint $\mathcal{Q}=0$ from \eqref{eq:hybrid_BRST} is of particular interest, along with the role of the additional ghost $\rho $ and the topologically twisted $\mathcal{M}_4$ sector. It has been suspected that there is a way of formulating the $k=1$ string with some form of (possibly twisted) ${\cal N}=2$ worldsheet supersymmetry \cite{Gaberdiel:2021qbb,Gaberdiel:2021jrv}. In this section, we show how this may come about. The left-moving sector of the $k=1$ string is written in terms of homogeneous coordinates $(\omega,\lambda)$, which transform linearly under $SL(2)$. A simple illustration of this is the linear transformation of the homogeneous coordinates $(\omega_+,\omega_-)$ on $\C\P^1$ given by
$$
\left(\begin{array}{c}
     \omega_+  \\
     \omega_-
\end{array}\right)\mapsto \left(\begin{array}{cc}
   1  & \alpha \\
0 & 1
\end{array}\right)\left(\begin{array}{c}
     \omega_+  \\
     \omega_-
\end{array}\right)=\left(\begin{array}{c}
     \omega_+ + \alpha \omega_-\\
     \omega_-
\end{array}\right).
$$
There is also a scaling symmetry that preserves the ratio $\gamma:=\omega_+/\omega_-$, which is identified with a holomorphic coordinate at the boundary. In terms of this scale invariant coordinate, this transformation is just the translation $\gamma\mapsto \gamma + \alpha$. The transformation acts linearly on the $\omega_{\alpha}$ but non-linearly on the scaling-invariant $\gamma$. We apply this mechanism to the global symmetry algebra and deduce that there is a partial breaking of the global supersymmetry, along the lines described in \cite{Hughes:1986dn}. We explain how the resulting effective theory is expected to have a local ${\cal N}=2$ symmetry which can be gauge-fixed. Without knowing the exact form of the effective theory, we can still say something about it. In \S\ref{sec:Twistor_string}, we argue that, under the assumption that there exists some form of picture changing operator (PCO) in the theory, symmetry constrains it to have a particular form. Furthermore, the existence of such a PCO requires additional ghost sectors which the $\rho$-ghost, supplemented with the topologically twisted $\mathcal{M}_4$ fields, naturally provide. Finally, we show that the existence of the PCO imposes an additional constraint on the physical space. We show that, subject to very mild assumptions, the constraint is precisely ${\cal Q}$ and that the BRST operator picks up an extra term $e^{-\rho}{\cal Q}$, as predicted by the hybrid formalism. Thus we show, with very mild assumptions, that the $k=1$ string is the natural Type IIB string theory embedding into the minitwistor space of $AdS_3$.

In \S\ref{sec:bulk_VO}, we describe how to write projective vertex operators in terms of the bulk (mini-)twistor variables and introduce a map from the twistor space to the bulk spacetime, using the bulk incidence relations. This demonstrates how information about the bulk geometry is encoded at the boundary of spacetime in the $k=1$ string.

\S\ref{sec:unflowed} is a standalone section, which sits somewhat outside the general development of the paper. It has been noted \cite{Eberhardt:2018ouy} that states with no spectral flow exist in the $k=1$ string that provide negative norm, ghost states. Utilising our knowledge of the PCOs and namely, $\mathcal{Q}$, we demonstrate that in fact, these states decouple from the theory. The result is of interest and fills a hole in the literature but the details of this section may be skipped by the overburdened reader.

We conclude with a discussion in \S\ref{sec:discussion}. The Appendices contain further background information and technical details.

\section{Twistor sigma models for complexified $\mathbf{AdS_3}$}\label{sec:twistor_sigma_model}

Good reviews of twistor theory can be found in \cite{Penrose:1986ca,Woodhouse:1985id, Adamo:2017qyl}; we briefly summarise only the most salient points here. The projective twistor space for four-dimensional complexified Minkowski space, $\C^4$, is given by an open set $\P\T = \C\P^3\backslash\C\P^1$, where the removed $\C\P^1$ corresponds to the point at infinity. Twistors $Z^I=(\omega_{\alpha},\pi^{\dot{\alpha}})$, where $\alpha,\dot{\alpha}=1,2$ transform as a $\mathbf{4}$ under the complexified conformal group $SU(4)$, are homogeneous coordinates on $\P\T$ and are related to points $X_{\alpha\dot{\alpha}}$ in complexified Minkowski space by the incidence relation
$$
\omega_{\alpha}=X_{\alpha\dot{\alpha}}\pi^{\dot{\alpha}}.
$$
Particular spacetime signatures can be achieved through imposing appropriate reality conditions on the twistor space and we review this procedure in Appendix \ref{sec:twistors}. The real utility of twistor theory is apparent in the Penrose transform --- a map that relates (\^Cech or Dolbeault) cohomology representatives on twistor space with solutions to massless equations of motion on spacetime. Whilst twistor space will play a major role throughout this text, the details of the Penrose transform will not be needed until \S\ref{sec:bulk_physics}, where the necessary background material will be introduced.

To see where $AdS_3$ fits in to this picture, we note that real $AdS_3$ can be realised as the universal cover of a quadric in $\R^{2,2}$  \cite{Maldacena:2000hw}. More generally, the quadric ${\cal X}_n=\{X\in\R^{4-n,n}|X^2=1\}$, where the inner product is given by the natural metric on $\R^{4-n,n}$, describes $S^3$, $H_3^+$ and $AdS_3$,  for $n=0$, $1$ and $2$ respectively. Here $H_3^+$ refers to three-dimensional hyperbolic space or Euclidean $AdS_3$. This is easily illustrated as follows. For a Lorentzian metric in the embedding space we can write $H_3^+$ as the set of matrices 
$$
X_{\alpha\dot{\alpha}}=\left(\begin{array}{cc}
  X_0+X_3   & X_1+iX_2 \\
   X_1-iX_2  & X_0-X_3
\end{array}\right),
$$
such that $\text{det}(X_{\alpha\dot{\alpha}})=X_0^2-X_1^2-X_2^2-X_3^2=1$. The $X_{\mu}$ are taken to be real and we note that $X\in SL(2;\C)/SU(2)$. By contrast, we find Lorentzian $AdS_3$ from the same constraint on the matrices
$$
X_{\alpha\dot{\alpha}}=\left(\begin{array}{cc}
  X_0+X_3   & X_1-X_2 \\
   X_1+X_2  & X_0-X_3
\end{array}\right),
$$
which we identify as elements of $SL(2;\R)$. The complexification to $SL(2;\C)$ thus includes both cases.

As $H_3^+$ (more generally $\mathcal{X}_n$) may be realised as a quadric inside a real slice of $\C^4$ there is a natural way of giving a twistorial description to $H_3^+$ in terms of this embedding \cite{Bailey:1998zif,Adamo:2016rtr,Bu:2023cef}. As such, we will generally work with the complexification of $H_3^+$ (i.e. $SL(2;\C)$) taking the view that our results may be applied to a family of real spacetimes. We will be explicit with our choice of reality condition if we want to discuss the specifics of, say, real $H_3^+$.

In this context, the relevant twistor space is often called the minitwistor space of complexified $H_3^+$, given by
$$
\M\T = (\C\P^1 \times \C\P^1) \backslash \C\P^1_{\mathfrak{b}}.
$$
This is the space of oriented geodesics in complexified $H_3^+$. In terms of real $H_3^+$, one can think of each $\C\P^1$ as a copy of the conformal boundary $S^2$ of $H_3^+$ and so $\C\P^1 \times \C\P^1$ describes the space of ordered pairs of end points of these geodesics\footnote{In Lorentzian $AdS_3$ with conformal boundary $\R \times S^1$, timelike geodesics do not extend out to the conformal boundary due to the negative curvature of $AdS_3$. Instead, they oscillate between a maximal and minimal value of the radial coordinate in the bulk \cite{Maldacena:2000hw}. However, in the compactified geometry with conformal boundary (Lorentzian) $S^2$, these geodesics start and end at the poles of the sphere. Indeed, all geodesics in the complexified, compactified geometry extend to the boundary.}. The removed $\C\P^1_{\mathfrak{b}}$ is where the endpoints of the geodesic coincide and so is a copy of the boundary. We parameterise the two $\C\P^1$s in $\M\T$ by projective coordinates $(\omega_{\alpha}, \pi^{\dot{\alpha}})$. On the real slice corresponding to real $H_3^+$, there is a natural complex conjugation and the boundary is identified as the locus where $\langle \omega\bar{\pi}\rangle:= \omega_{\alpha}\bar{\pi}^{\alpha} = \epsilon^{\alpha\beta}\omega_{\alpha} \bar{\pi}_{\beta} = 0$. We refer to Appendix \ref{sec:twistors} for further details. A closely related concept is the ambitwistor space of the conformal boundary,
$$
\mathbb{A}(S^2) = \C\P^1 \times \C\P^1.
$$
In two dimensions, a compactified, conformally flat spacetime with a choice of complex structure is equivalent to its twistor space. And so the ambitwistor space of the boundary (a $\C\P^1$) is just $\C\P^1 \times \C\P^1$, with coordinates on each factor identified as twistors and dual-twistors of the boundary, respectively.

The close relationship between $\M\T$ and $\mathbb{A}(S^2)$ has a higher-dimensional analogue associating the twistor space of $AdS_5$ with the ambitwistor space of $S^4$. In this setting, the quadric \cite{Adamo:2016rtr}
\begin{equation}\label{eq:AdS5_twistorspace}
    \P\T(AdS_5) = \mathbb{A}(S^4) = \{(\omega_A,\pi^B)\in \C\P^3\times \C\P^3| Z\cdot W = 0\},
\end{equation}
describes both the twistor space of complexified $AdS_5$ and the ambitwistor space of its boundary. It is in this sense that we view the sigma model on minitwistor space presented below as a toy model for $AdS_5/CFT_4$.
\\

In this section, we will propose a natural sigma model embedding into minitwistor space. Complexified $H_3^+$ is given by the group manifold $SL(2;\C)$ and this is the group we will have in mind, but much of what we discuss has a natural generalization to other groups. The left action of $SL(2;\C)$ acts on one $\C\P^1$ and the right action acts on the other. As with WZW models, we shall see a natural on-shell association of the left (right) group action to the left (right) moving sector on the worldsheet theory. We shall explicitly outline the nature of the incidence relations for complexified $H_3^+$, building on the work of \cite{McStay:2023thk}. See also \cite{Adamo:2016rtr} for the analogous twistor geometry of $AdS_5$. In \S\ref{sec:euclidean_AdS3} we will specialise to the case of real Euclidean $AdS_3$, i.e. the coset $H_3^+=SL(2;\C)/SU(2)$.

The sigma model that we shall build will ultimately be equivalent to the free field realisation of \cite{Dei:2020zui} for the $k=1$ string, clarifying the geometric interpretation of this theory as an embedding into minitwistor space. In particular, the complexification of the free field realisation of the $\mathfrak{sl}(2;\R)_1$ WZW model that is applied in \cite{Dei:2020zui} will be understood as components of the bulk minitwistor.

\subsection{The bulk target space}

Consider a group $G \leq SL(2;\C)$. The elements of the group $g_{\alpha\dot{\alpha}} \in G$ carry $SL(2;\C)$ spinor indices. Our conventions will be to raise and lower indices using the conventional antisymmetric tensors $\epsilon^{\alpha\beta}$ and $\epsilon^{\dot{\alpha}\dot{\beta}}$, with $\alpha,\beta=\pm$ and $\dot{\alpha},\dot{\beta}=\pm$, and where $\epsilon^{+-} = -\epsilon_{+-} = +1$. For undotted spinors, we define the contraction $\langle \omega \lambda \rangle = \omega_{\alpha} \lambda^{\alpha}$, whilst for dotted spinors we define $[\mu\pi] = \mu_{\dot{\alpha}}\pi^{\dot{\alpha}}$. In the contraction with a spacetime coordinate, we define $\langle \lambda g \pi ] = \lambda^{\alpha}g_{\alpha\dot{\alpha}}\pi^{\dot{\alpha}}$.\\

A group $G$ comes with a natural left- and right-action. In the sigma model we would like to construct, with embedding field $g: \Sigma \to G$, we shall associate these with global symmetries of the theory. We therefore denote the isometries by $G_L \times G_R$, with
$$
G_L:g\mapsto h_Lg,	\qquad		G_R:g\mapsto gh_R,
$$
for $h_L \in G_L$ and $h_R \in G_R$. More precisely, we write
\begin{equation}\label{eq:WZW_isometries}
    g_{\alpha\dot{\alpha}}\mapsto (h_L)_{\alpha}{}^{\beta}g_{\beta\dot{\beta}}(h_R)^{\dot{\beta}}{}_{\dot{\alpha}}.
\end{equation}
We define the doubled group to be $\mathcal{G} = G_L \times G_R$, written as
$$
\left(
\begin{array}{cc}
 h_L & 0 \\
  0 & h_R
\end{array}
\right)\in {\cal G},
$$
for which we may define a diagonal and anti-diagonal embedding of a copy of $G$ into ${\cal G}$ via
$$
H(h)=\left(
\begin{array}{cc}
h & 0 \\
0 & h
\end{array}
\right),	\qquad		\widetilde{H}(h)=\left(
\begin{array}{cc}
h^{-1} & 0 \\
0 & h
\end{array}
\right).
$$
We would now like to introduce a spinor bundle over spacetime for each of $G_L$ and $G_R$. These descend directly from the spinors in the ambient four-dimensional (complexified) Minkowski space. That is, we define a pair of correspondence spaces $F$ and $F'$ with local sections following the usual twistor constructions \cite{Adamo:2017qyl},
$$
(g_{\alpha\dot{\alpha}},\pi^{\dot{\alpha}})\in F,	\qquad		(g_{\alpha\dot{\alpha}},\lambda^{\alpha})\in F',
$$
and projection $F\rightarrow G$ is given by the forgetful map $(g,\pi)\mapsto g$ and similarly for $F'$ (recall that $G_L$, $G_R \cong G$). From this we build the bundle ${\cal F}\rightarrow G$ defined by sections $(g_{\alpha\dot{\alpha}},\pi^{\dot{\alpha}},\lambda^{\alpha}) \in \mathcal{F}$. That is, the fibres of $\mathcal{F}$ are the fibres of $F$ and $F'$ together. $\mathcal{F}$ then defines the doubled correspondence space.

With these ingredients we can write down two incidence relations in analogy with Minkowski space \cite{Bailey:1998zif,Adamo:2016rtr,Bu:2023cef},
\begin{equation}\label{eq:incidence}
    \omega_{\alpha}=g_{\alpha\dot{\alpha}}\pi^{\dot{\alpha}},	\qquad		\mu_{\dot{\alpha}}=-\lambda^{\alpha}g_{\alpha\dot{\alpha}}.
\end{equation}
This describes a natural pair of twistors and dual twistors
\begin{equation}\label{eq:twistor_and_dual}
    Z^I=\left(
\begin{array}{c}
\omega_{\alpha}	\\
\pi^{\dot{\alpha}}
\end{array}
\right),	\qquad	
W_I=\left(
\begin{array}{cc}
\lambda^{\alpha}	&,
\mu_{\dot{\alpha}}
\end{array}
\right).
\end{equation}
Thus far, this very much mirrors the usual twistor and dual twistor constructions for four-dimensional (complexified) Minkowski space. Compared to complexified Minkowski space, $\C^4$, for which we would have $X_{\alpha\dot
 \alpha} \in M(2;\C)$ (the space of $2\times 2$ complex matrices), there are additional constraints on $g_{\alpha\dot{\alpha}} \in G$. For $G =SL(2;\C)$ we have a constraint on the determinant, whilst if one was interested in subspaces of complexified $AdS_3$, further restrictions to $G< SL(2;\C)$ could be applied\footnote{For example, one could apply our general prescription here to other hyperbolic 3-manifolds that are quotients of (complexified) global $AdS_3$ by a discrete subgroup $\Gamma \leq PSL(2;\C)$. The minimal tension limit on such spaces was studied in \cite{Eberhardt:2021jvj}.}. This restricts $(\omega_{\alpha},\pi^{\dot{\alpha}})$ to live in a subspace of $\P\T = \C\P^3\backslash\C\P^1$. A simple way to determine this subspace for $G = SL(2;\C)$ is as follows. As described in \cite{Adamo:2016rtr}, instead of fixing the determinant of $g_{\alpha\dot{\alpha}}$ to be one, one can instead take $g_{\alpha\dot{\alpha}} \in GL(2;\C)$ (such that it has non-zero determinant) and impose a scaling condition via $g_{\alpha\dot{\alpha}} \sim r g_{\alpha\dot{\alpha}}$ for $r \in \C\backslash \{0\}$. This then allows $\omega_{\alpha}$ and $\pi^{\dot{\alpha}}$ to scale independently. Thus, rather than taking values in $\P\T$, our twistors and dual twistors for complexified $AdS_3$ take values in $(\C\P^1 \times \C\P^1)\backslash\C\P^1$. We can use these incidence relations to define a projection from the correspondence spaces $F$ and $F'$ onto the twistor and dual twistor spaces.

We can also write down equivalent incidence relations for the inverse of $g_{\alpha\dot{\alpha}}$,
\begin{equation}\label{eq:incidence_g_inverse}
\pi^{\dot{\alpha}} = \omega_{\alpha} g^{\alpha\dot{\alpha}}, \qquad \lambda^{\alpha} = -g^{\alpha\dot{\alpha}}\mu_{\dot{\alpha}},
\end{equation}
where $\left(g^{-1}\right)^{\dot{\alpha}\alpha} = g^{\alpha\dot{\alpha}}$. The raising of indices is defined using the $\epsilon$ tensors. In particular, one can check that $g^{\alpha\dot{\alpha}}g_{\alpha\dot{\beta}} =  \delta^{\dot{\alpha}}{}_{\dot{\beta}} \det g$ and similarly for the contraction on dotted indices. For $g_{\alpha\dot{\alpha}} \in SL(2;\C)$, we do not collect any factors of $\det g$. This means that the roles of $\omega$ and $\pi$ as well as $\mu$ and $\lambda$ can be simultaneously swapped.

Note that, on the support of the incidence relations,
\begin{equation}\label{eq:ambitwistor_quadric}
    W\cdot Z=\omega_{\alpha}\lambda^{\alpha}+\mu_{\dot{\alpha}}\pi^{\dot{\alpha}}=0,
\end{equation}
identically vanishes. $W\cdot Z = 0$ has a natural geometric interpretation as the quadric condition that defines the ambitwistor space of $\C^4$ (see \cite{Adamo:2017qyl}). $\C^4$ is the embedding space for $AdS_3$, so it is unsurprising that this constraint is satisfied by the (mini-)twistors and dual (mini-)twistors of $AdS_3$. We shall nevertheless refer to the bulk target space as the minitwistor space of $AdS_3$ in this work, even though we shall often be describing $Z$ and $W$. This is because a stronger version of the constraint $W\cdot Z = 0$ will appear when we consider the current algebra on the worldsheet in \S\ref{sec:bosonic_sector}. In that case, it will be natural to impose $\omega_{\alpha}\lambda^{\alpha} = \mu_{\dot{\alpha}}\pi^{\dot{\alpha}} = 0$, which identifies the $\C\P^1$s described by the twistors and dual twistors --- see \eqref{eq:U_constraint}.\\

These incidence relations describe the map from spacetime, $G$, to the doubled twistor space\footnote{A slight generalisation of (\ref{eq:incidence}) is $\omega=g_L\cdot\pi$ and $\mu=-\lambda\cdot g_R$, where we do not assume a priori that $g_L=g_R$. In the language of \cite{Hull:2009sg}, the spacetime is recovered by imposing the polarisation constraint $W\cdot Z=\langle \lambda(g_L-g_R)\pi]=0$, which selects the diagonal embedding of $G$ into the doubled group ${\cal G}$. }, defined by the coordinates $Z^I$ and $W_I$ such that $Z\cdot W = 0$. This map is non-local. As usual, a particular point $g \in G$ corresponds to two $\C\P^1$s in the doubled twistor space (one for $Z^I$ and the other for $W_I$). We can also understand the map the other way. Consider a point $Z^I \in \C\P^1\times \C\P^1$. The most general solution for $g_{\alpha\dot{\alpha}}$ satisfying the first incidence relation in \eqref{eq:incidence} is given by
\begin{equation}\label{eq:inverse_incidence} g_{\alpha\dot{\alpha}}=\frac{\omega_{\alpha}A_{\dot{\alpha}}+B_{\alpha}\pi_{\dot{\alpha}}}{[A\pi]},
\end{equation}
for arbitrary $A_{\dot{\alpha}}$ and $B_{\alpha}$, where $[A\pi] \neq 0$. As with $\C^4$ this appears to be the equation of a null, self-dual $\alpha$-plane, with variations in $A_{\dot{\alpha}}$ and $B_{\alpha}$ generating tangent vectors of the form $C_{\alpha}\pi_{\dot{\alpha}}$ for some spinor $C_{\alpha}$ \cite{Adamo:2017qyl}. However, we must also impose the constraint that $\det g = 1$ which removes an overall scale factor from this solution. As such, we restrict to the intersection of the $\alpha$-plane with the embedding of $AdS_3$ inside $\C^4$.

Similarly, a point $W^I \in \C\P^1 \times \C\P^1$ corresponds to a null, anti-self-dual $\beta$-plane in $\C^4$, which also restricts to the intersection with $AdS_3$. If we extend this to consider a point in the full doubled twistor space, a point $(Z^I,W_I)$ uniquely specifies a point in $G$, since $A_{\dot{\alpha}}$ and $B_{\alpha}$ are no longer arbitrary in \eqref{eq:inverse_incidence}, but are given by $\mu_{\dot{\alpha}}$ and $-\lambda_{\alpha}$. That is to say, the self-dual and anti-self-dual planes intersect in a null line, and this restricts to a point in $AdS_3$. Symmetrically, we may write
$$
g_{\alpha\dot{\alpha}}=\frac{-\omega_{\alpha}\mu_{\dot{\alpha}}+\lambda_{\alpha}\pi_{\dot{\alpha}}}{\langle\omega\lambda\rangle} = \frac{\omega_{\alpha}\mu_{\dot{\alpha}}-\lambda_{\alpha}\pi_{\dot{\alpha}}}{[\mu\pi]},
$$
since $-\mu_{\dot{\alpha}}\pi^{\dot{\alpha}} = \omega_{\alpha}\lambda^{\alpha} = \lambda^{\alpha}g_{\alpha\dot{\alpha}}\pi^{\dot{\alpha}}$. The denominator is non-vanishing in the bulk of $AdS_3$. We will see in \S\ref{sec:Boundary_spacetime} that, for Euclidean signature, the vanishing of $\langle\omega \lambda \rangle = [\mu\pi] = 0$ defines the boundary. \\

In order for the incidence relations (\ref{eq:incidence}) to be preserved under the left and right isometries of the group
\eqref{eq:WZW_isometries}, the spinors transform as
\begin{alignat*}{2}
   &\omega_{\alpha} \mapsto (h_L)_{\alpha}{}^{\beta}\omega_{\beta}, \qquad &&		\pi^{\dot{\alpha}} \mapsto (h^{-1}_R)^{\dot{\alpha}}{}_{\dot{\beta}}\pi^{\dot{\beta}},\\
   &\mu_{\dot{\alpha}} \mapsto \mu_{\dot{\beta}}(h_R)^{\dot{\beta}}{}_{\dot{\alpha}},	\qquad &&		\lambda^{\alpha} \mapsto \lambda^{\beta}(h^{-1}_L)_{\beta}{}^{\alpha}.
\end{alignat*}
In terms of the twistors, we write this as
\begin{eqnarray*}
Z^I=\left(
\begin{array}{c}
\omega	\\
\pi
\end{array}
\right)\mapsto \left(
\begin{array}{cc}
h_L & 0 \\
0 & h_R^{-1}
\end{array}
\right)\left(
\begin{array}{c}
\omega	\\
\pi
\end{array}
\right),	\qquad	
W_I=\left(
\begin{array}{cc}
\lambda	&,
\mu
\end{array}
\right)\mapsto \left(
\begin{array}{cc}
\lambda	&,
\mu
\end{array}
\right)\left(
\begin{array}{cc}
h_L^{-1} & 0 \\
0 & h_R
\end{array}
\right).	
\end{eqnarray*}
The natural inner product $W\cdot Z =W_IZ^I$ is invariant under the action of $G_L \times G_R$. As such, this inner product is well-defined both on $G$ and quotients of $G$ by subgroups of $G_L\times G_R$.

\subsection{The twistor sigma model}\label{sec:almost}

\subsubsection{The bosonic sector}\label{sec:bosonic_sector}

We now use the above framework to build a sigma model on complexified $AdS_3$. String theory embedding into the group manifold of $AdS_3$ is described by a WZW model \cite{DiFrancesco:1997nk} with conserved currents for both left and right group actions. In particular, for a group element $g(z,\bar{z}) \in G$, the equations of motion specify that on-shell $g(z,\bar{z}) = g_L(z)g_R(\bar{z})$. The conserved currents are then given by $J_L(z) = \p g g^{-1} = \p g_L g_L^{-1}$ and $J_R(\bar{z}) = g^{-1}\bar{\p}g = g_R^{-1}\bar{\p}g_R$. The incidence relations demonstrated that the twistor variables naturally transform under either the left- or right-action of the group. As such, we can define a free field realisation of these currents,
\begin{equation}\label{eq:J}
(J_L)_{\alpha}{}^{\beta}= \omega_{\alpha}\lambda^{\beta},	\qquad		(J_R)_{\dot{\alpha}}{}^{\dot{\beta}}=\mu_{\dot{\alpha}}\pi^{\dot{\beta}}.
\end{equation}
We will constrain these bilinears to be traceless, since this is the condition on the Lie algebra analogous to the fact that $g \in G$ has unit determinant. This means we impose the constraints
$$
U= \langle\omega\lambda \rangle=0, \qquad  \tilde{U}= [\mu\pi] = 0.
$$
Consistency with the WZW model suggests $\bar{\partial}J_L=0$ and $\partial J_R=0$. It is therefore natural to take $\omega_{\alpha}$ and $\lambda^{\alpha}$ to be holomorphic on-shell, whilst $\mu_{\dot{\alpha}}$ and $\pi^{\dot{\alpha}}$ should be anti-holomorphic on-shell. We are then lead to introduce the action
\begin{equation}\label{eq:FFR_action}
    S=\frac{1}{2\pi}\int_{\Sigma} \mathrm{d}^2 z \Big(\omega_{\alpha}\bar{\partial}\lambda^{\alpha} + \pi^{\dot{\alpha}}  \partial\mu_{\dot{\alpha}} + A\tilde{U} + \tilde{A}U\Big),
\end{equation}
where $A$ and $\tilde{A}$ are Lagrange multipliers that we gauge fix to zero\footnote{Strictly speaking, the $GL(1)$ gauge symmetries generated by $U$ and $\tilde{U}$ are anomalous at the quantum level \cite{Berkovits:2004jj}. This is not the case for the maximally supersymmetric extension to $AdS_3 \times S^3$, which is the main focus of the paper. We will need to take account of instanton numbers in this setting, however.}. Quantising, the corresponding OPEs are\footnote{
Relative to the typical conventions in the tensionless holography literature \cite{Dei:2020zui,McStay:2023thk}, we have that $\omega_{\alpha} = \xi_{\alpha}$, $\lambda^{\alpha} = \eta^{\alpha}$, $\mu_{\dot{\alpha}} = \bar{\xi}_{\dot{\alpha}}$ and $\pi^{\dot{\alpha}} = -\bar{\eta}^{\dot{\alpha}}$.
}
\begin{equation}\label{eq:bosonic_OPEs}
    \omega_{\alpha}(z)\lambda^{\beta}(w)=  \frac{\delta_{\alpha}{}^{\beta}}{z-w}+...,	\qquad		\mu_{\dot{\alpha}}(\bar{z}) \pi^{\dot{\beta}}(\bar{w})= -\frac{\delta_{\dot{\alpha}}{}^{\dot{\beta}}}{\bar{z}-\bar{w}}+...
\end{equation}

A few comments are in order. On the level of classical geometry, imposing the constraints $U = \tilde{U} = 0$ gives
\begin{equation}\label{eq:U_constraint}
    \gamma := \frac{\omega_+}{\omega_-} = - \frac{\lambda^-}{\lambda^+} \in \C\P^1, \qquad \tilde{\gamma} := \frac{\mu_+}{\mu_-} = -\frac{\pi^-}{\pi^+} \in \C\P^1.
\end{equation}
We think of $\gamma$ and $\tilde{\gamma}$ as local coordinates on each $\C\P^1$ in a chart where $\omega_- \neq 0$ and $\mu_- \neq 0$, respectively. Before imposing any reality conditions, we should view these as honestly independent copies of $\C\P^1$. However, we learn that on the solution of the constraints, our boundary twistors and dual twistors from the left-moving sector realise the same $\C\P^1$, and likewise for the right-moving sector. Overall, combining both sectors, we realise $\C\P^1 \times \C\P^1$. We will demonstrate in the next section (in the case of Euclidean signature) that the constraints $U = \tilde{U} = 0$ localise the physical states to live at the boundary and that \eqref{eq:U_constraint} are equivalent to the boundary incidence relations.

However, we must emphasise here that these are classical statements. In the case of the $k=1$ string on $AdS_3 \times S^3 \times T^4$, only the boundary incidence relations related to $\omega_{\alpha}$ and $\mu_{\dot{\alpha}}$ are promoted to Ward identities\footnote{We thank Vit Sriprachyakul for highlighting this point.}. These imply the localisation of correlation functions in moduli space to points where a covering map exists \cite{Eberhardt:2019ywk,Dei:2020zui,McStay:2023thk}. The distinction between which fields obey a Ward identity is inherently related to the picture changing operator we shall introduce in \S\ref{sec:ground_up}, which acts asymmetrically on twistors and dual twistors.

We should also highlight a rather peculiar feature of the bulk incidence relations,
$$
\omega_{\alpha}=g_{\alpha\dot{\alpha}}\pi^{\dot{\alpha}},	\qquad		\mu_{\dot{\alpha}} =-\lambda^{\alpha}g_{\alpha\dot{\alpha}}.
$$
In light of \eqref{eq:FFR_action}, it appears that they mix the left- and right-moving sectors of the theory. Yet, we emphasise that these incidence relations only hold off-shell in the bulk of $AdS_3$. We will take an appropriate near-boundary limit of these incidence relations in the next section, such that we recover the incidence relations that hold on-shell at $U= \tilde{U} = 0$. These will be purely (anti-)holomorphic statements analogous to \eqref{eq:U_constraint}. Nevertheless, the off-shell incidence relations will be crucial in \S\ref{sec:bulk_VO} where we will demonstrate how the bulk-boundary propagator can be recovered from on-shell boundary states.

As a final comment, the OPEs \eqref{eq:bosonic_OPEs} imply that the currents \eqref{eq:J} of the WZW model realise an affine extension of $G$ at some level $k \sim \frac{1}{\alpha'}$ related to the string tension. Specifically, the twistors realise $\mathfrak{sl}(2;\C)_1$ in the complexified geometry. Through an appropriate reality condition, we could instead find a free field realisation of $\mathfrak{sl}(2;\R)_1 \cong \mathfrak{su}(2)_{-1}$. In this setting, the twistor variables are typically referred to as ``symplectic bosons'' \cite{Gaberdiel:2018rqv,Goddard:1987td,Gaiotto:2017euk,Beem:2023dub}.

\subsubsection{Generalization to supertwistors}

The above construction describes a bosonic WZW model for the group $G$ in terms of twistor variables. We would like to generalise this construction to the supersymmetric theory on $AdS_3 \times S^3$. In the bosonic case, the relationship between the minitwistor space of the bulk and the ambitwistor space of the boundary was crucial. To the knowledge of these authors, however, the supersymmetric generalisation of this relationship has not been determined. Crucially, the incidence relations for the supertwistors of $AdS_3 \times S^3$ have not been determined. We will comment later in \S\ref{sec:further_examples} on a potential avenue for progress in this direction.

The geometry of the supersymmetric theory is therefore most transparently motivated in the boundary ambitwistor theory. We promote the boundary ambitwistors (equivalently the bulk minitwistors), $Z^I$, to superambitwistors, $\mathcal{Z}^m$, and similarly for $W_I$ to $\mathcal{W}_m$ \cite{Ferber:1977qx}, 
\begin{equation}\label{eq:supertwistors}
    \mathcal{Z}^m=\left(
\begin{array}{c}
\omega_{\alpha}	\\
\pi^{\dot{\alpha}} \\
\psi_{A} \\
\eta^{\dot{A}}
\end{array}
\right),	\qquad	
\mathcal{W}_m=\left(
\begin{array}{cccc}
\lambda^{\alpha}	&,
\mu_{\dot{\alpha}}  &,
\chi^{A}   &,
\zeta_{\dot{A}}
\end{array}
\right),
\end{equation}
which should each take values in $\C\P^{1|2} \times \C\P^{1|2}$ and $m$ runs over both the bosonic and fermionic indices.\footnote{
Strictly speaking, we are still referring to complexified geometry. When we later impose reality conditions on the bosons, we will assume there are appropriate reality conditions also for the fermions.
} Note that $\mathcal{Z}^m$ and $\mathcal{W}_m$ transform naturally under the supergroup $PSU(2,2|4)$ which describes the superconformal group of the bulk embedding space, $\C^{4|4}$. The fermions carry a $SU(2)$ fundamental index and can be seen to generate the current algebra $\mathfrak{su}(2)_1$ as fermion bilinears,
$$
K^{\pm}= \chi^{\pm}\psi_{\mp},   \qquad  K^3=\frac{1}{2}:(\chi^+\psi_+ -\chi^-\psi_-):,
$$
with similar expressions for the anti-holomorphic sector. This is the sense in which the $S^3$ of the bulk geometry is realised, since $S^3 \cong SU(2)$. We can extend this realisation to the full supergroup $\mathfrak{psu}(1,1|2)_1$ using bilinears of \eqref{eq:supertwistors} that combine bosons and fermions. The remaining holomorphic currents are
$$
S^{\pm A+}=\pm\omega_{\mp}\chi^A,    \qquad  S^{\alpha \pm -}=\mp\lambda^{\alpha}\psi_{\mp},
$$
with similar expressions in the anti-holomorphic sector. Even though a full understanding of the incidence relations for the bulk minitwistor supergeometry is not yet available, this natural supersymmetric generalisation of the boundary theory realises the supergeometry of $AdS_3\times S^3$, which is the group manifold $PSU(1,1|2)$. This is highly suggestive that $\C\P^{1|2}\times \C\P^{1|2}$ can be thought of as the supersymmetric analogue of the minitwistor space of $AdS_3$. Importantly, we see that the level of the $SU(2)$ current algebra is fixed by the natural generalisation of the twistors to supertwistors.

As with the bosons, we require that $(\psi_{A}, \chi^{A})$ are holomorphic and $(\eta^{\dot{A}},\zeta_{\dot{A}})$ are anti-holomorphic. The matter sector has the action
\begin{equation*}\label{eq:Matter_action}
    S=\frac{1}{2\pi}\int_{\Sigma} \mathrm{d}^2 z \Big(\omega_{\alpha}\bar{\partial}\lambda^{\alpha} + \pi^{\dot{\alpha}}  \partial\mu_{\dot{\alpha}} - \psi_{A}\bar{\p}\chi^{A} - \eta^{\dot{A}}\p \zeta_{\dot{A}} \Big),
\end{equation*}
with the associated OPEs
$$\psi_{A}(z) \chi^{ B }(w) = + \frac{\delta_{A}{}^{ B }}{z-w}+..., \qquad \zeta_{\dot{A}}(\bar{z}) \eta^{\dot{ B }}(\bar{w}) = + \frac{\delta_{\dot{A}}{}^{\dot{ B }}}{\bar{z}-\bar{w}}+...$$
Currently, these fields appear as coordinates on $\C^{2|2}$, so we need to add constraints to the action. We will take the convention that left moving twistors and dual twistors have conformal weight $(\frac{1}{2},0)$ and analogously weight $(0,\frac{1}{2})$ for the right-movers. Focusing on the holomorphic sector, the stress tensor generates a diagonal scaling of
\begin{equation*}\label{eq:holo_supertwistors}
    (\mathcal{Z}_L)^M = \left(
\begin{array}{c}
\omega_{\alpha}	\\
\psi_{A}
\end{array}
\right),	\qquad	
(\mathcal{W}_L)_M = \left(
\begin{array}{cc}
\lambda^{\alpha}	&,
\chi^{A}
\end{array}
\right),
\end{equation*}
and an anti-diagonal scaling is generated by $\mathcal{C}_L = \frac{1}{2}(\lambda^{\alpha}\omega_{\alpha} + \chi^{ A }\psi_{ A })$, where the index $M$ is the restriction of $m$ in \eqref{eq:supertwistors} to the holomorphic components. We therefore generate projective coordinates by gauging these two symmetries. Crucially, the choice to include as many fermions as bosons ensures that the $GL(1)$ gauge symmetry generated by $\mathcal{C}_L$ is non-anomalous. $\mathcal{Z}_L$ and $\mathcal{W}_L$ now take values in $\C\P^{1|2}$ and therefore describe the supertwistors and dual supertwistors of the boundary respectively (such that $\mathcal{Z}$ and $\mathcal{W}$ in \eqref{eq:supertwistors} are indeed the ambitwistors for the boundary superspace). 

Note that our bosonic currents in \eqref{eq:J} are invariant under $\mathcal{C}_L$ (and the anti-holomorphic counterpart $\mathcal{C}_R$). The bosonic part of the constraint $\mathcal{C}_L = 0$ precisely reduces to the constraint $U = 0$ that we imposed on the bosonic currents. We can now extend \eqref{eq:J} to include all bilinears of $(\mathcal{Z}_L)^M$ and $(\mathcal{W}_L)_M$ that are invariant under $\mathcal{C}_L$. We find that our free fields generate a realisation of $\mathfrak{u}(1,1|2)_1$ \cite{Eberhardt:2018ouy,Dei:2020zui}. There are two $\mathfrak{u}(1)$ currents that we would like to quotient out in order to get the correct sigma model for the superspace of $AdS_3 \times S^3$,
\begin{equation}\label{eq:psu_quotient}
    \mathfrak{psu}(1,1|2)_1 \cong \frac{\mathfrak{u}(1,1|2)_1}{\mathfrak{u}(1)_{\mathcal{C}} \oplus \mathfrak{u}(1)_{\mathcal{B}}},
\end{equation}
where $\mathcal{C}_L$ is the gauge symmetry we have already identified, whilst $\mathcal{B}_L = \frac{1}{2}( \lambda^{\alpha}\omega_{\alpha} -\chi^{ A }\psi_{ A })$. The $\mathfrak{psu}(1,1|2)_1$ currents are explicitly stated in Appendix \ref{sec:uniqueness}. Since $AdS_3 \times S^3$ is given by the supergroup manifold $PSU(1,1|2)$, this provides further evidence that $\C\P^{1|2} \times \C\P^{1|2}$ is indeed the supersymmetric generalisation of minitwistor space.

We should highlight that the current $\mathcal{B}_L$ does not generate a gauge symmetry of the theory since it does not commute with any of the supercurrents. Nevertheless, it can be shown that it decouples from the $\mathfrak{psu}(1,1|2)_1$ algebra, such that it contributes merely as an instanton number to the spectrum, labeling isomorphic copies of the theory. By contrast, imposing $\mathcal{C}_L$ as a gauge symmetry will introduce a fermionic ghost system $(u,v)$ where $h(u) = 1$ and $h(v) = 0$.\footnote{
A more thorough study of the ghost system required for imposing the quotient in \eqref{eq:psu_quotient} was given in \cite{Gaberdiel:2022als}, in the aim of realising an $\mathcal{N} = 4$ algebra. Applying techniques from \cite{Karabali:1989dk}, it was argued that two fermionic $(u,v)$ ghost systems of weight $h(u) = 1$ and $h(v) = 0$ as well as a bosonic $(\beta,\gamma)$ ghost system of weight $h(\beta) = 1$ and $h(\gamma) = 0$ should be included. This was as a consequence of the non-trivial algebra generated by the OPE between $\mathcal{C}$ and $\mathcal{B}$. Nevertheless, the precise mechanism for how physical state conditions were imposed on states was not elucidated. We shall ignore such technicalities here, since they remain unresolved.
}

\subsubsection{A comment on criticality}\label{sec:comment_on_criticality}

To elevate this sigma model to a string theory, we couple to worldsheet gravity in the usual way. Gauge-fixing reparameterisation invariance will introduce the usual $(b,c)$ ghosts associated to the stress tensor, $\mathcal{T}$, and we define the bosonisation of these ghosts as $b = e^{-i\sigma}$ and $c = e^{i\sigma}$. Combined with the ghost content for gauge-fixing $\mathcal{C}_L=0$, the holomorphic part of the BRST current is then given by
\begin{equation}\label{eq:BRST_CT}
    j_{BRST}^{\mathcal{C},\mathcal{T}} = e^{i\sigma}\mathcal{T} + v \mathcal{C}_L.
\end{equation}
We can now write down an action for the theory and in doing so, we \emph{almost} arrive at the action for string theory on the twistor space of $AdS_3 \times S^3$,
\begin{equation}\label{eq:CP1_string}
    S = \frac{1}{2\pi}\int_{\Sigma} \mathrm{d}^2 z \Big(\omega_{\alpha}\bar{\partial}\lambda^{\alpha} - \psi_{ A }\bar{\p}\chi^{ A } - b\bar{\p}c - u\bar{\p}v + c.c. \Big).
\end{equation}
Yet, the theory thus far has central charge $c=-28$ and so this cannot be the final story (for this reason, we shall refer to the action \eqref{eq:CP1_string} as the ``non-critical theory'' throughout the remainder of the text). A tempting resolution would be to extend our matter sector by coupling the theory to the current algebra of some group carrying $c = +28$, $C$. In fact, this is the resolution to the analogous problem in \cite{Berkovits:2004hg} for a string theory on the twistor space of $\C^4$, where a $c=-28$ conformal anomaly arises once again from the $(b,c)$-ghosts and the projective scaling of the twistors. If we were to double the number of bosons and fermions in \eqref{eq:CP1_string} and then couple to $C$, this would seemingly replicate the construction of \cite{Berkovits:2004hg}, the holomorphic sector of which is equivalent to the usual twistor string \cite{Witten:2003nn,Berkovits:2004jj}. However, as shall be explained in \S\ref{sec:PBGS} and \S\ref{sec:Twistor_string}, we claim that coupling to such a $C$ is not the correct thing to do in \eqref{eq:CP1_string}.

Ultimately, our alternative resolution will lead to the hybrid formalism of \cite{Berkovits:1999im} for string theory on $AdS_3 \times S^3 \times T^4$. As outlined in the introduction, at $k=1$ this consists of a WZW of $\mathfrak{psu}(1,1|2)_1$ along with a topologically twisted $T^4$ and ghosts. Curiously, in addition to the usual $(b,c)$ ghosts, there exists a further ghost field $\rho$ that is a chiral boson with $c = +28$. This is associated with gauging the pullback of a volume form on the boundary $\C\P^{1|2}$ onto the worldsheet \cite{McStay:2023thk},
\begin{equation}\label{eq:Q}
    \mathcal{Q} =  \frac{1}{2}\epsilon^{\alpha\beta}\epsilon^{ A  B } \chi_{ A } \chi_{ B } \omega_{\alpha} \p \omega_{\beta}.
\end{equation}
This suggests that, rather than adding a current algebra to the non-critical theory to cancel the conformal anomaly, there is instead an additional physical constraint. It is natural to ask why this is the case? That is, why do we not follow the same line of reasoning as in \cite{Berkovits:2004hg}? We will directly address this question in \S\ref{sec:PBGS} and \S\ref{sec:Twistor_string}, arguing why it is indeed necessary to impose the constraint $\mathcal{Q}=0$ and how it is associated with the $\rho$ ghost. Our argument will depend on the existence of a partially broken global supersymmetry algebra and the existence of a picture changing operator in the theory. This will provide intrinsic motivation for viewing the hybrid formalism at $k=1$ as the natural sigma model for describing superstring theory on minitwistor space. For now, we shall concentrate only on the matter sector of the non-critical theory \eqref{eq:CP1_string}. We will also delay a more in-depth discussion of the similarities of \eqref{eq:CP1_string} with the twistor string to \S\ref{sec:discussion}.

\subsection{On the naturalness of supertwistors}

To conclude this section, in light of all the machinery we have developed, we can now address why twistor theory is so useful for describing string theory on $AdS_3\times S^3\times T^4$ at $k=1$. The D1-D5 CFT \cite{Maldacena:1997re} comes with a 20-dimensional moduli space, describing deformations of the symmetric product orbifold CFT Sym$^N(T^4)$. For a long time, it was unclear which string theory on $AdS_3$ corresponds to the free CFT of Sym$^N(T^4)$ \cite{Seiberg:1999xz}. A key problem for the RNS string theory on $AdS_3 \times S^3 \times T^4$ with pure NS-NS flux is that it has a continuous spectrum \cite{Maldacena:2000hw}, whilst the proposed dual free CFT does not have this feature. Moreover, applying the RNS formalism to the minimal tension limit of $k=1$ units of NS-NS flux leads to a non-unitary theory. These problems were overcome in \cite{Gaberdiel:2018rqv,Eberhardt:2018ouy}, where the hybrid formalism of \cite{Berkovits:1999im} was used to define the theory at $k=1$ and it was proposed that, at $k=1$, there is a shortening of the spectrum and the vacua are described by a limited set of representations (see also \cite{Gaberdiel:2017oqg, Giribet:2018ada}). In these representations, all but the lowest lying state in the continuum is absent and, with appropriate physical state conditions, the spectrum is unitary. These representations are characterized by the vanishing of the quadratic Casimir of the supergroup $PSU(1,1|2)$, given by the zero modes of the generators,
\begin{align*}
    \mathfrak{C}^{\mathfrak{psu}(1,1|2)}&= \mathfrak{C}^{\mathfrak{sl}(2;\R)} + \mathfrak{C}^{\mathfrak{su}(2)} + \mathfrak{C}^{\mathfrak{psu}(1,1|2)}_{\text{ferm}}, &
    \mathfrak{C}^{\mathfrak{sl}(2;\R)} &=-J^3_0J^3_0+ \frac{1}{2}(J_0^+J_0^-+J_0^-J_0^+), \\
    \mathfrak{C}^{\mathfrak{su}(2)}&= K^3_0K^3_0+ \frac{1}{2}(K_0^+K_0^-+K_0^-K_0^+), &
    \mathfrak{C}^{\mathfrak{psu}(1,1|2)}_{\text{ferm}} &= \frac{1}{2}\epsilon_{IJ}\epsilon_{\alpha\beta}\epsilon_{AB}S_0^{\alpha AI}S_0^{\beta BJ} .
\end{align*}

This shortening condition is naturally described by the twistor variables in \eqref{eq:CP1_string}, which provide a free field realisation of $\mathfrak{u}(1,1|2)_1$ \cite{Dei:2020zui}. Consider promoting each of $\mathfrak{C}^{\mathfrak{sl}(2;\R)}$, $\mathfrak{C}^{\mathfrak{su}(2)}$ and $\mathfrak{C}^{\mathfrak{psu}(1,1|2)}_{\text{ferm}}$ to be operators defined by the twistor variables; for example,
$$\mathfrak{C}^{\mathfrak{sl}(2;\R)}(z) = -:\left(J^3J^3 + \frac{1}{2}(J^+J^-+J^-J^+)\right):(z),$$
with $J^a(z)$ defined as in \eqref{eq:psu_bilinears}. One can then show that \cite{Gaberdiel:2022als},
$$\mathfrak{C}^{\mathfrak{psu}(1,1|2)}(z) = T(z) + :\mathcal{C}_L\mathcal{B}_L:(z) - 2:\mathcal{C}_L^2:(z),$$
where $T(z)$ is the holomorphic stress tensor for the twistor variables. A physical state must satisfy
$$L_n|\varphi\rangle = (\mathcal{C}_L)_n|\varphi\rangle = 0,$$
for all $n\geq 0$, where $L_n$ are the modes of $T(z)$. Hence, at the level of zero modes,
$$\mathfrak{C}^{\mathfrak{psu}(1,1|2)}|\varphi\rangle = (\mathcal{C}_L\mathcal{B}_L)_0|\varphi\rangle = \sum_{n>0} (\mathcal{C}_L)_{-n} (\mathcal{B}_L)_n|\varphi\rangle.$$
However, $(\mathcal{C}_L)_{-n}$ descendants for $n>0$ describe null states, such that the quadratic Casimir vanishes on the restriction to $\mathfrak{psu}(1,1|2)_1$ as a quotient of $\mathfrak{u}(1,1|2)_1$ \cite{Dei:2020zui,Gaberdiel:2022als}. Hence, the sigma model \eqref{eq:CP1_string} with the constraint ${\cal C}_L=0$ imposed, presents natural variables to describe the appropriate $PSU(1,1|2)$ representations at $k=1$.

\section{The twistor sigma model for Euclidean $\mathbf{AdS_3}$}\label{sec:euclidean_AdS3}

The construction presented in the previous section focused on complexified $AdS_3$. A choice of signature follows from imposing appropriate reality conditions. We will consider the case of Euclidean $AdS_3$, $H_3^+ = SL(2;\C)/SU(2)$, the space of $2\times 2$ Hermitian matrices \cite{Gawedzki:1991yu,Teschner:1997ft,deBoer:1998gyt,Giveon:1998ns,Kutasov:1999xu,Maldacena:2000kv,Maldacena:2001km}. This is a coset rather than a group manifold, but our construction naturally generalises. We will show this using the Wakimoto representation of $\mathfrak{sl}(2;\C)_k$ \cite{Wakimoto:1986gf}, which provides a clear connection between elements of the coset and coordinates on $H_3^+$. It has been conjectured that the physics of the $k=1$ string, which has the same $AdS_3 \times S^3$ matter content as in \eqref{eq:CP1_string}, is restricted to the boundary of $AdS_3$ and does not probe the interior. Using the twistor construction of the previous section, we will justify this claim and demonstrate precisely why this occurs in Euclidean $AdS_3$.

\subsection{The Wakimoto construction}

An element $h \in SL(2;\C)$ can be parameterised as
$$
h=\left(
\begin{array}{cc}
1  & \gamma \\
0 & 1
\end{array}
\right)\left(
\begin{array}{cc}
e^{-\varphi}  & 0 \\
0 & e^{\varphi}
\end{array}
\right)M = \ell M,
$$
where $M \in SU(2)$. Since $MM^{\dagger} = 1$, we have that $g = hh^{\dagger} = \ell\ell^{\dagger} \in SL(2;\C)/SU(2)$. Defining $\phi = \varphi + \bar{\varphi}$, we find \cite{Giveon:1998ns,deBoer:1998gyt,Kutasov:1999xu} (see also \cite{Knighton:2023mhq}),
\begin{equation}\label{eq:g_wakimoto}
    g=hh^{\dagger}=\left(
\begin{array}{cc}
e^{-\phi}+\gamma\bar{\gamma}e^{\phi}  & e^{\phi}\gamma \\
e^{\phi}\bar{\gamma} & e^{\phi}
\end{array}
\right),
\end{equation}
which can be related to coordinates on Euclidean global $AdS_3$. In Poincar\'{e} coordinates, the metric is given by
$$\mathrm{d}s^2 = \frac{\mathrm{d}r^2 + \mathrm{d}\gamma \mathrm{d}\bar{\gamma}}{r^2},$$
where $r = e^{-\phi}$ is a radial coordinate. The conformal boundary is given by $\phi \to \infty$ ($r\to 0$) and the coordinates $(\gamma, \bar{\gamma})$ become parallel to the boundary $S^2$ in this limit. Note that, we are not suggesting that $\gamma$ or $\varphi$ need be holomorphic, nor $\Box\phi=0$.

A sigma model on $H_3^+$ may be given by inserting this coset representative into the $SL(2;\C)$ WZW model. This gives the Wakimoto action,
\begin{equation}\label{eq:H3_action}
    S= -\frac{k}{2\pi} \int_{\Sigma} \mathrm{d}^2z \, \left( \p\phi \bar{\p}\phi + e^{2\phi}\p\bar{\gamma}\bar{\p}\gamma \right),
\end{equation}
in agreement with the metric in Poincar\'{e} coordinates. It can further be written in a first order formalism as \cite{Knighton:2023mhq},
\begin{equation}\label{eq:Waki_action}
    S = -\frac{1}{2\pi} \int_{\Sigma} \mathrm{d}^2z \, \left( \frac{1}{2} \p\Phi \bar{\p}\Phi + \beta\bar{\p}\gamma + \bar{\beta}\p\bar{\gamma} - \frac{1}{k}\beta\bar{\beta}e^{-q\Phi} - \frac{q}{4}R\Phi \right),
\end{equation}
after the field redefinition $\Phi = \sqrt{2(k-2)}\phi$ and the inclusion of the background charge $q = \sqrt{2/(k-2)}$. Working in a patch of $\Sigma$ with a conformally flat metric, $R = 0$, and no punctures, the general solution to the classical equations of motion are given by \cite{deBoer:1998gyt,Eberhardt:2019ywk},
$$
\frac{q}{2}\Phi(z,\bar{z})= \Phi_0+\bar{\Phi}_0+\log(1+b\bar{b}),    \qquad  \gamma=a+\frac{e^{-2\Phi_0}\bar{b}}{1+b\bar{b}},  \qquad  \bar{\gamma}=\bar{a}+\frac{e^{-2\bar{\Phi}_0}b}{1+b\bar{b}},
$$
\begin{eqnarray}\label{solutions}
\frac{q^2}{2}\beta= e^{2\Phi_0}\partial b, \qquad \frac{q^2}{2}\bar{\beta}= e^{2\bar{\Phi}_0}\partial \bar{b},
\end{eqnarray}
where $a$, $b$, $\Phi_0$ are all holomorphic and $\bar{a}$, $\bar{b}$, $\bar{\Phi}_0$ are all anti-holomorphic. Since we are in Euclidean signature, we view $\gamma$ and $\bar{\gamma}$ as complex conjugates and $\Phi$ is real. $\beta$ is clearly holomorphic but $\gamma$ is only holomorphic in the limit of large $\Phi$, when $\Phi_0 \rightarrow\infty$. Nevertheless, one can show that $g(z,\bar{z})$ factorises arbitrarily in the bulk into a product of holomorphic and anti-holomorphic matrices. That is, $g = hh^{\dagger}$ where $h = h(z)$ and $h^{\dagger} = h^{\dagger}(\bar{z})$,
\begin{equation}\label{eq:g_factorised}
    g(z,\bar{z})= \begin{pmatrix}
    e^{-\Phi_0} + abe^{\Phi_0} & ae^{\Phi_0}\\
    be^{\Phi_0} & e^{\Phi_0}
\end{pmatrix} \begin{pmatrix}
    e^{-\bar{\Phi}_0} + \bar{a}\bar{b}e^{\bar{\Phi}_0} & \bar{b}e^{\bar{\Phi}_0}\\
    \bar{a}e^{\bar{\Phi}_0} & e^{\bar{\Phi}_0}
\end{pmatrix}.
\end{equation}
We can also take a large $\phi$ limit of this equation, where $\gamma$ is holomorphic, to find
$$X = e^{-\phi}g \to \left(
\begin{array}{cc}
\gamma\bar{\gamma}  & \gamma \\
\bar{\gamma} & 1
\end{array}
\right) = \left(
\begin{array}{cc}
0  & \gamma \\
0 & 1
\end{array}
\right)  \left(
\begin{array}{cc}
0  & 0 \\
\bar{\gamma} & 1
\end{array}
\right).$$

Since $g$ factorises, the conserved currents are purely (anti-)holomorphic throughout the bulk. They are given by the WZW currents, $J_L = \p g g^{-1}$, evaluated on our coset representative, $g = hh^{\dagger}$. With an appropriate choice of basis, this gives
$$
J^+= \beta, \qquad  J^3= -\frac{1}{q} \partial \Phi + \beta\gamma,  \qquad  J^-= -\frac{2}{q} \gamma\partial\Phi + \beta\gamma^2 - (k-2)\p\gamma.
$$
Substituting in the classical solutions (\ref{solutions}),
$$J^+ = \beta, \qquad J^3 = -\frac{2}{q^2}\p\Phi_0 + \beta a, \qquad J^- = -\frac{4}{q^2}a\p\Phi_0 + \beta a - (k-2)\p a,$$
which are manifestly holomorphic, in agreement with the WZW model equations of motion $\bar{\p}J^a = 0$. This current algebra (at $k=1$) can be described using chiral twistors as in \S\ref{sec:twistor_sigma_model}. Despite the non-linear equations of motion for $(\Phi,\gamma,\bar{\gamma},\beta,\bar{\beta})$ that mix holomorphic and anti-holomorphic variables, we can recast this sigma model in terms of chiral twistors precisely because the currents remain holomorphic.

In the quantum theory, there is a normal ordering correction to the currents. In the near-boundary regime, they are given by
$$
J^+= \beta, \qquad  J^3= -\frac{1}{q} \partial \Phi + \beta\gamma,  \qquad  J^-= -\frac{2}{q} \gamma\partial\Phi + \beta\gamma^2 - k\p\gamma,
$$
which forms the Wakimoto representation of $\mathfrak{sl}(2;\R)_k$ \cite{Wakimoto:1986gf}, as can be checked using the OPEs
\begin{equation}\label{eq:Waki_OPEs}
    \beta(z)\gamma(w) \sim -\frac{1}{z-w}, \qquad \Phi(z)\Phi(w) \sim -\log (z-w).
\end{equation}
These OPEs are seen to hold in the regime of large $\Phi$, when the interaction term in \eqref{eq:Waki_action} vanishes, and they have been applied widely throughout the literature at $k=1$. This is one reason why it has been suspected that the worldsheet of the $k=1$ string is effectively pinned to the boundary \cite{Eberhardt:2019ywk,McStay:2023thk,Dei:2023ivl}. Indeed, only the bottom of the continuum of states (and its spectral flow) contributes at $k=1$ \cite{Eberhardt:2018ouy} which corresponds to worldsheet instantons that wrap the boundary \cite{Maldacena:2001km}. Yet, the fact that we can recast the $H_3^+$ sigma model at $k=1$ in terms of chiral twistors at generic bulk points provides evidence that the $k=1$ string really does capture all of the physics of $AdS_3$, not merely a near-boundary subsector.\footnote{Further evidence of this was given in \cite{McStay:2023thk,Sriprachyakul:2024gyl}. In \cite{McStay:2023thk}, an alternative prescription for how the Wakimoto free field OPEs arise was presented, without the need for a large $\Phi$ limit, based off \cite{Gerasimov:1990fi}. Moreover, in \cite{Sriprachyakul:2024gyl}, it was shown that the interaction term can be perturbatively reintroduced to correlation functions and gives a vanishing contribution at all orders when $k=1$.} We will see in \S\ref{sec:Boundary_spacetime}, nevertheless, the mechanism by which all of the physics at $k=1$ becomes localised to the boundary.
\\

The twistor sigma model for $H_3^+$ follows from generalising the construction introduced in the previous section from groups to cosets. The bulk incidence relations are of the form \eqref{eq:incidence}, with $g=hh^{\dagger}$. The model is not invariant under $g \mapsto A(z)gB(\bar{z})$ for $A,B\in SL(2;\C)$, as this does not preserve the Hermitian property of our coset representative. Rather, we are restricted to $B(\bar{z}) = A^{\dagger}(\bar{z})$ \cite{deBoer:1998gyt}. The preservation of the incidence relations requires
\begin{alignat*}{2}
   &\omega_{\alpha} \mapsto A_{\alpha}{}^{\beta}\omega_{\beta}, \qquad &&		\pi^{\dot{\alpha}} \mapsto (B^{-1})^{\dot{\alpha}}{}_{\dot{\beta}}\pi^{\dot{\beta}},\\
   &\mu_{\dot{\alpha}} \mapsto \mu_{\dot{\beta}}B^{\dot{\beta}}{}_{\dot{\alpha}},	\qquad &&		\lambda^{\alpha} \mapsto \lambda^{\beta}(A^{-1})_{\beta}{}^{\alpha}.
\end{alignat*}
Our left- and right-conserved WZW currents are constructed as usual and transform as $J_L \mapsto A J_L A^{-1}$ and $J_R \mapsto B^{-1}J_R B$ under the global isometries. We therefore define our free field realisation as
\begin{equation*}
(J_L)_{\alpha}{}^{\beta}= \omega_{\alpha}\lambda^{\beta} ,	\qquad		(J_R)_{\dot{\alpha}}{}^{\dot{\beta}}=\mu_{\dot{\alpha}}\pi^{\dot{\beta}},
\end{equation*}
in precisely the same way as before, with the constraint that we require the traces to vanish on-shell.

In our coset, we have restricted ourselves to Hermitian $g$, which corresponds to a Lorentzian slice of the $\C^4$ embedding space and hence Euclidean $AdS_3$. Taking the Hermitian conjugate of the incidence relations, we find that
\begin{align*}
    \omega = g\cdot \pi \quad &\mapsto \quad \omega^{\dagger} = \pi^{\dagger}\cdot g,\\
    \mu = -\lambda \cdot g \quad  &\mapsto \quad \mu^{\dagger} = -g \cdot \lambda^{\dagger}.
\end{align*}
This suggests that $\omega^{\dagger} = \pm\mu$ and $\pi^{\dagger} = \mp\lambda$, where we take the opposite sign for the two constraints. Without loss of generality, we shall take
\begin{equation}\label{eq:spinor_reality_condition}
    \bar{\omega}_{\dot{\alpha}} = \mu_{\dot{\alpha}}, \qquad \bar{\pi}^{\alpha} = -\lambda^{\alpha}.
\end{equation}
Note that a quaternionic action on the twistors can be defined through
$$
Z^I=\left(\begin{array}{c}
\omega_{\alpha}	\\
\pi^{\dot{\alpha}}
\end{array}
\right) \mapsto \tilde{Z}_I = (\bar{\pi}^{\alpha},-\bar{\omega}_{\dot{\alpha}})
$$
and $Z\cdot \tilde{Z} = 0$ in Euclidean signature for $AdS_3$, as is explained in Appendix \ref{sec:twistors} for our conventions. That is to say, the reality condition of Euclidean $AdS_3$ is
\begin{equation}\label{eq:euclidean_signature}
    Z\cdot \tilde{Z} = \langle \omega \bar{\pi}\rangle - [\bar{\omega} \pi] = -U - \tilde{U} = 0,
\end{equation}
where we use \eqref{eq:spinor_reality_condition}.
This is equivalent to requiring that the trace of $J_L + J_R$ vanishes. Note that we can also rewrite this condition as $W\cdot Z = U + \tilde{U} = 0$, which generates the projective scaling of \eqref{eq:twistor_and_dual}. Since we are understanding our minitwistors as descending from four-dimensional twistor space, we can identify $W\cdot Z=0$ as the quadric condition for the ambitwistor space of $\C^4$, as alluded to in \eqref{eq:ambitwistor_quadric}.

We  propose the following action for the worldsheet theory,
\begin{equation}\label{eq:bosonic_twistor_action}
    S=\frac{1}{2\pi}\int_{\Sigma} \mathrm{d}^2 z \Big(\omega_{\alpha}\bar{\partial}\lambda^{\alpha} + \pi^{\dot{\alpha}}  \partial\mu_{\dot{\alpha}} \Big).
\end{equation}
This precisely agrees with the usual free field realisation of the $SL(2;\C)/SU(2)$ coset model at level 1 (as is utilised in \cite{Dei:2020zui}), making explicit its twistorial origin. Unlike \eqref{eq:Waki_action}, this action splits into free holomorphic and anti-holomorphic sectors. As we have already mentioned, this is to be expected, since the currents of the $H_3^+$ coset model are generically (anti-)holomorphic. Moreover, the bulk incidence relations relate the left- and right-moving sectors, so it is not surprising that the $AdS_3$ coordinate fields $(\Phi, \gamma,\bar{\gamma})$ satisfy more complicated equations of motion.

\subsection{Dependence on the boundary coordinates}

The zero modes of the $SL(2;\C)/SU(2)$ currents form the M\"{o}bius generators on the boundary \cite{Brown:1986nw,Maldacena:1997re,Maldacena:1998bw}, such that the location of a boundary coordinate $(x,\bar{x})$ can be shifted by the action of $(J^+_0,\bar{J}^+_0)$. Since we are restricted to transformations of the form $g \mapsto AgA^{\dagger}$, we must translate both coordinates simultaneously. The incidence relations then give the corresponding action on twistor space. Acting with $h_L=e^{xJ^+_0}$ on the left gives a right action of $h_R = \left( e^{xJ_0^+} \right)^{\dagger} = e^{-\bar{x}J_0^-}$. Using the parameterization of $g$ given in (\ref{eq:g_wakimoto}),
\begin{align*}
g \mapsto e^{xJ_0^+}g e^{-\bar{x}J_0^-}  &= e^{\phi} \left(
\begin{array}{cc}
1  & -x \\
0 & 1
\end{array}
\right) \left(
\begin{array}{cc}
\gamma\bar{\gamma}+e^{-2\phi} & \gamma \\
\bar{\gamma} & 1
\end{array}
\right)
\left(
\begin{array}{cc}
1  & 0 \\
-\bar{x} & 1
\end{array}
\right)\\
&= e^{\phi}\left(
\begin{array}{cc}
(\gamma - x) (\bar{\gamma} - \bar{x}) +e^{-2\phi} & \gamma - x \\
\bar{\gamma} - \bar{x} & 1
\end{array}
\right)
\end{align*}
We see that this results in the translation $(\gamma, \bar{\gamma}) \mapsto (\gamma-x, \bar{\gamma} - \bar{x})$, as is consistent with \eqref{eq:Waki_OPEs}. To preserve the incidence relations, the twistors transform as
\begin{align*}
Z^I=\left(
\begin{array}{c}
\omega	\\
\pi
\end{array}
\right) \mapsto \left(
\begin{array}{cc}
h_L & 0 \\
0 & h_R^{-1}
\end{array}
\right)\left(
\begin{array}{c}
\omega	\\
\pi
\end{array}
\right),    \qquad  W_I=\left(
\begin{array}{cc}
\lambda	&,
\mu
\end{array}
\right)
\mapsto \left(
\begin{array}{cc}
\lambda	&,
\mu
\end{array}
\right)\left(
\begin{array}{cc}
h_L^{-1} & 0 \\
0 & h_R
\end{array}
\right).
\end{align*}
Thus,
$$
\omega_+\mapsto \omega_+ - x\omega_-,   \qquad  \pi^-\mapsto \pi^- + \bar{x}\pi^+,  \qquad  \mu_+\mapsto \mu_+ - \bar{x}\mu_-,\qquad \lambda^+\mapsto \lambda^+ + x\lambda^-,
$$
with all other components invariant.

\subsection{Recovering the boundary spacetime}\label{sec:Boundary_spacetime}

Part of the principal motivation for studying AdS/CFT in terms of twistor theory was that the bulk minitwistor space (the space of oriented geodesics) is closely related to the boundary ambitwistor space \cite{Adamo:2016rtr,Bu:2023cef,McStay:2023thk}, see Appendix \ref{sec:twistors} for details. We would like to demonstrate here how the boundary incidence relations can be derived from the bulk incidence relations as a large radius limit. This will illuminate how the bulk and boundary theories are related and explain how the incidence relations simplify at large radius to give purely holomorphic and anti-holomorphic statements. We will also see the reason for why and in what sense the $k=1$ string is localised to the boundary.

\subsubsection{Boundary incidence from bulk incidence}

Since $AdS_3$ has a conformal boundary, $\C\P^1$, at spatial infinity, any action we write down for the theory must be combined with boundary conditions for the fields. The same is true for our twistors and dual twistors, which we view as fibres over spacetime in the doubled correspondence space, $\mathcal{F}$. As we approach the boundary, the fields living in these fibres must have appropriate fall off conditions to achieve Dirichlet boundary conditions. This implies that $(\omega_{\alpha}, \pi^{\dot{\alpha}}, \lambda^{\alpha},\mu_{\dot{\alpha}})$ all tend to zero as $\phi \to \infty$. In fact, our (massless) spinors fall off as $r = e^{-\phi}$ near the boundary \cite{Witten:1998qj,Henningson:1998cd,Henneaux:1998ch}. We recover a finite boundary twistor as $\tilde{\omega}_{\alpha} = e^{\phi}\omega_{\alpha}$ and similarly for the other spinors. We can accordingly rewrite our incidence relations as
$$
\tilde{\omega}_{\alpha}(z)=g_{\alpha\dot{\alpha}}(z,\bar{z})\tilde{\pi}^{\dot{\alpha}}(\bar{z}),	\qquad		\tilde{\mu}_{\dot{\alpha}}(\bar{z}) = -\tilde{\lambda}^{\alpha}(z)g_{\alpha\dot{\alpha}}(z,\bar{z}).
$$

The $\phi \to \infty$ limit of $g$ in \eqref{eq:g_wakimoto} is somewhat singular. It is helpful to consider the unnormalised combination $X = e^{-\phi}g$ \cite{Adamo:2016rtr,McStay:2023thk}, for which the limit is well-defined. To be explicit, we define
\begin{equation*}
    X_{\alpha\dot{\alpha}}=\left(
\begin{array}{cc}
e^{-2\phi}+\gamma\bar{\gamma}  & \gamma \\
\bar{\gamma} & 1
\end{array}
\right), \qquad
    X^{\alpha\dot{\alpha}}=\left(
\begin{array}{cc}
1  & -\bar{\gamma} \\
-\gamma & e^{-2\phi}+\gamma\bar{\gamma}
\end{array}
\right).
\end{equation*}
One can check that $X^{\alpha\dot{\alpha}}g_{\beta\dot{\alpha}} = X^{\alpha\dot{\alpha}} (g^T)_{\dot{\alpha}\beta} = e^{-\phi}\delta^{\alpha}{}_{\beta}$ and similarly $X^{\alpha\dot{\alpha}}g_{\alpha\dot{\beta}} = (X^T)^{\dot{\alpha}\alpha}g_{\alpha\dot{\beta}} = e^{-\phi}\delta^{\dot{\alpha}}{}_{\dot{\beta}}$. Therefore,
$$X^{\alpha\dot{\alpha}}\tilde{\omega}_{\alpha} = e^{-\phi}\tilde{\pi}^{\dot{\alpha}}, \qquad X^{\alpha\dot{\alpha}}\tilde{\mu}_{\dot{\alpha}} = -e^{-\phi}\tilde{\lambda}^{\alpha}.$$
Since we have rescaled the spinors to be finite in the large $\phi$ limit, we deduce that
$$X^{\alpha\dot{\alpha}}\tilde{\omega}_{\alpha} \to 0, \quad  X^{\alpha\dot{\alpha}}\tilde{\mu}_{\dot{\alpha}} \to 0,$$
at the boundary.

The boundary incidence relations may then be written as $(X_{bdy})^{\alpha\dot{\alpha}}\tilde{\omega}_\alpha = (X^T_{bdy})^{\dot{\alpha}\alpha}\tilde{\omega}_{\alpha} = 0$. That is,
$$\left(
\begin{array}{cc}
1  & -\gamma \\
-\bar{\gamma} & \gamma\bar{\gamma}
\end{array}
\right)
\left(
\begin{array}{c}
 \tilde{\omega}_+     \\
 \tilde{\omega}_-
\end{array}
\right) = 0,$$
from which we deduce $\tilde{\omega}_+ = \gamma \tilde{\omega}_-$. $\gamma$ is purely holomorphic at the boundary so this really is a holomorphic statement. This is the incidence relation that has appeared widely in the $k=1$ string literature \cite{Dei:2020zui,Knighton:2020kuh,Bhat:2021dez,Knighton:2023mhq}. Of course, this is the expected result, since the twistor space of the boundary $\C\P^1$ (parameterised by $\gamma$) is just another $\C\P^1$ (parameterised by homogeneous coordinates $\tilde{\omega}_{\alpha}$). In general, the tilde will be dropped on these fields and whether we refer to minitwistors in the bulk or twistors on the boundary is implied by which set of incidence relations we apply.

To be precise, this canonical association between the bulk minitwistors and boundary twistors and dual twistors is as follows. We have boundary twistors given by $\omega_{\alpha}$ or $\mu_{\dot{\alpha}}$ and boundary dual twistors given by $\lambda^{\alpha}$ or $\pi^{\dot{\alpha}}$. The bulk minitwistors are given by $Z^I = (\omega_{\alpha},\pi^{\dot{\alpha}})^T$, whilst the dual bulk minitwistors are given by $W_I = (\lambda^{\alpha},\mu_{\dot{\alpha}})$.

\subsubsection{Boundary reconstruction}\label{sec:boundary}

We will now show how the physical boundary of $AdS_3$ arises in the sigma model and explain why the physical states are constrained to live at the boundary. We confine our discussion to Euclidean $AdS_3$, where the reality condition is defined by $g=g^{\dagger}$. This implies that complex conjugation on the spinors relates the minitwistors $Z^I$ and the dual minitwistors $W_I$ as in \eqref{eq:spinor_reality_condition}. The minitwistor space of $AdS_3$ is defined by
$$Z^I \in \{(\omega_{\alpha},\pi^{\dot{\alpha}})\in \C\P^1\times\C\P^1|\langle\omega\bar{\pi}\rangle\neq 0\}.$$
Yet, in this signature, $\langle\omega \bar{\pi}\rangle = -\langle\omega\lambda\rangle = -U$ and the sigma model we have constructed requires $U=0$. Hence, our sigma model strictly embeds into the completion of minitwistors space (including the points where $\langle\omega\bar{\pi}\rangle =0$) and the constraint $U=0$ localises the physics to points where $\langle\omega\bar{\pi}\rangle =0$.

Specifically, consider the patch $\lambda^+ \neq 0$ (which is equivalent to $\omega_- \neq 0$ when $U=0$). We projectively define
$$\lambda^{\alpha} = \left(\begin{array}{cc}
    1  &,  -x
 \end{array} \right), \qquad \bar{\lambda}^{\dot{\alpha}} = \left( \begin{array}{c}
      1  \\
      -\bar{x} 
 \end{array} \right), $$
where $x$ is a coordinate on $\C\P^1$. Then, one finds that
\begin{equation}\label{para}
\langle\omega\bar{\pi}\rangle = -\langle\omega\lambda\rangle = -\langle\lambda g\pi] = \langle\lambda g\bar{\lambda}] = e^{\phi}|\gamma-x|^2+e^{-\phi},
\end{equation}
where we have used the reality conditions in the first and third steps, the incidence relation $\omega_{\alpha} = g_{\alpha\dot{\alpha}}\pi^{\dot{\alpha}}$ was used in the second step, and the parameterisation of $g$ in \eqref{eq:g_wakimoto} was used in the final step. We see that the condition $U=-\langle\omega\bar{\pi}\rangle=0$ holds when we take the limit $\phi\rightarrow\infty$ and identify $(x,\bar{x})=(\gamma,\bar{\gamma})$ as a point on the boundary of Euclidean $AdS_3$.

Furthermore, since complex conjugation relates $\bar{\pi}^{\alpha} = -\lambda^{\alpha}$, we deduce that $\tilde{\gamma}$ as defined in \eqref{eq:U_constraint} satisfies $\tilde{\gamma} = \bar{\gamma}$. Hence the sigma model encodes the boundary in the following way. The $\C\P^1$s defined by $\omega_{\alpha}$ and $\lambda^{\alpha}$ are identified by $U =0$ and similarly for the $\C\P^1$s defined by $\mu_{\dot{\alpha}}$ and $\pi^{\dot{\alpha}}$, which are identified by $\tilde{U}=0$. Then the further Euclidean reality condition relates each of these $\C\P^1$s through complex conjugation. Hence the twistor sigma model for $H_3^+$ encodes a single copy of $\C\P^1$, which is the boundary of $H_3^+$.
\\

From the insights gleaned above, we can now ask to what extent the sigma model is constrained to live at the boundary. We have seen that, in Euclidean signature, the locus of $U = \tilde{U}=0$ describes the boundary. Hence, the physical state conditions immediately imply that all vertex operators of the sigma model must live at the boundary of spacetime. This is an on-shell statement and is consistent with the observation that the correlation functions have support on the boundary, rather than the bulk, incidence relations. Yet it remains to understand whether the worldsheet, away from vertex operator insertion points, is also fixed to the boundary. In terms of path integrals, this can be phrased more succinctly as follows: to what extent does the path integral receive contributions from worldsheets that probe the bulk? The expectation of some observable $\cal O$ can be written schematically as
$$
\int{\cal D}\omega{\cal D}\lambda{\cal D}\tilde{A}\;{\cal O}\;e^{-\frac{1}{2\pi}\int_{\Sigma}\rd^2 z \;(\omega_{\alpha}\bar{\partial}\lambda^{\alpha}+\tilde{A}U+...)}=\int{\cal D}\omega{\cal D}\lambda\;{\cal O}\;\delta[U]\,e^{-\frac{1}{2\pi}\int_{\Sigma}\rd^2z \; (\omega_{\alpha}\bar{\partial}\lambda^{\alpha}+...)},
$$
where we explicitly write down only the holomorphic sector for simplicity. On the left-hand side we integrate over all embeddings $(\omega_{\alpha},\lambda^{\beta}):\Sigma\rightarrow \C\P^1\times\C\P^1$, compatible with the chosen reality condition. On the right-hand side, where the $\tilde{A}$ integral has been performed and the $U=0$ constraint imposed, the integral is over only those embeddings compatible with both the reality condition and $U=0$ (similarly for $\tilde{U}=0$). As we have seen, this restricts the embeddings to the boundary such that the entire worldsheet is localised at the boundary. We will see further evidence in \S\ref{sec:bulk_VO} for how $U=\tilde{U}=0$ localises the physics to the boundary --- there we shall relax the constraint to only impose the reality condition $U + \tilde{U}=0$, and we will show how information in the bulk can then be recovered.

Two caveats to our analysis here should be given. The first is that our analysis only applies to Euclidean $AdS_3$, whilst the Lorentzian analogue appears less clear to the authors. The second is that we have restricted ourselves to only considering the bosonic sector. In the full supersymmetric theory, we actually impose $\mathcal{C}_L = 0$, which also contains a fermionic contribution. Nevertheless, the correlation functions of the supersymmetric theory also have support on the boundary incidence relations, suggesting that the above argument naturally generalises. One would need a clearer understanding of the incidence relations for $AdS_3 \times S^3$ superspace to directly reconstruct this argument in the presence of supersymmetry.

\subsection{The supersymmetric case}\label{sec:further_examples}

We shall briefly comment on the supersymmetric extension to $AdS_3 \times S^3$ superspace. We will assume that appropriate reality conditions can be applied to the complexified geometry (as we have explicitly demonstrated for the bosons). In \cite{Berkovits:1999im}, it was argued that the starting point for such a theory is given by a WZW model for the supergroup $PSU(1,1|2)$ (see also \cite{Gotz:2006qp,Troost:2011fd,Gaberdiel:2011vf,Gerigk:2012cq}). We stated in \S\ref{sec:almost} that a natural candidate for the corresponding supertwistor space is given by an embedding into $\C\P^{1|2} \times \C\P^{1|2}$, by comparison with the boundary superambitwistors. Indeed, bilinears of the free fields in this description are known to reproduce $\mathfrak{psu}(1,1|2)_1$ \cite{Dei:2020zui}; however, it is desirable to give this proposal a clear geometric underpinning by formulating the incidence relations in the bulk superspace, an outstanding problem that we highlighted in \S\ref{sec:almost}.

A potential way forward may come through the work of \cite{Ito:1998vd,Ito:1999pk}, where an action for the Euclidean superspace $SL(2|2;\C)/SU(2|2)$ was proposed. This involved six bosonic coordinates describing both $AdS_3$ and $S^3$ in Wakimoto descriptions analogous to \eqref{eq:H3_action} and \eqref{eq:Waki_action}, along with four fermionic coordinates in each of the left- and right-moving sectors. The WZW currents reproduced the same $\mathfrak{psu}(1,1|2)_k$ currents as were proposed in the hybrid formalism of \cite{Berkovits:1999im}. Namely, the fermionic coordinates in spacetime were the same as the ``decoupled fermions'' that appear in conventional hybrid string constructions (see \cite{Giveon:1998ns} and also \cite{Demulder:2023bux} for a recent review). In this sense, it may well be possible to view the hybrid formalism as a ``super-Wakimoto representation'' of the supergeometry, making clear the connection between superspace coordinates and the current algebra.

One interesting feature of \cite{Ito:1998vd,Ito:1999pk} is that only four fermionic coordinates of the target space are included, which is half of the fermionic dimension for the full superspace. In \cite{Berkovits:1999im}, it was highlighted that the remaining four fermionic coordinates can be realised in the theory, but that they are not independent fields and instead are functions of the other field content. In some sense, this was to be expected. In conventional approaches to Green-Schwarz superstrings with manifest spacetime supersymmetry, there naively appears to be twice as many fermions as required for supersymmetry \cite{Green:1983wt}. Yet, one finds an additional local fermionic symmetry on the worldsheet, called $\kappa$-symmetry, that removes half of the fermionic degrees of freedom. This suggests that the hybrid formalism is the $\kappa$-gauge fixed description of some full superspace covariant description of $AdS_3 \times S^3$, as is implied in \cite{Berkovits:1999im}. This observation will be important in the following chapters.

\section{Non-linearly realised supersymmetry}\label{sec:PBGS}

In \S\ref{sec:almost}, we saw that a sigma model embedding into $\C\P^{1|2} \times \C\P^{1|2}$ combined with the usual reparametrisation ghosts gave a conformal anomaly of $c = -28$. As mentioned in \S\ref{sec:comment_on_criticality}, a known resolution to this problem follows from Berkovits' description of the twistor string by coupling the theory to a current algebra carrying $c=+28$ \cite{Berkovits:2004hg}. We will, however, argue in this section and the next that this is not the correct resolution for our superstring theory on minitwistor space, but rather that the hybrid formalism of \cite{Berkovits:1999im} is the correct prescription to take. Indeed, our sigma model with the action \eqref{eq:CP1_string} is almost the $AdS_3\times S^3$ sector of the hybrid formalism of \cite{Berkovits:1999im} at $k=1$. What is missing is a $c=+28$ scalar field $\rho$ and a physical constraint ${\cal Q}:=\chi_+\chi_-\epsilon^{\alpha\beta}\omega_{\alpha}\p\omega_{\beta}=0$. We would now like to give intrinsic motivation for this additional field content of the hybrid string.

We conjecture that there exists a formulation of the theory in which all eight target space fermions, conjugate to $S^{\alpha A I}$, appear independently. This is in contrast to the hybrid formalism of \cite{Berkovits:1999im} where only four appear independently.\footnote{A formulation of the hybrid string where all eight of the fermionic coordinates for the target space are manifestly realised was given in \cite{Berkovits:1999du} (see also \cite{Daniel:2024ymb}). It would be interesting to see whether this formulation can be related to the proposed doubly supersymmetric constructions discussed here.} To simplify matters, we will work with respect to our sigma model embedding into the ambitwistor space of the boundary, where there is a much clearer understanding of the supergeometry. In this setting, we conjecture that there exists a formulation of the theory in which four target space fermions are manifest and that the correct number of physical degrees of freedom (two independent fermions) is recovered by imposing an additional constraint associated to a local symmetry on the worldsheet. This is reminiscent of $\kappa$-symmetry in the Green-Schwarz formalism. The key difference here is that the proposed formulation also has a twisted ${\cal N}=2$ \emph{worldsheet} supersymmetry algebra and so the theory is in some sense ``doubly supersymmetric'' (with worldsheet and spacetime supersymmetry). Our goal in this section and the next is to motivate the existence of this formulation of the theory and to argue that fixing the local fermionic symmetry gives rise to the conventional description of the hybrid formalism at $k=1$ \cite{Berkovits:1999im}. We will argue that the missing constraint is provided by $\mathcal{Q}=0$.

\subsection{The boundary theory and non-linear symmetries}\label{sec:boundary_pbgs}

It will be simplest to phrase this discussion on the constraint surface
$$
\mathcal{C}_L := \frac{1}{2} \left( \lambda^{\alpha}\omega_{\alpha} + \chi^{ A }\psi_{ A } \right)=0.
$$
In light of our discussion in the previous section, this places us directly at the boundary. In particular, the current $\mathcal{B}_L = \frac{1}{2}(\lambda^{\alpha}\omega_{\alpha} - \chi^A\psi_A)$ acts as an instanton number, so we can view $U$ as having been gauge fixed. This means that we can work with the boundary supergeometry, whose super-incidence relations are known, unlike the $AdS_3 \times S^3$ counterpart (for which only the $AdS_3$ sector has been identified).

In terms of the boundary twistors, the Lagrangian for the matter sector takes the same form as before,
$$\mathcal{L} = \omega_{\alpha}\bar{\partial}\lambda^{\alpha}+ \chi_{ A }\bar{\p}\psi^{ A }+...,$$
combined with a right-moving sector. To solve the scaling constraint, we shall work with manifestly ${\cal C}_L$-invariant combinations of fields. Following the incidence relation on the boundary $\omega_+-\gamma \omega_-=0$, and its natural fermionic counterpart, we introduce the scaling-invariant fields,
$$
\gamma:=\frac{\omega_+}{\omega_-},  \qquad  \theta_{ A } := \frac{\psi_{ A }}{\omega_-},
$$
on a patch where $\omega_-\neq 0$. $\mathcal{C}_L$ scales the dual supertwistors in the opposite direction, and so natural scaling-invariant fields of weight one are\footnote{On this patch, we can classically solve the scaling constraint to give $\lambda^-=-\frac{1}{\omega_-}(\lambda^+\omega_++\chi^A\psi_A)$. The corresponding incidence relation is then seen to be $\lambda^-=-\gamma\lambda^+-\chi^A\theta_A$.}
$$\beta := \lambda^+\omega_-, \qquad p^{ A } := \chi^{ A }\omega_{-}.$$
Direct substitution of these incidence relations into the holomorphic Lagrangian yields
\begin{equation}\label{eq:boundary_theory}
    {\cal L}= - \beta\bar{\partial}\gamma - p^{ A } \bar{\partial} \theta_{ A } - 
2\mathcal{C}_L \frac{\bar{\partial}\omega_-}{\omega_-}.
\end{equation}
The fields $(\gamma, \theta_{ A })$ are embedding coordinates on the boundary $\mathcal{N}=2$ superspace, a $\C\P^{1|2}$ \cite{Dei:2023ivl}. On the constraint $\mathcal{C}_L = 0$, we have a free theory embedding into the boundary $\C\P^{1|2}$, which is described by a bosonic $\beta\gamma$-system and two fermionic $p\theta$-systems, with OPEs
$$\beta(z)\gamma(w) = - \frac{1}{z-w} + \dots, \qquad p^{ A }(z) \theta_{ B }(w) = + \frac{\delta_{ B }{}^{ A } }{z-w} + \dots$$
These boundary free fields can be related to a free field realisation of $\mathfrak{psu}(1,1|2)_1$ \cite{Dei:2023ivl,Beem:2023dub},
\begin{align*}
    J^+ &= \beta, & J^3 &= \beta\gamma + \frac{1}{2}p^{ A }\theta_{ A }, & J^- &= (\beta\gamma)\gamma + \left( p^{ A }\theta_{ A } \right) \gamma,\\
    K^+ &= p^+\theta_-, & K^3 &= \frac{1}{2}\left(p^+\theta_+\right) - \frac{1}{2}\left(p^-\theta_-\right), & K^- &= p^-\theta_+,
\end{align*}
and
\begin{align*}
    S^{+++} &= p^+, & S^{+-+} &= p^-, & S^{-++} &= -\gamma p^+, & S^{--+} &= -\gamma p^-,\\
    S^{++-} &= -\beta\theta_-, & S^{+--} &= \beta\theta_+, & S^{-+-} &= \left(\beta\gamma + p^+\theta_+\right)\theta_-, & S^{---} &= -\left(\beta\gamma + p^-\theta_- \right)\theta_+,
\end{align*}
where brackets refer to normal ordering. The indices in $J^a$ and $K^{\tilde{a}}$ refer to adjoint indices for $SL(2)$ and $SU(2)$, respectively, whilst $S^{\alpha A I}$ has a spinor index for $SL(2)$, a fundamental index for $SU(2)$ and $I$ labels an $SU(2)$ outer automorphism of the theory.\\

\subsubsection{Non-linear realisations of $PSU(1,1|2)$}

Imposing the constraint ${\cal C}_L=0$ and so passing from the projective coordinates to $(\gamma,\theta_A)$, our realisation of the $\mathfrak{psu}(1,1|2)_1$ algebra becomes non-linear. $PSU(1,1|2)$ acts linearly on the boundary supertwistors, $(\mathcal{Z}_L)^M$. Infinitesimally, we have
$$
\delta(\mathcal{Z}_L)^M = U^M{}_N (\mathcal{Z}_L)^N,  \qquad  U=\left(\begin{array}{cc}
  J^{\alpha}{}_{\beta}   & S^{\alpha}{}_B{}^+ \\
  S_{\beta}{}^{A-}   & K^A{}_B
\end{array}\right)\in \mathfrak{psu}(1,1|2),
$$
where $M$ and $N$ run over the bosonic and fermionic indices. This is true essentially by the definition of the currents that we constructed in \S\ref{sec:twistor_sigma_model}. Alternatively, this could be taken as the definition of our supertwistor variables. To see why solving the scaling constraint leads to non-linear realisations, consider the following example. Under the symmetry generated by $J^+$, the $\omega_{\alpha}$ transform as
$$
\left(
\begin{array}{c}
     \omega_+  \\
     \omega_- 
\end{array}
\right)\rightarrow \left(\begin{array}{cc}
    1 & -\alpha \\
    0 & 1
\end{array}\right) \left(
\begin{array}{c}
     \omega_+  \\
     \omega_- 
\end{array}
\right),
$$
which is clearly linear. If we pull out a factor of $\omega_-$, which is invariant under $J^+$, and define $\gamma=\omega^+/\omega^-$, we find
$$
\left(
\begin{array}{c}
     \gamma  \\
     1 
\end{array}
\right)\rightarrow \left(\begin{array}{cc}
    1 & -\alpha \\
    0 & 1
\end{array}\right) \left(
\begin{array}{c}
     \gamma  \\
     1 
\end{array}
\right),
$$
i.e. $\gamma\rightarrow \gamma-\alpha$, which is a \emph{non}-linear transformation. The reverse of this process is the well-known augmentation of a translation to an affine transformation. Exactly the same thing occurs for the transformations generated by $S^{+A+}=p^A$, under which the fermions transform non-linearly as $\theta_A\rightarrow \theta_A+\varepsilon_A$. To summarize; the twistor variables transform linearly under $PSU(1,1|2)$ but, after solving the projective constraint ${\cal C}_L=0$, some of the symmetries act non-linearly. In particular, $J^+$ and $S^{+A+}$ generate shift symmetries of $\gamma$ and $\theta_A$, respectively.\\

One of the advantages of describing $\mathfrak{psu}(1,1|2)_1$ in terms of the projectively rescaled coordinates is that the interpretation of many of the fields becomes clear. Working at the level of zero modes, we first notice that $J^+_0$ acts as a translation operator for $\gamma$, whilst $S^{+ A +}_0$ is a translation operator for $\theta_{ A }$. The boundary Hamiltonian is given by $J^3_0$ since it generates dilations, and our coordinates have the desired scaling dimensions of $-1$ for $\gamma$ and $-\half$ for $\theta_{ A }$ under $J_0^3$. Moreover, $K^{\tilde{a}}_0$ rotate the fermions, generating an $R$-symmetry on the boundary.

The remaining generators are less simple to interpret and fall into two categories. We firstly have the generators that carry a minus index in the $SL(2)$ index (adjoint or spinor representations). This includes $J^-_0$ and $S^{- A I}_0$. In gauge fixing $\mathcal{C}_L = 0$, we made a choice of patch $\omega_- \neq 0$. Had we made the alternative choice $\omega_+ \neq 0$, then the interpretation of these generators would be clear. They would describe translations of the boundary spacetime coordinates in this new patch. Therefore, in our chosen patch of $\omega_- \neq 0$, it is unsurprising that the interpretation of $J^-_0$ and $S^{-AI}_0$ is less clear, since they involve a translation of the point at infinity and so are related to special (super)conformal transformations. 

The second category are the supersymmetries which carry a minus index in the $SU(2)$ outer automorphism index. This includes $S^{+ A -}_0$, as well as $S^{- A -}_0$ which also fell into the first category. We therefore expect that $S^{- A -}_0$ has a more natural interpretation when viewed from the $\omega_+\neq 0$ patch. The $S^{+A-}_0$ are realised linearly in the theory, generating the supersymmetries that relate $\gamma$ and $\theta_A$.

In light of this, we shall focus our attention on the subset of generators $\{J^+, J^3, S^{+A\pm}\}$, noting that the $K^{\tilde{a}}$ are internal symmetries.

\subsubsection{Broken symmetries}

If we introduce a vacuum for the coordinate fields,
$$
\langle\gamma\rangle = \langle\theta_{ A }\rangle = 0,
$$
then this vacuum is spontaneously broken by $S^{+ A +}_0$ and $J^+_0$, but is preserved by the $S^{+ A -}_0$ and $J^3_0$. In this patch, the $S^{+ A +}_0$ correspond to broken supersymmetries and $\theta_{ A }$ are the corresponding Goldstinos. By contrast, the supersymmetries $S^{+ A -}_0$, being linearly realised, are unbroken symmetries of the theory. There are no corresponding physical string excitations corresponding to these supersymmetries. Likewise, we can interpret $\gamma$ as the Goldstone boson associated to the broken translation symmetry $J^+_0$, whilst the boundary Hamiltonian $J^3_0$ and the $R$-symmetry are unbroken. We see that the broken symmetries are realised non-linearly in the theory, whilst the unbroken symmetries are realised linearly.\footnote{$J^-_0$ and $S^{- A -}_0$ are realised quadratically and correspond to the other patch.}\\

The fact that some of the global symmetries are broken by a choice of vacuum and others are unbroken implies the existence of a partially broken global supersymmetry (PBGS) algebra. This is also true of the hybrid formalism \cite{Berkovits:1999im}, in which only four of a possible eight spacetime fermions for $AdS_3 \times S^3$ are manifest. The remaining four fermions are functions of the manifestly realised fields. It is also noted in \cite{Berkovits:1999im} that the spacetime supersymmetries $S^{\alpha A +}_0$ are broken, whilst $S^{\alpha A -}_0$ are unbroken. One way to think of this is by analogy with the Green-Schwarz superstring \cite{Green:1983wt}, in which a local fermionic symmetry called $\kappa$-symmetry halves the number of fermionic degrees of freedom. Indeed, a GS superstring in $D=6$ dimensions would naively have eight fermionic coordinates, with only four physical degrees of freedom.

In our discussion here, we are instead viewing our string as embedding into the boundary theory, where we have a better grasp on the supergeometry. Given the global symmetries we have described, one would naively guess that a spacetime supersymmetric theory would have four fermionic coordinates, conjugate to the four $S^{+ A \pm}$. Instead we have only two fields $\theta_A$, conjugate to the $S^{+A+}$. Combined with $\gamma$, these describe one copy of $\C\P^{1|2}$. As such, the presence of this PBGS algebra in the non-critical theory \eqref{eq:CP1_string} is significant, since it suggests that what we have is something akin to a $\kappa$-gauge fixed phase of a fully target space covariant description of the theory. This then suggests that there should be compensating ghost terms included in the theory. We will determine these ghost terms for our string theory on minitwistor space below, but it would also be interesting to ask whether such an argument could be made to construct novel twistor string theories akin to \cite{Witten:2003nn,Berkovits:2004hg} in four dimensions; we will delay a discussion of this to \S\ref{sec:discussion}.

Before we can discuss what ghost terms may arise, we must first explain a general prescription for constructing spacetime supersymmetric actions in the presence of PBGS alegbras. We can then apply this to \eqref{eq:CP1_string}, to outline how one could construct a string theory embedding into the full superspace, including all fermionic degrees of freedom.

\subsection{Partially broken global supersymmetry}\label{sec:PBGS_subsection}

Non-linear realisations of symmetries and their role in symmetry breaking has long been a topic of interest \cite{Callan:1969sn}. This topic was studied by Witten in \cite{Witten:1981nf}, laying significant foundations for this area of research within the context of supersymmetry. It was initially assumed that if some supersymmetry was broken by the vacuum, then in fact all of the supersymmetry was broken. However, Hughes and Polchinski \cite{Hughes:1986dn} demonstrated a more general framework for realising supersymmetry algebras in which it was possible to break only some of the supersymmetry. This gave rise to the subject of PBGS. We will briefly review the basic ideas, following \cite{Hughes:1986dn} and how those ideas apply to the model under consideration. We refer the reader to the references for further details.

The starting point is a spacetime symmetry group $G$, built from a collection of broken and unbroken symmetries generated by $\{Z_a\}$ and $\{U_s\}$ respectively. The unbroken symmetries form a subgroup of $G$, with commutation relations
$$[U_s,U_t] = C_{st}{}^u U_u, \qquad [U_s,Z_a] = C_{sa}{}^b Z_b,$$
and are linearly realised in the theory. We can split the unbroken generators up in to two types of symmetry: those corresponding to translations or supersymmetries, $\{P_s\}$, and those corresponding to internal symmetries, $\{V_s\}$. The goal is to construct a non-linear realisation of the coset $G/U$. The key insight of \cite{Hughes:1986dn} was to construct a group element of $G$ as
\begin{equation}\label{eq:group_realisation}
    g(x,\xi) = e^{ix^sP_s}e^{i\xi^a(x)Z_a}.
\end{equation}
We introduce a coordinate $x^s$ for each unbroken $P_s$ and a Goldstone \emph{field} $\xi^a(x)$, taken to be functions of these coordinates, associated to $Z_a$. That is $g(x,\xi)$ a \emph{function} of $x^s$ and \emph{functional} of $\xi^a$. In general, the form of $g$ is not preserved under the action of a general group element $g' \in G$, but instead
$$g' g(x,\xi) = e^{i\tilde{x}^sP_s}e^{i\tilde{\xi}^a(\tilde{x})Z_a} e^{iu^s(\xi,g';x)V_s},$$
where internal symmetries, generated by $V_s$, may be introduced. This provides a realisation of $G$ and the fact that the parameter fields $\xi^a(x)$ (which parameterise $G/U$) also transform, due to their dependence on $x^s$, means the associated realisation of $G/U$ will in general be non-linear. Since $g$ is a functional of the Goldstone fields $\xi^a(x)$, the left-invariant one-forms $g^{-1}dg$ contain derivatives of these fields. Importantly, one can then construct natural candidates for covariant derivatives for the Goldstone fields from the components of these left-invariant one-forms\footnote{In the case where there are no internal symmetries, a covariant derivative for the Goldstone fields is $\tilde{\mathscr{D}}_s\xi^a:=(A^{-1})_s{}^tB_t{}^a$, where $g^{-1}dg=i\rd x^tA_t{}^sP_s+i\rd x^tB_t{}^aZ_a$. For covariantly transforming fields, a covariant derivative is given by $\mathscr{D}_s:=(A^{-1})_s{}^t\partial_t$. An invariant measure is given by $\rd\mu=\text{sdet}(A)\rd^dx$, where we write $\text{sdet}(A)$ to emphasise that $\{x^s\}$ can include bosons and fermions.}. The details of this construction are not important here but the interested reader may wish to consult \cite{Hughes:1986dn} and references therein.

Building on the seminal work of \cite{Salam:1969bwb,Volkov:1972jx,Volkov:1973vd,Ogievetsky:1974}, Hughes and Polchinski \cite{Hughes:1986dn} then present an algorithm for constructing $G$-invariant actions for such a realisation. Crucially, the volume element for these actions involves an integral over $\mathrm{d}^d x$, where $s = 1, \dots, d$. Since $x^s$ are the coordinates associated to $P_s$, which are unbroken translations and supersymmetries, they can be either bosonic or fermionic. The action is not uniquely determined but takes the general form
$$
S = \int \mathrm{d}^d x \, \alpha(x) \, \mathcal{L}(\xi(x),x),
$$
where $\alpha(x)$ depends on the group generated by $\{P_s\}$, such that $d\mu(x) = \mathrm{d}^dx \, \alpha(x)$ is defined to be the Haar measure, the group-invariant measure. $\mathcal{L}(\xi(x),x)$ is any function built out of the covariantly transforming quantities, leading to manifest $G$-invariance.

This algorithm has been applied to the GS superstring with $\mathcal{N}=2$ worldsheet supersymmetry, as well as in other contexts, including higher-dimensional membrane theories \cite{Hughes:1986fa,Townsend:1987yy,Achucarro:1988qb,deAzcarraga:1989mza}. The key observation is that, by inserting a membrane in spacetime (such as the worldsheet), translations transverse to the membrane are broken, and the embedding fields associated with these directions are the corresponding Goldstone modes. The superpartners of these Goldstone modes are associated with  broken supersymmetry. To write down a fully $G$-invariant action, we would like to include the fermionic modes associated to both the broken and unbroken supersymmetries. To do so, we must construct a supermembrane theory, where the broken supersymmetries act transversely to the supermembrane and the unbroken supersymmetries act longitudinally. We require that the tangential modes are pure gauge for consistency with the PBGS construction. That is, they are not physical excitations. This is a feature of the construction in \cite{Hughes:1986dn} --- the additional fermionic degrees of freedom introduced in $\{x^s\}$ turn out to be pure gauge.

\subsection{Doubly supersymmetric boundary theory}\label{sec:doubly_susy}

How might we apply this procedure to a string theory on the minitwistor space of $AdS_3 \times S^3$? We shall work at the level of algebra for now. That is, we can forget about the underlying worldsheet sigma model and focus solely on the PBGS algebraic structure. Implicitly, that means we are working just with the global part of $\mathfrak{psu}(1,1|2)_1$, the zero modes. Moreover, since we are focusing on the boundary theory in a patch ($\omega_-\neq 0$), we need only consider the subalgebra generated by $\{J^+, J^3, K^{\tilde{a}}, S^{+ A I} \}$. We identify the operators as follows:
\begin{align*}
    \{P_s\} &= \{J^3, S^{+ A -}\} &&\text{Unbroken dilations and supersymmetries.}\\
    \{V_s\} &= \{K^{\tilde{a}}\} &&\text{Unbroken $R$-symmetry}.\\
    \{Z_a\} &= \{J^+,S^{+ A +}\} &&\text{Broken translations and supersymmetries.}
\end{align*}
One can check that the algebra of unbroken symmetries closes and that they are indeed linearly realised in the theory.

Following the prescription of \cite{Hughes:1986dn}, we introduce parameters $(y,\vartheta_{ A })$, which we shall think of as coordinates, which correspond to the unbroken $(J^3, S^{+ A -})$. We emphasise that these do not correspond to physical degrees of freedom. We also introduce Goldstone modes $(\gamma,\theta_{ A })$ for each of the broken symmetries such that
$$\gamma \xmapsto{\alpha J^+_0} \gamma - \alpha, \quad \theta_+ \xmapsto{\epsilon S^{+++}_0} \theta_+ + \epsilon, \quad \theta_- \xmapsto{\epsilon S^{+-+}_0} \theta_- + \epsilon.$$
These generate non-linear realisations of the broken symmetries. The Goldstone fields are functions of the parameters so that
$$
\gamma=\gamma(y,\vartheta_+,\vartheta_-),   \qquad  \theta_A=\theta_A(y,\vartheta_+,\vartheta_-).
$$
Thus, as in \eqref{eq:group_realisation}, to construct a general group element, we promote our Goldstone modes to be superfields of the coordinates $(y,\vartheta_{ A })$. That is, our superfields $(\gamma, \theta_A)$ now describe maps from the space spanned by $(y, \vartheta_A)$ to superspace, and we can therefore view $(\gamma, \theta_A)$ as embedding superfields. 

Moreover, we are actually interested in taking two copies of this PBGS algebra, since the sigma model that we hope to build has a left- and right-action. As such, there also exist Goldstone modes $(\bar{\gamma}, \bar{\theta}_{ A })$ which are functions of the coordinates $(\bar{y},\bar{\vartheta}_{ A })$. For now, these are just algebraic entities, but the notation is suggestive of the fact that we will be able to identify the two copies of the Goldstone modes as holomorphic and anti-holomorphic fields and the coordinates for unbroken symmetries as complex conjugates. For the most part, we will not explicitly write down the barred fields to simplify the notation.

We can then propose an action for these superfields and we require it to be $G$-invariant. The most general $G$-invariant action is of the form
\begin{equation}\label{eq:superRS}
    S = \int \mathrm{d}^2y \mathrm{d}^2\vartheta_+ \mathrm{d}^2 \vartheta_- \, \alpha(x) \, \mathcal{L}\left(\gamma, \bar{\gamma}, \theta_{ A }, \bar{\theta}_{ A }; x \right),
\end{equation}
where $x$ is shorthand for $(y, \bar{y}, \vartheta_{ A }, \bar{\vartheta}_{ A })$. A crucial point here is that the general structure of this action has been determined purely on algebraic grounds, by studying the symmetry structure as done in \cite{Hughes:1986dn}. Yet, it now appears to be of the form of a sigma model action, with worldsheet coordinates given by the $(y,\vartheta_A)$. It is natural to view $(y,\bar{y})$ as complex coordinates on a patch in $\C$. We initiated this discussion in \S\ref{sec:boundary_pbgs} by restricting to a patch on the boundary $\C\P^{1|2}$, so more generally we promote $(y,\bar{y})$ to be coordinates on a Riemann surface. We introduce a canonical choice of complex structure along with appropriate gluing conditions for patches. Moreover, the $\vartheta_{ A }$ can now be thought of as worldsheet fermions, parameterising a worldsheet with supercoordinates $(y, \bar{y}, \vartheta_{ A }, \bar{\vartheta}_{ A })$.

In summary, consideration of the non-linear realisations appearing due to imposing the scaling constraint ${\cal C}_L=0$ lead us to conjecture that, for fully manifest spacetime supersymmetry in \eqref{eq:boundary_theory}, we should strictly embed a super-Riemann surface into the target superspace. The fermionic worldsheet coordinates are identified with the spacetime pure gauge fermionic degrees of freedom that are missing in conventional descriptions. Typically, the procedure of \cite{Hughes:1986dn} introduces the bosonic worldsheet coordinates $(y,\bar{y})$ in static gauge. Upon performing an appropriate conformal transformation $(y,\bar{y}) \mapsto (z,\bar{z})$, we could instead describe this string theory in the usual complex coordinates. By integrating out the $\vartheta_{ A }$, we can in principle reduce this string theory to an ordinary Riemann surface embedding into supersymmetric target space.\\

The above prescription gives a new outlook for constructing a string theory on minitwistor space. It says that we should view \eqref{eq:boundary_theory} as the gauge-fixed version of a theory involving both spacetime and worldsheet supersymmetry. Given that the generators $S^{\alpha AI}$ are weight one, we anticipate that the coordinates conjugate to the $S^{\alpha AI}$ are weight zero, giving a twisted form of the usual ${\cal N}=2$ supersymmetry. Therefore, we have a string theory with $\mathcal{N} = 2$ supersymmetry on the worldsheet (from the inclusion of $\vartheta_A$) and $\mathcal{N} = 2$ supersymmetry in spacetime (from $\gamma$ and $\theta_A$). Gauging the worldsheet $\mathcal{N} = 2$ superconformal algebra will give the ghost content of the theory.

At present, we do not have an explicit construction for this gauge fixing process and the reduction to the usual matter content of the $k=1$ string, and we leave this to future work. We expect both of the usual gauge symmetries --- the stress tensor and $GL(1)$ scaling of the homogeneous twistor coordinates --- to be encoded in this $\mathcal{N} = 2$ algebra, whilst additional gauge symmetries from the two worldsheet supersymmetries are now also important.

Whilst this conjecture is new to the twistor literature, it is not entirely new in the context of tensionless holography. Indeed, the idea that the $k=1$ string could be expressed in terms of an $\mathcal{N} = 2$ string also appeared in \cite{Gaberdiel:2021qbb,Gaberdiel:2021jrv}. In that case, the $k=1$ ghost content was observed to naturally describe the ghosts associated to a topologically twisted $\mathcal{N} = 2$ algebra, although the details of where this $\mathcal{N} = 2$ came from were not given. This conjecture marries up nicely with the degree of freedom counting in the $k=1$ string. The associated ghost sector would have central charge $c=0$, with four bosonic and four fermionic degrees of freedom being removed by the constraints. This leaves the topologically twisted $T^4$ as the only physical degrees of freedom, combined with an essentially topological theory on $AdS_3 \times S^3$ \cite{Eberhardt:2018ouy}. Our PBGS arguments provide further evidence for this claim, as well as machinery for how one could derive this action in principle (see \S\ref{sec:discussion} for a further discussion of this).

To further support the nature of our claim, we should highlight the analogous story in the four-dimensional hybrid formalism \cite{Berkovits:1994wr,Berkovits:1996bf} (see also \cite{Demulder:2023bux}), where a four-dimensional GS superstring was constructed, combined with a Calabi-Yau compactification. Unlike the six-dimensional hybrid formalism, all of the spacetime supersymmetries are already manifestly realised in the sigma model and one finds that there is a gauged $\mathcal{N} = 2$ superconformal algebra on the worldsheet. Upon transforming to twistor variables, it is also observed that the $GL(1)$ part of the constraint algebra precisely generates the scaling of the twistor variables.\\

Despite the lack of an explicit expression for the action \eqref{eq:superRS}, its existence allows us to deduce further properties of the string theory. Indeed, in \S\ref{sec:Twistor_string}, we will explain how by viewing \eqref{eq:boundary_theory} as a gauged $\mathcal{N} = 2$ string, we must introduce picture changing operators (PCOs) into the theory. We will be able to study the properties of such PCOs on fairly general grounds, and this will be sufficient to uniquely determine their form in the gauge-fixed theory \eqref{eq:boundary_theory} that we are well-acquainted with. We will determine the ghost content that must be combined with such PCOs and demonstrate the need for the additional physical constraint ${\cal Q}=0$ (\ref{eq:Q}).

\section{The hybrid formalism as a string theory in twistor space}\label{sec:Twistor_string}

We shall now explore the consequences of formulating our minitwistor space string theory in terms of an $\mathcal{N} = 2$ string. The motivation for doing so will be to demonstrate that the non-critical theory constructed in \eqref{eq:CP1_string} is incomplete and must be augmented by additional matter and a constraint condition. To this end, we will briefly review the role of picture changing operators (PCOs) in superstring theory, focusing on $\mathcal{N} = 2$ correlation functions. By consideration of the required PCOs for \eqref{eq:CP1_string}, we will be able to identify an additional constraint that must be imposed on the physical spectrum of the theory. We will also motivate the inclusion of a topologically twisted $\mathcal{M}_4$.

\subsection{From PBGS to ${\cal N}=2$}\label{sec:n=2_alg}

The action \eqref{eq:superRS} contains $\mathcal{N}=2$ worldsheet supersymmetry and so we will assume that the gauge constraints of our string theory on minitwistor space form an $\mathcal{N} = 2$ algebra, $(\mathcal{T, G^{ A }, J} )$. One might suspect that this algebra should be topologically twisted \cite{Gaberdiel:2021jrv,Gaberdiel:2021qbb}, but we will keep our assumptions sufficiently general. We will write the conformal weights as in Table \ref{tab:1}, where the topologically twisted $\mathcal{N} = 2$ algebra is given by $\lambda = \half$. Compared to \eqref{eq:CP1_string}, there are now additional constraints. Nevertheless, after gauge fixing $\mathcal{G}^A$ and $\mathcal{J}$ and integrating out the $\vartheta_A$, we should land on a theory given by the Goldstone modes, which is \eqref{eq:boundary_theory} with $\mathcal{C}_L = 0$.
\begin{table}[h]
    \centering
    \begin{tabular}{|c|c|}
        \hline
        Constraint & $h$\\
        \hline
       $\mathcal{T}$  & 2 \\
        $\mathcal{G}^+$ & $\frac{3}{2} - \lambda$ \\
        $\mathcal{G}^-$ & $\frac{3}{2} + \lambda$ \\
        $\mathcal{J}$ & 1\\
        \hline
    \end{tabular}
    \caption{The $\mathcal{N}=2$ worldsheet constraint algebra.}
    \label{tab:1}
\end{table}

The $\mathcal{G}^A$ generate superconformal transformations associated with shifts in $\vartheta_A$. These fermionic coordinates on the worldsheet are defined to be conjugate to $S^{+A-}_0$ and therefore we expect that $\mathcal{G}^A$ and $S^{+A-}_0$ do not commute. Yet, since the $\mathcal{G}^A$ generate gauge transformations, they must commute with global symmetries of the physical spectrum. As such, we conjecture that $S_0^{+A-}$ does not generate a symmetry of the physical spectrum. Extending this to the other patch of $\C\P^{1|2}$ (or equivalently lifting to projective coordinates), we anticipate that all $S_0^{\alpha A-}$ do not commute with $\mathcal{G}^A$.\footnote{In fact, if $[\mathcal{G}^A,J^+_0]=\{\mathcal{G}^A,S^{-A-}_0\} = 0,$ then $\{\mathcal{G}^A,S^{+A-}_0\} = 0$ from $[J^+_0,S^{-A-}_0] = S^{+A-}_0$. We have assumed $[\mathcal{G}^A,J^+_0]=0$ whilst $\{\mathcal{G}^A,S^{+A-}_0\} \neq 0$ and hence $\{\mathcal{G}^A,S^{-A-}_0\} \neq 0$.]} It was precisely these supersymmetries that were seen in \cite{Berkovits:1999im} to form unbroken symmetries, whose conjugate fermionic coordinates were not manifestly realised in that description.

By contrast, we conjecture that all other generators of $\mathfrak{psu}(1,1|2)$ should commute with $\mathcal{G}^A$. We saw that $J^+_0$ and $S^{+A+}_0$ were associated to broken symmetries in the previous section, which have associated local fields on the worldsheet. Lifting to projective coordinates we therefore expect $J^{\pm}_0$ and $S^{\alpha A+}_0$ to commute with $\mathcal{G}^A$. Moreover, the $R$-symmetry generated by $K^{\tilde{a}}_0$ and the boundary Hamiltonian $J^3_0$, conjugate to the worldsheet coordinate, should also commute with $\mathcal{G}^A$.

With these mild assumptions, we will be able to derive additional field content that must be added to \eqref{eq:CP1_string}, leading to the sigma model of \cite{Berkovits:1999im}. We hope that an explicit construction of \eqref{eq:superRS} will elucidate on the details of our proposal.\footnote{The role of the generator $\mathcal{J}$ is less clear and will not play an important role in the details of our argument. One might hope that it is naturally associated with the quotient imposed by $\mathcal{C}$. This is indeed the case for the four-dimensional hybrid formalism \cite{Berkovits:1994wr,Berkovits:1996bf}, where there is a gauged $\mathcal{N} = 2$ SCA whose $GL(1)$ component generates a scaling of the twistor variables.}

\subsection{$\mathcal{N}=2$ correlation functions and pictures}

In RNS superstring theory, there is a natural coupling between the gravitino and the supercurrent. The presence of this coupling leads to the insertion of picture changing operators, denoted $P_+$, on the worldsheet \cite{Witten:2012bh,Verlinde:1987sd}. These PCOs interpolate between the inequivalent vacua of the $\tilde{\beta}\tilde{\gamma}$ ghost system associated to gauge fixing the gravitino \cite{Friedan:1985ge,Polchinski:1998rr}.\footnote{We use tildes to distinguish these superconformal ghosts from the matter fields in (\ref{eq:boundary_theory}).} The number of PCOs that must be inserted is fixed by the requirement that the ghost current is not anomalous. This is significant, since it implies that if the non-critical theory \eqref{eq:CP1_string} is the gauge-fixed form of a theory with worldsheet supersymmetry, then we must include PCOs.

For the case of $\mathcal{N} = 2$, we require a PCO associated to each supersymmetry $\mathcal{G}^A$, which we denote by $P_+^A$. Moreover, there is a quantity called the instanton number associated to $\mathcal{J}$, which is analogous to picture numbers in the sense that it also labels the physical states.\footnote{Specifically, the current $\mathcal{J}$ couples to a holomorphic $(0,1)$-form, $a$, in the action with gauge equivalence $a \sim a + \mathrm{d}\Lambda$. That is to say, $a \in \text{Jac}(\Sigma_g)$, the space of flat $U(1)$ connections on a genus $g$ Riemann surface. The instanton number is defined as the degree of $a$, $\frac{1}{2\pi i}\int_{\Sigma}\mathrm{d}a \in \Z$ and hence, the integral over $\text{Jac}(\Sigma_g)$ yields a sum over instanton number in the path integral.}
There exist corresponding instanton number changing operators (ICOs), which are defined by \cite{Berkovits:1993he,Berkovits:1993xq,Berkovits:1994vy}
$$I = e^{\int \{Q_{BRST}, \tilde{b} \}}, \quad I^{-1} = e^{-\int \{Q_{BRST}, \tilde{b}\}},$$
where $\tilde{b}$ is the weight 1 fermionic ghost from the $(\tilde{b},\tilde{c})$ system introduced by the gauging of $\mathcal{J}$.

These PCOs and ICOs arrange themselves in $\mathcal{N} = 2$ correlation functions in the following way. We will work on the reduced space of the super Riemann surface implicit in \eqref{eq:superRS}, defined as the bosonic Riemann surface attained by setting the odd coordinates to zero (or equivalently integrating them out) \cite{Witten:2012bh,Witten:2012ga}. For a conventional $\mathcal{N}= 2$ algebra (where $\lambda = 0$ in Table \ref{tab:1}), the partition function on a genus $g$ Riemann surface of instanton number $n_I$ is \cite{Berkovits:1994vy},
$$A^{n_I}_g = \prod_{i=1}^g \int \mathrm{d}^2 M_i \, \prod_{j=1}^{3g-3} \int \mathrm{d}^2 m_j \, \left\langle  \left| \left( \int_{a_i} \tilde{b} \right)  \left( \int_{\Sigma} \mu_j b\right) (P_+^-)^{2g-2+n_I} (P_+^+)^{2g-2-n_I} I^{n_I} \tilde{c}(z_0) \right|^2 \right\rangle.$$
The $M_i$ are $U(1)$ moduli in Jac$(\Sigma_g)$, $m_j$ parameterise the moduli space of $\Sigma_g$, $a_i$ is the $i$-th $a$-cycle of $\Sigma_g$, $\mu_j$ are Beltrami differentials and $z_0$ is an arbitrary point on $\Sigma_g$ \cite{Nakahara:2003nw}. The full path integral involves a sum over integers $n_I$ and at the level of the partition function, we have non-vanishing contributions only for $|n_I| \leq 2g-2$ (similar restrictions will exist for correlation functions once vertex operators have been inserted). Moreover, topologically twisting the $\mathcal{N} = 2$ algebra ($\lambda \neq 0$) is equivalent to switching on a background gauge field for $\mathcal{J}$ in the untwisted theory (see \S16 of \cite{Hori:2003ic}). Taking account of this in the partition function removes $(g-1)$ of the insertions of $I$, but the overall structure is the same as above.

$\mathcal{N}=2$ correlation functions that are defined directly on the reduced space also require a sum over instanton numbers. We need to be able to construct this same sum in the non-critical theory \eqref{eq:CP1_string}, also defined on the reduced space. This requires the inclusion of at least one PCO, to play the equivalent role to
$P_+^- (P_+^+)^{-1} I,$ which we call $\mathcal{P_+}$,
to interpolate between the different sectors.\footnote{In perhaps more familiar language to the $k=1$ string literature, this sum over instanton number is equivalent to the sum over choices of Beltrami differentials when we recast the $\mathcal{N} = 2$ string as a small $\mathcal{N} = 4$ topological string \cite{Berkovits:1994vy}, which is a consequence of the $SU(2)$ outer automorphism of the small $\mathcal{N} = 4$ algebra.}

\subsection{From picture changing to the hybrid formalism at $k=1$}\label{sec:ground_up}

To recap, in \S\ref{sec:PBGS} we presented an argument for a reformulation of \eqref{eq:CP1_string} in which the $S^{\alpha A\pm}$ are realised through a combination of worldsheet and target space supersymmetries. Given the central importance of PCOs in constructing correlation functions in worldsheet supersymmetric theories, we now turn to the question of what operator might play the role of a PCO, $\mathcal{P}_+$, in our string theory on minitwistor space? That is to say, we would like to construct the possible candidates for $\mathcal{P}_+$ built from the field content of \eqref{eq:CP1_string}, that satisfy the desired properties of a PCO. We shall not restrict ourselves to working solely on the patch $\omega_- \neq 0$, but instead work with the homogeneous coordinates in \eqref{eq:CP1_string}. Moreover, we need only focus on the holomorphic sector, with matter content
\begin{equation*}
    (\mathcal{Z}_L)^M = \left(
\begin{array}{c}
\omega_{\alpha}	\\
\psi_{ A }
\end{array}
\right),	\qquad	
(\mathcal{W}_L)_M = \left(
\begin{array}{cc}
\lambda^{\alpha}	&,
\chi^{ A }
\end{array}
\right).
\end{equation*}

Consider the ansatz
$$\mathcal{P}_+ = f_{ghost} \mathcal{O}_{matter},$$
for the PCO in \eqref{eq:CP1_string}. Since this operator should be inherently related to the supercurrents $\mathcal{G}^A$, we propose that
\begin{equation}\label{eq:O_constraints}
    \left[ \mathcal{O}_{matter}(z), A_0 \right]_{\pm} = 0, \qquad \left[ \mathcal{O}_{matter}(z), S^{\alpha A -}_0 \right]_{\pm} \neq 0,
\end{equation}
for $A = J^a, K^{\tilde{a}}$ or $S^{\alpha A +}$. We use the anticommutator if both operators are fermionic, since we a priori make no assumption about the statistics of $\mathcal{O}_{matter}$. We expect this behaviour because the supercurrents $\mathcal{G}^A$ should (anti-)commute with all global $\mathfrak{psu}(1,1|2)_1$ generators, except for $S^{\alpha A -}_0$, the unbroken supersymmetry generators (see \S\ref{sec:n=2_alg}). Moreover, we require that
\begin{equation}\label{eq:O_scaling_constraint}
    \left[ \mathcal{O}_{matter}(z), (\mathcal{C}_L)_n \right]_- = 0
\end{equation}
for all $n \geq 0$, such that $\mathcal{P}_+$ commutes with the usual $GL(1)$ scaling physical state condition of \eqref{eq:CP1_string}, in order to map between physical states. This will additionally impose that $\mathcal{O}_{matter}$ is well-defined in terms of the $\mathfrak{psu}(1,1|2)_1$ WZW model from which the $k=1$ string is traditionally defined.

In Appendix \ref{sec:uniqueness}, we classify the possible conformal primaries that satisfy \eqref{eq:O_constraints} and \eqref{eq:O_scaling_constraint} in the space of fields constructed from finite sums and products of the free fields (i.e. the supertwistors) and their derivatives. The solution is almost unique, with all solutions being constructed from
$$\mathcal{Q}= \frac{1}{2}\epsilon^{\alpha\beta}\epsilon^{ A  B } \chi_{ A } \chi_{ B } \omega_{\alpha} \p \omega_{\beta}$$
and its derivatives. More precisely, there is a family of solutions given by
$$\mathcal{O}^{(\ell)} = \mathcal{Q}\p^2 \mathcal{Q} \cdots \p^{2(\ell-1)}\mathcal{Q},$$
for all $\ell \geq 1$. We deduce that the PCO must take the form $\mathcal{P}_+ = f_{ghost}^{(\ell)} \mathcal{O}^{(\ell)}$ for some $\ell$. Up to an overall factor that we shall ignore, this can be rewritten in terms of the free fields as
$$
\mathcal{O}^{(\ell)} = \Big(\omega_+\p\omega_- - \omega_-\p\omega_+\Big)^{\ell}\,\prod_{i=0}^{\ell - 1} \left( \p^{i}\chi^+ \p^i \chi^- \right) ,
$$
or even more helpfully, by bosonising $\chi^+ = e^{iq_1}$, $\chi^- = e^{-iq_2}$,
$$\mathcal{O}^{(\ell)} = e^{i\ell(q_1-q_2)}\Big(\omega_+\p\omega_- - \omega_-\p\omega_+\Big)^{\ell}.$$
This has weight $h\left( \mathcal{O}^{(\ell)} \right) = \ell(\ell + 2)$ and OPE
$$\mathcal{O}^{(\ell_1)}(z) \mathcal{O}^{(\ell_2)}(w) \sim (z-w)^{2\ell_1\ell_2} \mathcal{O}^{(\ell_1+\ell_2)} + \dots$$
As such, $\mathcal{O}^{(\ell)}$ is the leading order contribution from fusing $\ell$ copies of $\mathcal{Q}$ together.\\

Our next task is to determine the form of $f_{ghost}^{(\ell)}$. Let us assume that the ghost part can be written in the form $f_{ghost}^{(\ell)} = e^{\varphi}$, where $\varphi$ is a linear dilaton with some background charge that we have to determine \cite{DiFrancesco:1997nk}. This is in fact uniquely determined for fixed $\ell$, by requiring that $\mathcal{P}_+$ has the expected behaviour of a PCO: vanishing conformal weight and a regular (but non-vanishing) OPE with itself,
$$\mathcal{P}_+(z) \mathcal{P}_+(w) = \text{ const.} + \dots$$
We deduce that
$$h(e^{\varphi}) = -\ell(\ell+2), \qquad e^{\varphi(z)}e^{\varphi(w)} = (z-w)^{-2\ell^2}e^{\varphi(z)+\varphi(w)},$$
by comparison with the conformal weight and OPE structure of $\mathcal{O}^{(\ell)}$. In fact, one can determine that
$$h(e^{m\varphi}) = -m\ell(m\ell + 2),$$
which fixes the background charge.

It is possible to rewrite this ghost CFT in language adapted to the hybrid formalism of \cite{Berkovits:1999im}. Consider two linear dilaton systems, $\rho$ and $iH$, where
\begin{equation}\label{eq:rho_iH_systems}
\begin{aligned}
    \rho(z)\rho(w) &\sim -\log (z-w), & h(e^{m\rho}) &= -\frac{1}{2}m(m - 3),\\
    iH(z) iH(w) &\sim +2\log(z-w), & h(e^{imH}) &= m(m-1). 
\end{aligned}
\end{equation}
One can verify that $\varphi = -\ell(2\rho + iH)$ precisely satisfies the required properties for our ghost contribution in $\mathcal{P}_+$. That is,
$$f_{ghost}^{(\ell)}\mathcal{O}^{(\ell)} = e^{-\ell(2\rho + iH)}\mathcal{Q}\p^2\mathcal{Q} \cdots \p^{2(\ell-1)} \mathcal{Q} = \left( e^{-2\rho-iH} \mathcal{Q} \right)^{\ell}.$$
Hence, the simplest PCO of this kind is given by $\ell=1$ and so we propose that the PCO for \eqref{eq:CP1_string} is
\begin{equation}\label{eq:hybrid_PCO}
    \mathcal{P}_+ = e^{-2\rho-iH}\mathcal{Q}.
\end{equation}
Indeed, this is precisely the operator that has been observed to be so crucial in the $k=1$ string for generating non-vanishing correlation functions \cite{Dei:2020zui,McStay:2023thk}.\\

With minimal  assumptions, we have been able to show that the simplest form of the picture changing operator, consistent with the PBGS we see in the boundary theory is given by \eqref{eq:hybrid_PCO}. The existence of a PCO tells us something about the BRST operator and hence the physical constraints in the theory, since a PCO should map between physical states. Introducing a conjugate pair of fermionic ghosts $(\eta, \xi)$ where $h(\eta) = 1$ and $h(\xi) = 0$, the picture changing and BRST operators are conventionally related by their action on a physical state as $\mathcal{P}_+ \Psi_{phys} = [Q_{BRST},\xi \Psi_{phys}]_{\pm}$. The BRST current then takes the form
$$j_{BRST} = \eta e^{-2\rho-iH}\mathcal{Q} + \dots,$$
where there will be additional terms that commute with $\xi$. That $\mathcal{P}_+ \Psi_{phys}$ is not BRST-exact stems from the fact that the zero mode of $\xi$ is not in the (small) physical Hilbert space. This is ensured by the supplementary constraint $$\eta_0 \Psi_{phys} = 0.$$ Thus, the existence of a PCO and requiring that it maps between physical states implies that an additional term is included in the BRST operator.\footnote{Ignoring the $GL(1)$ scaling constraint for now, the worldsheet symmetry algebra is given by
$$
    [L_m,L_n] = (m-n)L_{m+n} + \frac{c}{12}(m^3-m)\delta_{m+n,0},\quad
    [L_m,\mathcal{Q}_n] = (2m-n) \mathcal{Q}_{m+n}, \quad [\mathcal{Q}_m, \mathcal{Q}_n] = 0,
$$
where the central charge, $c$, will vanish when the theory is critical. This Abelian extension of the Virasoro algebra is very similar to the algebra $\mathcal{W}(0,-2)$, a deformation of the BMS$_3$ algebra (see \cite{Figueroa-OFarrill:2024wgs} and references therein). In contrast to traditional Abelian extensions, $\mathcal{Q}$ has a double zero in the OPE with itself. As such, the connection between the $k=1$ string and $\mathcal{W}(0,\lambda)$ algebras may just be circumstantial. Nevertheless, it is interesting that the minimal tension string on $AdS_3$ admits such an algebra, since the BMS$_3$ algebra has played a prominent role in the study of tensionless strings on flat space \cite{Bagchi:2013bga}.
}

Additional fields are not required for this construction since the $\rho$ and $iH$ fields can again be used to describe an appropriate candidate for the $(\eta,\xi)$ system. We define
$$\eta = e^{\rho + iH}, \qquad \xi = e^{-\rho-iH}.$$
Indeed, this combination is independent of $2\rho + iH$, such that we can effectively define $\varphi$ and $(\eta,\xi)$ using $\rho$ and $iH$. We deduce that
\begin{equation}\label{eq:BRST_Q}
    j_{BRST} = e^{-\rho}\mathcal{Q} + \dots
\end{equation}
whilst we also impose
\begin{equation}\label{eq:G_tilde}
    (e^{\rho+iH})_0 \Psi_{phys} = 0,
\end{equation}
on physical states. The picture number can simply be defined as the charge under $J_{pic} = \p\rho$, such that $P(e^{m\rho}) = -m$ and $\mathcal{P}_+ = e^{-2\rho-iH}\mathcal{Q}$ clearly raises the picture.\\

To summarise; we have argued that when we impose the constraint ${\cal C}=0$, PBGS suggests a description of the theory in which all $S^{\alpha AI}$ global supersymmetries are realised and there is some, possibly twisted, worldsheet supersymmetry. Natural assumptions lead to a minimal form of a picture changing operator (\ref{eq:hybrid_PCO}), which in turn implies that an additional term must be added to the BRST operator of \eqref{eq:BRST_CT}, along with a set of additional linear dilaton fields $\rho$ and $iH$, which we choose to interpret as ghosts. The suggestion then is that the initial description of the $AdS_3\times S^3$ worldsheet theory in \S\ref{sec:almost} was incomplete and requires additional constraints and ghost fields. These modifications bring the theory almost to the form dictated by the hybrid formalism. The $\rho$ ghost carries a central charge of $c = +28$, which cancels the conformal anomaly of the theory in \S\ref{sec:almost}. The inclusion of $iH$ then reintroduces a conformal anomaly once again, which by \eqref{eq:rho_iH_systems} has a background charge of $-\sqrt{2}$ and hence a central charge of $c=-5$. A natural  fix is to identify this as the $U(1)$-potential for a topologically twisted, hyperk\"{a}hler $\mathcal{M}_4$, such as $T^4$. That is, we choose to extend the $iH$ CFT to include a sector that is $c=0$ overall.\footnote{To give an explicit example, consider a topologically twisted $\mathcal{M}_4 = T^4$. This is constructed from complex bosons $(\p X^i,\p\bar{X}^i)$ and complex fermions $(\lambda^i,\bar{\lambda}^i)$ for $i = 1,2$. Bosonising the fermions as $\lambda^i = e^{iH^i}$ and $\bar{\lambda}^i = e^{-iH^i}$ for $h(\lambda^i) =1-h(\bar{\lambda}^i)= 0$, we identify $iH = iH^1 + iH^2$. In this sense, the $c=-5$ contribution from the $iH$ CFT is cancelled by contributions from $iH^1-iH^2$ as well as the bosons. See, for example, Appendix B of \cite{Gaberdiel:2021njm}.}

Some comments are in order. Firstly, it is significant that our gauge fixing procedure and the search for PCOs identified that the $U(1)$ potential $iH$ was necessary for constructing a string theory on the minitwistor space of $AdS_3 \times S^3$. It suggests that it is not sufficient to only consider the $c=0$ sector of $\mathfrak{psu}(1,1|2)_1$ plus $(b,c)$ and $\rho$ ghosts, but non-vanishing correlation functions require that further ghost and matter content is included in the theory. Moreover, we have only provided arguments for minimal field content that must be included in such a string theory. If we had an explicit realisation of \eqref{eq:superRS}, it should be possible to trace through the gauge fixing process and precisely identify how the ghost content for gauging the $\mathcal{N} = 2$ algebra maps onto the $\rho$ and $iH$ ghosts we have seen here, as well as how $\mathcal{Q}$ appears as a gauge constraint. In doing so, extra constraints on the form of $\mathcal{M}_4$ may arise and we expect that the ghosts for gauging $\mathcal{C}$ in the twistor description should also be built into the formalism. We hope that the extension of $iH$ to include $\mathcal{M}_4$ will then become more natural, in a similar way to the four-dimensional hybrid string where a Calabi-Yau compactification is required \cite{Berkovits:1994wr}. It is also possible that an explicit construction will lead to alternative models, other than that given by the hybrid formalism.\\

We now have all of the ingredients of the hybrid formalism at $k=1$. The form of the matter sector has been deduced from the twistor geometry. A $(b,c)$ ghost sector is introduced in the usual way. The $\rho$ ghost and the ${\cal Q}$ constraint have been motivated from the PBGS induced by the twistor scaling constraint ${\cal C}_L=0$. The BRST operator is built from several pieces: \eqref{eq:BRST_CT}, \eqref{eq:BRST_Q} and any contribution from the compact $\mathcal{M}_4$, $j_C$. We deduce that
$$j_{BRST} = e^{i\sigma}\mathcal{T} + e^{-\rho}\mathcal{Q} + v \mathcal{C}_L + j_{C},$$
up to $\p$-exact terms. Moreover, our restriction to the small Hilbert space through \eqref{eq:G_tilde} can be interpreted as a second BRST operator and our physical states will be defined in double cohomology. A natural picture for compactification independent vertex operators is given by $\Psi = e^{2\rho+i\sigma+iH}\Phi$, for which
$$L_n\Phi = \mathcal{Q}_n\Phi = (\mathcal{C}_L)_n\Phi = 0,$$
for all $n\geq 0$.

In principle, we could rewrite this sigma model in the language of $\mathcal{N} = 4$ topological strings \cite{Berkovits:1993xq,Berkovits:1994vy,Gaberdiel:2022als} and land precisely on the hybrid formalism of \cite{Berkovits:1999im}.

\section{Vertex operators and bulk physics}\label{sec:bulk_VO}

We have now shown that the hybrid string at $k=1$ is a natural prescription for describing superstring theory on minitwistor space. In this section and the next, we shall study the vertex operators of the $k=1$ string. The spectrum and correlation functions of the $k=1$ string on $AdS_3 \times S^3 \times T^4$ have been shown to precisely match the single-cycle twisted sector of the symmetric orbifold Sym$^N(T^4)$ in the large $N$ limit \cite{Eberhardt:2018ouy,Eberhardt:2019ywk} and a key feature of the analysis in \cite{Eberhardt:2018ouy,Eberhardt:2019ywk} is the spectral flow, which generates an automorphism of $\mathfrak{psu}(1,1|2)_1$. The spectral flow is a generic feature of strings on $AdS_3$, since it generates inequivalent representations of $SL(2;\R)$ and ensures modular invariance of the theory \cite{Henningson:1991jc,Maldacena:2000hw}. The amount of spectral flow is labelled by the parameter $w \in \mathbb{Z}$ which for the continuous representations of $SL(2;\R)$ (the only representations present at $k=1$) can be interpreted as long strings near the boundary of $AdS_3$, winding around the boundary $w$ times. Without loss of generality, we may restrict to the case of $w\geq 0$ as is the case in the analysis of \cite{Eberhardt:2018ouy}. They observed that the unitary representation of $\mathfrak{psu}(1,1|2)_1$ can be built from the continuous representations of $\mathfrak{sl}(2,\R)$, however they were only unitary for $w>0$. Such representations describe long strings that wind the boundary with positive orientation \cite{Knighton:2024pqh}. We will discuss this unitary sector of the theory in this section, whilst the $w=0$ sector is discussed in \S\ref{sec:unflowed}.
\\

There are many expressions for the vertex operators of the $k=1$ string in the literature \cite{Naderi:2022bus,Dei:2023ivl,Knighton:2023mhq}. We have shown in \S\ref{sec:euclidean_AdS3}, that in the case of Euclidean $AdS_3$, our vertex operators are inserted at the boundary which is a $\C\P^1$. The fact that the boundary twistor space may be identified with the boundary spacetime directly means that the interpretation of the vertex operators in terms of the boundary ambitwistor space is straightforward. Our interest in this section is to better understand how these objects can be understood in terms of the bulk variables. We begin by recasting the vertex operators with spectral flow $w\geq 1$ in terms of naturally projective objects that may be understood in terms of bulk minitwistors and then we propose a simple analogue of the Penrose transform to write the vertex operators in terms of bulk spacetime variables.

The $k=1$ string has been extensively studied in the near-boundary approximation, where the interaction term in \eqref{eq:Waki_action} is neglected. It has been suggested \cite{Dei:2023ivl,McStay:2023thk,Sriprachyakul:2024gyl} that this near-boundary approximation is exact at $k=1$ and that the worldsheet is pinned to the boundary \cite{Eberhardt:2019ywk}. This fact was explicitly demonstrated in \S\ref{sec:boundary} for Euclidean signature, through the imposition of the constraint $U = \tilde{U} = 0$. As a consequence of this localisation to the boundary, the $k=1$ string is only sensitive to the asymptotic geometry, as shown in \cite{Eberhardt:2021jvj}. Given the role of the scaling constraints in localising the physics to the boundary, it is interesting to ask whether bulk information can be recovered by loosening the condition $U=\tilde{U}=0$. We will demonstrate how, using the bulk incidence relations introduced in \S\ref{sec:euclidean_AdS3} for Euclidean $AdS_3$, the $k=1$ string vertex operators do indeed still encode information about the bulk. We do not mean this in the sense that one can identify on-shell bulk observables in the Hilbert space, but rather that the vertex operators encode important information about structures in the bulk, such as the bulk-boundary propagator. In order to recover spacetime objects we will need an analogue of the Penrose transform in field theory. A crucial step is to rewrite the vertex operators in terms of the bulk minitwistor variables, which we do in \S\ref{sec:examples_of_proj_VO} and \S\ref{sec:general_proj_VO}. In \S\ref{sec:bulk_physics} we adapt the Penrose transform in \cite{Bu:2023cef} to show how the bulk-boundary propagator can be recovered from the $w=1$ vertex operators.\footnote{One might object that we are far from the supergravity limit here so it would be surprising to recover the bulk-boundary propagator. These propagators can also be thought of as the $H_3^+$ analogue of plane waves. Just as translation invariance gives rise to plane waves of the form $e^{ik_L\cdot X_L+ik_R\cdot X_R}$ for all values of $\alpha'$, so we expect the basic structure $A_{\alpha}g^{\alpha\dot{\alpha}}\bar{A}_{\dot{\alpha}} = \langle Ag\bar{A}]$ to be present in $H_3^+$ at all values of $k$, with $A_{\alpha}$ defined as in \eqref{data}. This is discussed explicitly in \cite{Teschner:1997fv}, where the kernel $\langle Ag \bar{A}]$ plays the role in $SL(2;\C)/SU(2)$ that $e^{ik\cdot X}$ plays in the Fourier transform on flat spacetime.}

Another motivation for studying how the worldsheet theory encodes bulk physics is because understanding how to relate twistor and spacetime descriptions is important for any attempt to build a string theory in the twistor space of $AdS_5$. In that case, to recover spacetime physics in the bulk and boundary, one would expect to have to perform a map from the bulk twistor space to $AdS_5$ \emph{and} also a map from the boundary ambitwistor space to the boundary spacetime to recover the conventional description of ${\cal N}=4$ SYM. The map from (ambi)twistor space to spacetime objects in this case is therefore unavoidable. This is not such a direct problem for $AdS_3$, where the twistor space of the boundary $\C\P^1$ is just the spacetime $\C\P^1$.

\subsection{Examples of projective vertex operators in the string theory}\label{sec:examples_of_proj_VO}

We shall focus on the vertex operators for the ground states of the theory. Excited states may be found by applying the DDF operators given in \cite{Naderi:2022bus,Dei:2023ivl}. Our task in this subsection will be to rewrite these vertex operators in terms of the bulk supertwistor variables, rather than the usual formulation in terms of the boundary spacetime fields $(\gamma,\theta_A)$. The form of the ground states of the theory split into two categories, depending on if they live in a sector with an odd or even amount of spectral flow, $w \in \N$. For $w$ odd, the states transform as singlets under the $\mathfrak{su}(2)_1$ $R$-symmetry, whilst they transform as doublets for even $w$ \cite{Dei:2020zui}. Following \cite{Dei:2023ivl}, we bosonise the boundary spacetime fermions of \S\ref{sec:boundary_pbgs} as
\begin{align*}
    \theta_- &= \frac{\psi_-}{\omega_-} = e^{if_1},     &  \theta_+ &= \frac{\psi_+}{\omega_-}= e^{-if_2}, \\
    p^- &= \chi^-\omega_- =e^{-if_1},   &   p^+ &= \chi^+ \omega_- = e^{if_2}.
\end{align*}
In what follows, when we move between free field variables and their bosonisation, we will not be careful with cocycle factors \cite{Kostelecky:1986xg}. As such, our vertex operator expressions will all be correct only up to an overall scale factor. 

The ground states with $w$ units of spectral flow are given by \cite{Dei:2023ivl}
 $$
 \Omega_w(x,z)= \exp\left[\frac{w+1}{2}(if_1-if_2)\right]\left(\frac{\p^w\gamma}{w!}\right)^{-m_w}\delta_w(\gamma(z)-x)e^{2\rho+i\sigma+iH},
 $$
for $w$ odd, whilst
$$
 \Omega^{\pm}_w(x,z)= \exp\left[\pm \frac{(if_1+if_2)}{2}\right]\exp\left[\frac{w+1}{2}(if_1-if_2)\right]\left(\frac{\p^w\gamma}{w!}\right)^{-m^{\pm}_w}\delta_w(\gamma(z)-x)e^{2\rho+i\sigma+iH},
 $$
for even $w$, where
 \begin{equation}\label{m}
 m_w=-\frac{(w-1)^2}{4w},   \qquad  m^{\pm}_w=-\frac{w-2}{4}.
 \end{equation}
Unintegrated vertex operators are given by taking holomorphic and anti-holomorphic combinations in the usual way. The vertex operators are labelled by their location on the worldsheet $(z,\bar{z})\in\Sigma$ and their location on the boundary $(x,\bar{x})\in\C\P^1$. The notation for the delta-functions follows \cite{Dei:2023ivl,Witten:2012bh},
$$
\delta_w(\gamma(z)-x):=\delta(\gamma(z)-x)\prod_{j=1}^{w-1}\delta(\partial^j\gamma(z)).
$$
This ensures $\gamma(z)$ takes the form
$$
\gamma(z)=x_i+a_i(z-z_i)^w+...
$$
near $z=z_i$. In what follows, we will find it helpful to repackage the boundary insertion point, $x$, into a homogeneous coordinate as 
\begin{equation}\label{data}
A_{\alpha}(x)=\left(\begin{array}{c}
     A_+  \\
     A_- 
\end{array}\right):=A_-\left(\begin{array}{c}
     x  \\
     1 
\end{array}\right).
\end{equation}
The point at infinity is given by
$$
(A_{\infty})_{\alpha}:=A_-\left(\begin{array}{c}
     1  \\
     0 
\end{array}\right).
$$
We could choose to set $A_-=1$, but we shall keep the $A_-$ dependence explicit in much of what follows to help clarify the calculations.

\subsubsection*{The $w=1$ case}

As a warm-up, let us look at the $w=1$ sector. In this case $m_w=0$ and so the derivative term drops out leaving
$$
 \Omega_1(x,z)= \frac{\psi_-\psi_+}{(\omega_-)^2}\delta(\gamma(z)-x)e^{2\rho+i\sigma+iH}.
 $$
This definition for the operator is well-defined only in a patch of the boundary theory. We would therefore like to recast this operator in homogeneous coordinates so that it can be understood in terms of the bulk minitwistor coordinates. This, in turn, will allow us to impose the bulk incidence relations \eqref{eq:incidence_g_inverse} and give a bulk interpretation for expressions involving the vertex operators.

A useful tool to consider is the projective delta-function,\footnote{In contrast to the projective delta functions defined in say \cite{Adamo:2016rtr}, the one used here is a function, not a $(0,1)$-form.}
 $$
 \delta_m(A(x),\omega(z))=\int \mathrm{d}s\, s^{-m-1}\delta^2(A_{\alpha}-s\omega_{\alpha}),
 $$
 where $\delta^2(A_{\alpha}-s\omega_{\alpha})=\delta(A_+-s\omega_+) \delta(A_--s\omega_-)$. To account for the fact that $\omega$ has scaling weight $-1/2$ under ${\cal C}_L$ and $+1/2$ under $L_0$ and $A$ is invariant under both, we introduce the parameter $s$ which is defined to have scaling weight $(1/2,-1/2)$ under ${\cal C}_L$ and $L_0$. $\delta_m(A(x),\omega(z))$ then has weight $(-m/2,m/2)$. Given the boundary incidence relation $\omega_+ - \gamma\omega_-=0$ and $A_+-xA_-=0$ and assuming we work in a patch where $\omega_-,A_-\neq 0$, the integral can be evaluated to give
 \begin{eqnarray}
     \delta_m(A(x),\omega(z))= 
\frac{\omega_-^m}{A_-^{m+2}} \delta(\gamma(z)-x).
 \end{eqnarray}
Any observable must be invariant under ${\cal C}_L$, so it is safe to set $A_-=1$. It is then not hard to see that a candidate for the  operator $\Omega_1(x,z)$ is
$$
 \Omega_1(x,z)= C_1 \, \psi_-\psi_+\,\delta_{-2}(A(x),\omega(z))\,e^{2\rho+i\sigma+iH},
 $$
where we have chosen the ${\cal C}_L$-weight of the projective delta-function to balance that of the fermionic terms and $C_1$ is an arbitrary constant. Note that the ghost terms have conformal weight $h\left( e^{2\rho+i\sigma+iH} \right)=0$ and are invariant under ${\cal C}_L$, so $\Omega_1(x,z)$ is scaling and conformally invariant.

\subsubsection*{The $w=2$ case}

The first task is to write $\delta_w(\gamma-x)$ projectively in terms of the twistor coordinates. The new feature in this pair of operators is the $\delta(\partial\gamma)$ in $\delta_2(\gamma(z)-x)$. It is useful to define
$$
 \delta_m(A(x),\partial^w\omega(z))=\int ds_w\, s_w^{-m-1}\delta^2(A_{\alpha}-s_w\,\partial^w\omega_{\alpha}),
 $$
where $s_w$ has scaling and conformal weights $(1/2, -w-1/2)$. It is not hard to show that\footnote{It is important to note that, by $\partial\gamma/\partial\omega_-$ we intend
$$
\frac{\partial\gamma}{\partial\omega_-}=\frac{\partial\gamma}{\partial z}\left(\frac{\partial\omega_-}{\partial z}\right)^{-1},
$$
\emph{not} the derivative of $\gamma$ with respect to $\omega_-$.} 

$$
\delta_m(A(x),\partial\omega(z))= \frac{(\partial \omega_-)^m}{A_-^{m+2}}\delta\left(\frac{A_+}{A_-} - \omega_-\frac{\partial\gamma}{\partial\omega_-} - \gamma\right).
$$
Restricting to the support of $\delta_{n}(A(x),\omega(z))$, this simplifies to
$$
\delta_m(A(x),\partial\omega(z))\sim \frac{(\partial \omega_-)^m}{A_-^{m+2}}\frac{\partial \omega_-}{\omega_-}\delta\left(\partial\gamma\right).
$$
The scaling and conformal weights of $\delta_m(A(x),\partial\omega(z))$ are $(-m/2,3m/2)$. We then see that
\begin{eqnarray}
    \Omega_2^{\pm}&=&\frac{\psi_-\partial\psi_{\mp}\psi_+}{(\omega_-)^3}\delta(\gamma-x)\delta(\partial\gamma)\,e^{2\rho+i\sigma+iH}\nonumber\\
    &=&\frac{\psi_-\partial\psi_{\mp}\psi_+}{(\omega_-)^3} \frac{A_-^{m+n+4}}{(\partial\omega_-)^{n+1}(\omega_-)^{m-1}}\delta_m(A,\omega)\delta_n(A,\partial\omega)\,e^{2\rho+i\sigma+iH}.
\end{eqnarray}
Note that, in passing from the bosonised variables to $\theta_A = \psi_A /\omega_-$, the derivative must act on the fermionic numerator to achieve a non-vanishing contribution. We can remove the pre-factors of $\omega_-$ and $\partial \omega_-$ in this expression by setting $m=-2$ and $n=-1$. The resulting object
$$
\Omega_2^{\pm}= C_2^{\pm} \,\psi_-\partial\psi_{\mp}\psi_+\,\delta_{-2}(A,\omega)\delta_{-1}(A,\partial\omega)\,e^{2\rho+i\sigma+iH},
$$
is invariant under projective scaling and conformal transformations.

\subsubsection*{The $w=3$ case}

The last novel ingredient, the $(\partial^w\gamma)^{m_w}$ term, first appears when $w=3$. It is not hard to show that, on the support of $\delta_m(A,\omega)\delta_n(A,\partial\omega)\delta_p(A,\partial^2\omega),$
$$
\partial^3\gamma= \frac{\langle A,\partial^3\omega\rangle}{\langle A_{\infty},\omega\rangle}.
$$
This has scaling and conformal weight $(0,3)$ and $m_3=-\frac{1}{3}$. Similar to the $w=2$ case above, it is straightforward to show
$$
\delta_m(A,\omega)\delta_n(A,\partial\omega)\delta_p(A,\partial^2\omega)= \frac{1}{A_-^{m+n+p+6}}\frac{(\omega_-)^{m+1}}{\omega_-}\frac{(\partial\omega_-)^{n+1}}{\omega_-}\frac{(\partial^2\omega_-)^{p+1}}{\omega_-}\delta_3(\gamma-x).
$$
Moreover, by considering a few examples, one can see that this result generalises. Consider
\begin{eqnarray}
\delta_{-m_0}(A,\omega)\delta_{-m_1}(A,\partial \omega) \cdots \delta_{-m_{w-1}}(A,\partial^{w-1} \omega) &=& {\cal F}\,\delta\left(\frac{A_+}{A_-}-\frac{\omega_+}{\omega_-}\right)\delta\left(\frac{A_+}{A_-}-\frac{\partial\omega_+}{\partial\omega_-}\right)\cdots \nonumber\\
&&\cdots\delta\left(\frac{A_+}{A_-}-\frac{\partial^{w-1}\omega_+}{\partial^{w-1}\omega_-}\right),\nonumber
\end{eqnarray}
where ${\cal F}$ is a rational function of $A_-$, $\omega_-$ and its derivatives that depends on the parameters $m_n$. Imposing the boundary incidence relation $\omega_+ - \gamma \omega_-=0$ and $A_+ - xA_-=0$,
\begin{eqnarray}
&&\delta_{-m_0}(A,\omega)\delta_{-m_1}(A,\partial \omega)\cdots\delta_{-m_{w-1}}(A,\partial^{w-1} \omega)\nonumber\\
&&={\cal F}\,\delta\left(x - \gamma\right)\delta\left(x - \gamma - \omega_-\frac{\partial \gamma}{\partial\omega_-}\right)\cdots\delta\left(x - \frac{1}{\partial^{w-1}\omega_-}\sum_{n=0}^{w-1}\left(\begin{array}{c}
     w-1  \\
     n 
\end{array}\right)\partial^n\gamma\partial^{w-1-n}\omega_- \right).\nonumber
\end{eqnarray}
On the support of the first delta function, the second delta function becomes $\frac{\p\omega_-}{\omega_-} \delta(\partial\gamma)$. Similarly, on the support of the first two delta functions, the third delta function is proportional to $\delta(\partial^2\gamma)$ and so on, giving
$$
\delta_{-m_0}(A,\omega)\delta_{-m_1}(A,\partial \omega) \cdots \delta_{-m_{w-1}}(A,\partial^{w-1} \omega)\propto \delta_w(\gamma-x).
$$
This will be useful in the general case. For $\Omega_3$, the obvious choices for $\{m_n\}$ give
$$
 \Omega_3=C_3\, (\psi_-\partial\psi_-)(\psi_+\partial\psi_+) \,\left( \frac{\langle A,\partial^3\omega\rangle}{\langle A_{\infty},\omega\rangle} \right)^{\frac{1}{3}}\,\delta_{-2}(A,\omega)\prod_{n=1,2}\delta_{-1}(A,\partial^n\omega)\,e^{2\rho+i\sigma+iH}.
$$

\subsection{General projective ground state vertex operators}\label{sec:general_proj_VO}

The above examples can be straightforwardly generalised. For $w$ odd we have
$$
    \Omega_w=C_{w}\prod_{j=0}^{\frac{1}{2}(w-1)}\partial^j\psi_- \prod_{k=0}^{\frac{1}{2}(w-1)}\partial^k\psi_+\,\left( \frac{\langle A,\partial^w\omega\rangle}{\langle A_{\infty},\omega \rangle} \right)^{-m_w}\,\delta_{-2}(A,\omega)\prod_{n=1}^{w-1}\delta_{-1}(A,\partial^n\omega)\,e^{2\rho+i\sigma+iH},
$$
where $C_{w}$ is a normalization constant, which depends on $A_-$, and $m_w$ is given above in (\ref{m}). It is simple to check that the expression has overall scaling charge zero, as it should. The conformal weights may be simply calculated as
$$
h[\Omega_w]=\left(\frac{(w+1)^2}{4}\right)+\left(\frac{(w-1)^2}{4}\right)+\left(-\frac{1}{2}(w^2+1)\right)=0,
$$
where the terms in brackets correspond to the conformal weights of the fermion, $\partial^w\omega$, and projective delta-function terms in $\Omega_w$.

For $w$ even we have
$$
    \Omega^{\pm}_w=C_{w}^{\pm}\prod_{j=0}^{\frac{w}{2}}\partial^j\psi_{\mp} \prod_{k=0}^{\frac{w}{2} - 1}\partial^k\psi_{\pm}\,\left( \frac{\langle A,\partial^w\omega\rangle}{\langle A_{\infty},\omega\rangle} \right)^{-m_w^{\pm}}\,\delta_{-2}(A,\omega)\prod_{n=1}^{w-1}\delta_{-1}(A,\partial^n\omega)\,e^{2\rho+i\sigma+iH},
$$
where $C_{w}^{\pm}$ is normalization constant and $m^{\pm}_w$ is given above. The conformal weights may be simply calculated as
$$
h[\Omega^{\pm}_w]=\left(\frac{(w+1)^2}{4}+\frac{1}{4}\right)+\left(\frac{(w-1)^2}{4}-\frac{1}{4}\right)+\left(-\frac{1}{2}(w^2+1)\right)=0,
$$
where, as before, the terms in brackets correspond to the conformal weights of the fermions, $\partial^w\omega$, and projective delta-function terms in $\Omega_w$.

\subsection{Bulk physics}\label{sec:bulk_physics}

With the projective description of the vertex operators, we can now apply the bulk incidence relation and attempt to give a bulk interpretation of the physics. The construction presented here bears a superficial resemblance to the Penrose transform for the minitwistor space of complexified $AdS_3$, $\M\T (AdS_3)$ \cite{Bu:2023cef}. An important distinction should be made between a genuine Penrose transform of the kind discussed in the literature and the construction presented here. For example, a genuine Penrose transform for complexified $AdS_3$ maps cohomology representatives,
\begin{equation}\label{eq:cohom_representative}
    f(\omega_{\alpha}, \pi^{\dot{\alpha}}) \in H^{(0,1)}\left( \M\T ,{\cal O}(\Delta-2, - \Delta) \right),
\end{equation}
into scalar solutions $\phi(\underline{x})$ of the equation of motion,
$$\Box_{H_3^+}\phi(\underline{x})=\Delta(\Delta+2)\phi(\underline{x}),$$
where $\underline{x}$ represents coordinates on $AdS_3$ and $\Box_{H_3^+}$ is the Laplacian on $AdS_3 \cong H_3^+$. The notation in \eqref{eq:cohom_representative} refers to the fact that $f(\omega_{\alpha},\pi^{\dot{\alpha}})$ is a 
$\bar{\p}$-closed $(0,1)$-form on $\M\T(AdS_3)$, with scaling weights $(\Delta - 2)$ and $-\Delta$ with respect to the projective coordinates $\omega_{\alpha}$ and $\pi^{\dot{\alpha}}$, respectively. That is, $f(s\omega_{\alpha},t\pi^{\dot{\alpha}}) = s^{\Delta - 2}t^{-\Delta} f(\omega_{\alpha},\pi^{\dot{\alpha}})$.
The Penrose transform is an integral transform
$$
\phi(\underline{x})=\int_{\C\P^1_{\underline{x}}} D\omega\wedge f(\omega,\pi),
$$
where the integral is taken over the line in minitwistor space corresponding to the point $\underline{x}$ in spacetime, which corresponds to an element $g \in SL(2;\C)$. That means that the integral is evaluated on the incidence relation $\pi^{\dot{\alpha}}= \omega_{\alpha} g^{\alpha\dot{\alpha}}$ and $D\omega = \epsilon^{\alpha\beta}\omega_{\alpha}d\omega_{\beta}$ is the (holomorphic) projective measure. The generalization to spacetime fields of general spin follows straightforwardly and is discussed in \cite{Bu:2023cef}.

The discussion for the string theory below will differ in important ways. In particular, the vertex operators are left-right symmetric, unlike the usual twistor constructions which deal with purely holomorphic variables. Indeed, since we are far from the supergravity limit, we do not a priori expect solutions to the supergravity equations of motion to have particular relevance. As such, it is not immediately clear that the basic feature of the Penrose transform --- relating cohomology representatives on twistor space to solutions to equations of motion --- should be at play here.\\

To generalise this construction to the $k=1$ string, we are looking for objects living in the $\C\P^1$ bundle over $H_3^+$, which we can then descend to the spacetime by integrating over the $\C\P^1$ fibres. In the language of \S\ref{sec:twistor_sigma_model}, we shall work with the inverse metric, where $H_3^+$ is parameterised by $g^{\alpha\dot{\alpha}}$, and we consider two correspondence spaces, where $\omega_{\alpha}$ and $\mu_{\dot{\alpha}}$ are fibred over $H_3^+$. The incidence relations are given by
$$
\pi^{\dot{\alpha}} = \omega_{\alpha}g^{\alpha\dot{\alpha}},    \qquad  \lambda^{\alpha} = -g^{\alpha\dot{\alpha}} \mu_{\dot{\alpha}}.
$$
We therefore propose that a natural bulk object, derived from the bulk vertex operators with one unit of spectral flow and inspired by the field theory analysis of \cite{Bu:2023cef}, is
\begin{equation}\label{O}
{\cal O}_{\Delta,\bar{\Delta}}(g(z_i,\bar{z}_i)|x)=\int_{\C\P^{1}_g} D\omega D\mu  \frac{1}{  [ \bar{\omega}\pi ]^{\Delta} \langle\bar{\mu}\lambda\rangle^{\bar{\Delta}}}{\cal V}(\omega,\mu|x).
\end{equation}
We define $\langle \bar{\mu}\lambda\rangle:= 
\bar{\mu}_{\alpha} \lambda^{\alpha}$, $[ \bar{\omega}\pi] := \bar{\omega}_{\dot{\alpha}}\pi^{\dot{\alpha}}$ and $\mathcal{V}(\omega,\mu|x)$ is a vertex operator lifted from the boundary to the bulk,
$$
\mathcal{V}(\omega,\mu|x)=\Omega_1(\omega|x)\bar{\Omega}_1(\mu|\bar{x}).
$$
$\Omega_w(\omega|x)$ are the projective ground state vertex operators we have been studying, now explicitly stating their dependence on $\omega_{\alpha}$ and $x$. $\bar{\Omega}_w (\mu|\bar{x})$ is the natural anti-holomorphic counterpart. The integral in \eqref{O} is not a path integral over the fields $\omega(z)$ and $\mu(\bar{z})$. Instead, we fix $(z_i,\bar{z}_i)$ and then interpret $\omega(z_i)$ and $\mu(\bar{z}_i)$ as coordinates on the target space $\C\P^1$. We therefore integrate over the values of $\omega(z_i), \mu(\bar{z}_i) \in \C\P^1$. We will show how (\ref{O}) encodes bulk physics below and provide some speculations as to how one could make sense of (\ref{O}) in the full string theory in the discussion section.

There are several points worth noting. The first is that, in Euclidean signature, $\bar{\omega}_{\dot{\alpha}} = \mu_{\dot{\alpha}}$ as we saw in \S\ref{sec:Boundary_spacetime}, such that the projective delta functions naturally associate $\bar{\omega}_{\dot{\alpha}}$ with the boundary data. The second is that the factors of $\langle \bar{\mu} \lambda \rangle$ and $[\bar{\omega}\pi]$ vanish on the boundary and so their appearance in the denominator hints at delta-function support at the boundary. Finally, the integral is taken over the $\C\P^1$ corresponding to the point $g(z_i,\bar{z}_i)$ in the bulk spacetime, on which the bulk incidence relations \eqref{eq:incidence_g_inverse} are imposed. For this reason, we explicitly include the label of $g$ in the integration domain, $\C\P^1_g$.

The presence of the $\langle \bar{\mu}\lambda\rangle$ and $ [ \bar{\omega}\pi]$ terms requires further comment. We assume that we are working in Euclidean $AdS_3$ which requires the reality condition $Z\cdot \tilde{Z} = -U - \tilde{U} = 0$. This reality condition is manifestly satisfied throughout $H_3^+$ on the solution to the incidence relations \eqref{eq:incidence},
$$U + \tilde{U} = \langle\omega\lambda\rangle + [\mu\pi] = \langle \lambda g\pi] - \langle\lambda g \pi] = 0.$$
Equivalently, we can rephrase this constraint as $W\cdot Z = U + \tilde{U} = 0$, which also vanishes on the solution to the incidence relations and holds for any choice of signature as in \eqref{eq:ambitwistor_quadric}. This is the quadric condition that defines the ambitwistor space of $\C^4$ into which complexified $AdS_3$ embeds. By contrast, on the support of the incidence relations \eqref{eq:incidence}, the linear complement $U - \tilde{U}$ may be written as
\begin{equation}
\frac{1}{2}(U - \tilde{U}) = \langle \lambda g\pi] = -(e^{\phi}|\gamma-x|^2+e^{-\phi}),
\end{equation}
as in \eqref{para}. This vanishes only on the boundary of $H_3^+$, where we identify the boundary point $x$ with $\gamma$ and $\phi \to \infty$. Hence, if we require that $U$ and $\tilde{U}$ vanish independently as in the physical constraints, then we localise to the boundary of $H_3^+$. Instead, we shall impose here only the $U + \tilde{U}=0$ condition, but loosen the condition that $U$ and $\tilde{U}$ must vanish independently.
It is in this sense that we take the theory off-shell in order to evaluate the integral. One might think of this as similar to the loosening of the condition that $L_0$ and $\bar{L}_0$ individually vanish in string field theory \cite{Zwiebach:1992ie}; instead, one takes $L^-_0:=L_0-\bar{L_0}=0$ to be a background-independent condition and allows $L_0^+=L_0+\bar{L}_0$ (with $L_0^+=0$ interpreted as the target space equation of motion) to be unconstrained off-shell. Here we take the view that the quadric condition $U+\tilde{U}=0$ cannot be relaxed, but the condition $U-\tilde{U}=0$ can be relaxed to allow us to move away from the boundary. This is because $U + \tilde{U} = 0$ holds by virtue of the incidence relations, but $U - \tilde{U}=0$ only holds at the boundary.\\

To demonstrate the utility of \eqref{O}, consider $\mathcal{V}(\omega,\mu|x)=\delta_{-2}(A(x),\omega)\delta_{-2}(\bar{A}(\bar{x}),\mu)$, where we only include the bosonic part of the vertex operator. We can evaluate the corresponding bulk expression as follows 
\begin{equation*}
\begin{aligned}
{\cal O}_{\Delta,\bar{\Delta}}(g|x,\bar{x})&=\int_{\C\P^1_g} D\omega D\mu \frac{1}{  [ \bar{\omega}\pi ]^{\Delta} \langle\bar{\mu}\lambda\rangle^{\bar{\Delta}}}\delta_{-2}(A,\omega)\delta_{-2}(\bar{A},\mu)\nonumber\\
&=\int_{\C\P^1_g} D\omega D\mu \frac{(-1)^{\bar{\Delta}}}{ \langle  \omega g \bar{\omega}]^{\Delta}\langle \bar{\mu}g\mu]^{\bar{\Delta}}}\delta_{-2}(A,\omega)\delta_{-2}(\bar{A},\mu)\nonumber\\
&=\int_{\C\P^1_g} D\omega D\mu \frac{(-1)^{\bar{\Delta}}}{\langle Ag\bar{A}]^{\Delta+\bar{\Delta}}}\delta_{-2}(A,\omega)\delta_{-2}(\bar{A},\mu)\nonumber\\
&= \frac{(-1)^{\bar{\Delta}}}{\langle Ag \bar{A}]^{\Delta+\bar{\Delta}}}
\end{aligned}
\end{equation*}
where, as stated above, the expression is understood to be integrating only over the values of $\omega(z_i)$ and $\mu(\bar{z}_i)$ in the target $\C\P^1_g$. The incidence relations have been used in going from the first to the second line and the delta-functions have been used to go from the second to the third line. Finally, the delta functions have been used to do the integral and then evaluated explicitly in the last line using a Wakimoto coset representative (\ref{eq:g_wakimoto}), where we have also introduced the notation $\langle Ag \bar{A} ] = A_{\alpha} g^{\alpha\dot{\alpha}} \bar{A}_{\dot{\alpha}}$. Using the parameterisation of $g$ given in \S\ref{sec:euclidean_AdS3} and the expression for $A_{\alpha}$ given in (\ref{data}) with $A_- = 1$, this expression can be evaluated to give
$$
{\cal O}_{\Delta,\bar{\Delta}}(g(z_i,\bar{z}_i)|x)= (-1)^{\bar{\Delta}}\left(\frac{e^{-\phi(z_i,\bar{z}_i)}}{e^{-2\phi(z_i,\bar{z}_i)}+|\gamma(z_i,\bar{z}_i)-x|^2}\right)^{\Delta+\bar{\Delta}}
$$
This is analogous to the computation we performed in \S\ref{sec:boundary}. The result is the bulk-boundary propagator \cite{Teschner:1997ft,deBoer:1998gyt,Kutasov:1999xu}, the analogue of a momentum eigenstate for $AdS_3$. We would also like to include the fermionic directions: for example, the measure should be generalised to $D\omega D\mu d^4\psi$, the fermionic directions of which would cancel against the fermionic directions in the vertex operators, such as
$\mathcal{V}_1(\pi,\lambda|x)=\psi_-\psi_+\bar{\psi}_-\bar{\psi}_+\delta_{-2}(A,\omega)\delta_{-2}(\bar{A},\mu)$. The measure $D\omega D\mu d^4 \psi$ is the natural projective measure for $\C\P^{1|2}$. More general vertex operators for $w>1$ will carry additional delta-functions which require $\gamma$ to take the form $\gamma(z_i)=x_i+a_i(z-z_i)^w+...$ near the $i$'th vertex operator insertion, for some constant $a_i$. We discuss potential generalisations of \eqref{O} for $w>1$ to incorporate this behaviour in \S\ref{sec:discussion}, speculating how vertex operators with larger spectral flow might be associated to bulk objects. This raises the interesting question of whether such considerations lead to a stringy generalisation of the bulk-boundary propagator, and what implications this would have for studying bulk physics at $k>1$ (where the theory is not localised to the boundary on-shell).

\section{A comment on the unflowed sector}\label{sec:unflowed}

In the previous section, we studied the vertex operators of the $k=1$ string with spectral flow $w\geq 1$. Yet, there also exist vertex operators with $w=0$ that live in non-unitary representations of $\mathfrak{psu}(1,1|2)_1$ \cite{Eberhardt:2018ouy}. These states satisfy all of the physical state conditions, so in principle could contribute to the theory. The proposed resolution to this in \cite{Eberhardt:2018ouy} was to simply ignore this unflowed sector, which drops out from the fusion rules. From the perspective of holography, it is important that such states do not contribute to the physical spectrum. The spectral flow, $w$, corresponds to the length of the single cycle $w$-twisted sector of the dual symmetric product orbifold CFT and as such, $w=0$ has no meaning in the CFT. This section therefore fills a small gap in our understanding of the $k=1$ string and presents a more satisfying resolution to this problem; the $w=0$ states decouple from the theory. That is, whenever an unflowed physical state is inserted inside a correlation function, that correlator identically vanishes. We shall use the full machinery of the $k=1$ string as described by the hybrid formalism to show this \cite{Berkovits:1999im, Berkovits:1994vy}, since we have by now justified that this is the natural prescription for studying string theory at $k=1$. The PCO we encountered in \S\ref{sec:Twistor_string}, and namely $\mathcal{Q}$, will play a key role.

In terms of the connections between the $k=1$ string and twistor geometry, this section lies somewhat outside the main development of the paper and may be skipped.\\

The continuous representations of $\mathfrak{sl}(2;\R)$, $\mathscr{C}^j_{\lambda}$, carry two labels. A state can be written as $|m_1,m_2\rangle$, where $m_1 + m_2 = \lambda$ (mod 1) and $m_1-m_2 = j$. The global $\mathfrak{sl}(2;\R)$ generators act as
$$
J^+_0|m_1,m_2\rangle = 2m_1 |m_1 + 1/2,m_2 + 1/2 \rangle,\qquad
        J^-_0|m_1,m_2\rangle = 2m_2 |m_1 - 1/2 , m_2 - 1/2 \rangle,
$$
\begin{equation*}\label{eq:ground_states}
J_0^3|m_1,m_2\rangle = (m_1+m_2)|m_1,m_2\rangle,    
\end{equation*}
and the $\mathfrak{sl}(2;\R)$ quadratic Casimir is $\mathcal{C} = -j(j-1)$. The Fock space on $\mathscr{C}^j_{\lambda}$ is then built by acting with the negative modes of the $\mathfrak{psu}(1,1|2)_1$ currents, whilst the positive modes annihilate $|m_1,m_2\rangle$. This can be reinterpreted in terms of the modes of the twistor variables \cite{Dei:2020zui} in the R-sector. For the bosons,
\begin{align*}
    (\omega_-)_0|m_1,m_2\rangle &= |m_1,m_2 + 1/2\rangle,  & \lambda^+_0|m_1,m_2\rangle &= 2m_1|m_1 + 1/2, m_2 \rangle,\\
    (\omega_+)_0|m_1,m_2\rangle &= |m_1-1/2,m_2\rangle, & \lambda^-_0|m_1,m_2\rangle &= -2m_2|m_1, m_2 - 1/2 \rangle,
\end{align*}
whilst the positive modes annihilate. For the fermions,
$$(\psi_-)_r|m_1,m_2\rangle = \chi^+_r|m_1,m_2\rangle = 0,$$
for all $r \geq 0$ and
$$(\psi_+)_r|m_1,m_2\rangle = \chi^-_r|m_1,m_2\rangle = 0,$$
for all $r >0$.

The ground states of the unflowed sector can now be written down. There are two $\mathfrak{su}(2)$ singlets of the form
$$(\psi_+)_0|m_1,m_2\rangle, \qquad \chi^-_0|m_1,m_2\rangle.$$
To be physical, the first of these has $j = 1$, whilst the second has $j=0$. There is also an $\mathfrak{su}(2)$ doublet given by
$$|m_1,m_2\rangle, \quad \chi^-_0 (\psi_+)_0|m_1,m_2\rangle,$$
for $j = \frac{1}{2}$. By acting with the supersymmetries $S^{\alpha A I}_0$, we can move between the singlets and doublet.

This is a non-unitary representation of $\mathfrak{psu}(1,1|2)_1$ because of the singlet living in $\mathscr{C}^0_{\lambda}$, which forms a non-unitary representation for $\lambda \neq 0$ (see \cite{Eberhardt:2018ouy} for more on the subtlety at $\lambda = 0$). That is, we would like to consider states
$$\chi^-_0|m,m\rangle,$$
where $2m = \lambda \neq 0$ (mod 1). Then, one can check that
\begin{align*}
    \left|\left| J^+_0 \chi^-_0 |m,m\rangle \right|\right|^2 &= \epsilon \, 2m(2m + 1) \left|\left| \chi^-_0 |m,m\rangle \right|\right|^2 \\
    &=  \epsilon \left[\left(2m + \frac{1}{2}\right)^2 - \frac{1}{4}  \right] \left|\left| \chi^-_0 |m,m\rangle \right|\right|^2,
\end{align*}
where $\epsilon$ is a sign that depends on your choice of signature, $\left(J_0^+\right)^{\dagger} = \epsilon J_0^-$. But since this representation contains states for all $m$ satisfying $2m = \lambda \neq 0$ (mod 1), we are guaranteed to have states with both positive and negative signs for this norm. All of these states satisfy the physical state conditions, so this implies the existence of ghost states (unless the norm is vanishing). Thankfully, we can demonstrate that such states decouple from the $w>0$ sector.
\\

We have seen that the $k=1$ string is naturally described by the hybrid formalism of \cite{Berkovits:1999im}. It's correlation functions can therefore be defined using the machinery of $\mathcal{N} = 4$ topological strings \cite{Berkovits:1994vy,Berkovits:1999im}. We can construct a measure on moduli space, $\mathcal{M}_{g,n}$, using insertions of the usual $b$-ghost and its picture-raised version using \eqref{eq:hybrid_PCO}. With all vertex operators $\Psi_i$ in picture $P=-2$ and $g \geq 1$, the non-vanishing contribution to the holomorphic part of the correlation function takes the form \cite{Dei:2020zui,McStay:2023thk},
\begin{equation}\label{eq:correlator}
    \int_{\mathcal{M}_{g,n}} \left\langle \prod\limits_{I=1}^{g-1} \mathbf{b}(\mu_I) \prod\limits_{I=g}^{n+3g-3} \widetilde{\mathbf{b}}(\mu_I) \left[ \int_{\Sigma} e^{\rho+iH} \right]^{g-1} \int_{\Sigma} J \; \prod\limits_{r=1}^n \Psi_i \right\rangle,
\end{equation}
where $\{\mu_I\}$ are a basis of Beltrami differentials for $\mathcal{M}_{g,n}$ and\footnote{The necessity of $\tilde{\mathbf{b}}$, essentially a picture-raised version of $\mathbf{b}$, in building non-trivial correlation functions highlights the importance of ${\cal P}_+$.}
$$\mathbf{b}(\mu_I) = \int_{\Sigma} \mathrm{d}^2z \; e^{-i\sigma}(z) \mu_I(z), \quad \widetilde{\mathbf{b}}(\mu_I) = \int_{\Sigma} \mathrm{d}^2z \; e^{-2\rho-i\sigma-iH} \mathcal{Q}(z) \mu_I(z).$$
$J$ is the ghost current, given by $J = \p(\rho + i\sigma + iH)$.

At genus zero, we have to be a little more careful, since we must invert $e^{\rho+iH}$ and naively we have a negative number of $\mathbf{b}(\mu_I)$ insertions. We also require that $n \geq 3$ to fix the global $SL(2;\C)$ transformation. We fix the first of these problems by noting that, since the $(e^{\rho+iH})_0$ cohomology is trivial, we can rewrite one of the physical states as
$$\Psi_i = e^{2\rho+i\sigma+iH}\Phi_i = \left( e^{\rho+iH} \right)_0 e^{\rho+i\sigma}\Phi_i = \left( e^{\rho+iH} \right)_0 \tilde{\Psi}_i.$$
We interpret $(e^{\rho+iH})_0^{-1}$ as replacing $\Psi_i$ by $\tilde{\Psi}_i$. We deal with the notionally negative number of $\mathbf{b}(\mu_I)$ insertions in a similar way; we replace one of the $\widetilde{\mathbf{b}}(\mu_I)$ by $\mathbf{b}(\mu_I)$, such that our measure is built from only $(n-3)$ $\widetilde{\mathbf{b}}(\mu_I)$ and no $\mathbf{b}(\mu_I)$. We then picture raise one of the vertex operators, which amounts to the replacement
$$\Psi_j = e^{2\rho+i\sigma+iH}\Phi_j \quad \mapsto \quad \Psi_j' = e^{i\sigma}\mathcal{Q}_{-1}\Phi_j.$$
Moreover, it is no longer necessary to insert $\int_{\Sigma}J$ inside the correlation function. The locations of which $\Psi_i$ is replaced by $\tilde{\Psi}_i$ and which $\Psi_j$ is replaced by $\Psi_j'$ do not matter. Given any other insertion with $l\neq i$,
$$\Psi_l = \left( e^{\rho+iH} \right)_0 \tilde{\Psi}_l,$$
and so we may deform the contour away from $z_l$ to the rest of the worldsheet $\Sigma$. The action of $\left(e^{\rho+iH}\right)_0$ is trivial everywhere, except on $\tilde{\Psi}_i$. This argument would still hold for $l=j$, so the location of where we ``inverted'' $\left(e^{\rho+iH}\right)_0$ did not matter. Similarly,
$$\Psi_j' = \left(e^{-\rho}\mathcal{Q}\right)_0 e^{\rho+i\sigma}\Phi_j.$$
Deforming this contour away from $z_j$, it annihilates all physical states in $P=-2$ but has a non-trivial action on $\tilde{\Psi}_i$. This effectively swaps the locations of $\tilde{\Psi}_i$ and $\Psi_j'$.
\\

With the above considerations in mind, it is simple to see that correlation functions with an unflowed $w=0$ state inserted vanish. Consider the state $|\Phi_0\rangle = \chi^-_0 |m_1,m_2\rangle$. To be a physical state, we require that $\mathcal{Q}_n|\Phi_0\rangle = 0$ for all $n \geq 0$ (along with the stress tensor and scaling constraints). However, $|\Phi_0\rangle$ satisfies the stronger condition that $\mathcal{Q}_n|\Phi_0\rangle = 0$ for all $n \geq -2$. This guarantees a trivial OPE between $\mathcal{Q}(z)$ and $\Phi_0(z_i)$. In the genus zero case, we see that the correlation function immediately vanishes, as we may choose the location of $\mathcal{Q}_{-1}$ to act on $\Phi_0(z_i)$, referring to the simple pole which vanishes.

For $g\geq 1$, we return to the expression \eqref{eq:correlator}. We have $(n+2g-2)$ insertions of $\widetilde{\mathbf{b}}(\mu_I)$ in this expression. As is usual with Beltrami differentials, we can choose to remove $n$ of these by instead acting with $\left(e^{-2\rho-i\sigma-iH}\mathcal{Q}\right)_{-1} \Psi_i$ on the vertex operators. However, in picture $P=-2$, this requires a double pole in the OPE $\mathcal{Q}(z)\Phi(z_i)$, which is not present in the case of $\Phi_0$. Hence, we deduce that all correlation functions containing $\Phi_0$ vanish.

We can generate all remaining unflowed states by acting on $|\Phi_0\rangle$ with the supersymmetries $S^{\alpha A I}_0$. In general, these do not all commute with $\mathcal{Q}$ as this was, in many ways, its defining property. As such, the global symmetries $S_0^{\alpha A -}$ are replaced by the zero modes of \cite{Berkovits:1999im,Gaberdiel:2022als}
$$\tilde{S}^{\alpha A -} = S^{\alpha A -} - e^{-\rho-i\sigma}S^{\alpha A +}.$$
Focusing on the $P=-2$ case for simplicity, one finds that
$$\tilde{S}^{\alpha A -}_0  \left(e^{2\rho+i\sigma+iH}|\Phi_0\rangle\right) = S^{\alpha A -}_0  \left(e^{2\rho+i\sigma+iH}|\Phi_0\rangle\right),$$
and similarly for acting with $\tilde{S}^{\alpha A -}_0$ twice. This is sufficient to generate all the ground states of the unflowed sector using global symmetries. We deduce that whenever a $w=0$ state is inserted inside a correlation function, we may view it as an insertion of $\Phi_0$ being acted on by global symmetries. Deforming the contours of these global symmetries away from $\Phi_0$ to the rest of $\Sigma$, we return to the case of a correlation function with an insertion of $\Phi_0$, which vanishes. We conclude that the $w=0$ sector decouples from the theory and we need not consider it further.

\section{Discussion}\label{sec:discussion}

In this paper, we have made progress in presenting the $k=1$ sigma model for the minimal tension string on $AdS_3 \times S^3 \times T^4$ \cite{Gaberdiel:2018rqv,Eberhardt:2018ouy} as a natural construction for a critical sigma model for superstring theory on the minitwistor space $\M\T(AdS_3)$. The proposed twistor construction in \S\ref{sec:twistor_sigma_model} gives a natural realisation of the $G_L\times G_R$ symmetry of the underlying group manifold and leads to novel incidence relations \eqref{eq:incidence} that are quite different from those seen in usual twistor constructions as it mixes left- and right-moving degrees of freedom. The minitwistors in this sigma model were seen to reproduce the free field realisation of \cite{Dei:2020zui}, whilst the incidence relations of \cite{Dei:2020zui} were observed to be the near-boundary limit of the incidence relations \eqref{eq:incidence} for the bulk minitwistors. We have also elucidated the relationship between the scaling constraint of the minitwistors (i.e. $U=0$) and the requirement that the gauge-invariant observables of the theory be constrained to live at the boundary. In \S\ref{sec:euclidean_AdS3}, the case of Euclidean $AdS_3$ was studied in detail and it was argued that the $k=1$ sigma model does indeed describe the full physics of $AdS_3$ at minimal tension, but that the constraint $U=0$ implied that the path integral only contained contributions where the entire worldsheet was localised to the boundary of $AdS_3$.

We have gone some way towards showing that the more exotic constraints arising in the hybrid formalism may owe their origins to a formulation of the theory which has a, probably twisted, form of ${\cal N}=2$ worldsheet supersymmetry. Such a suggestion has been made before \cite{Gaberdiel:2021jrv,Gaberdiel:2021qbb} but the novelty here is two-fold. Firstly, the existence of an ${\cal N}=2$ string was shown to be motivated by the partial breaking of global supersymmetry arising from imposing the $GL(1)$, ${\cal C}=0$ constraint. Secondly, by a consideration of the very general properties such a formulation is expected to have (namely the form of its PCOs) it was shown that an additional physical constraint --- the ${\cal Q}$ constraint --- needs to be applied, that is not present in traditional twistor string constructions \cite{Berkovits:2004hg}. It would clearly be desirable to have a concrete construction of the conjectured ${\cal N}=2$ theory.

The physical states of the theory were studied in the final two sections. \S\ref{sec:unflowed} was somewhat tangential to the main focus of the paper, filling in a gap in the literature by demonstrating that the unflowed, $w=0$ sector of the theory decouples as one would hope. Meanwhile, in \S\ref{sec:bulk_VO}, we recast the usual $w\geq 1$ ground state vertex operators in terms of the twistor variables and our novel incidence relations were used to explore the bulk structure of the theory in \S\ref{sec:bulk_physics}, through an integral transform from minitwistor space to spacetime. It is intriguing that, by relaxing a constraint on the value of $U - \tilde{U}$, one can recover off-shell information that seems to directly probe the bulk spacetime. It would be interesting to develop these ideas further and to see if any light can be thrown on the problem of bulk reconstruction in $AdS_3$.

We shall finish by discussing a number of interesting avenues for future work.

\subsection*{An $\mathcal{N}=2$ formulation of the $k=1$ sigma model}

As mentioned above, it would be highly desirable to explicitly construct the conjectured $\mathcal{N} = 2$ description of the $k=1$ string, which is manifestly spacetime supersymmetric. Our expectations for how one would likely construct such a theory are as follows.

We claimed earlier that we should view our string as embedding into the ambitwistor space of the boundary, $\C\P^{1|2} \times \C\P^{1|2}$, which has four fermionic degrees of freedom, as well as two bosonic degrees of freedom. In applying the work of \cite{Hughes:1986dn} in \S\ref{sec:doubly_susy}, we introduced physical degrees of freedom corresponding to the Goldstone modes $(\gamma,\theta_A)$ as well as parameters $(y,\vartheta_A)$ for the super-Riemann surface. In this sense, we have a $\C\P^{1|2}$ of physical degrees of freedom for the string. In view of \cite{Hughes:1986dn}, we anticipate that the covariant description is to be interpreted as embedding into $\C\P^{1|2} \times \C\P^{1|2}$, but that the worldsheet has been taken in a static gauge aligned with the second copy of $\C\P^{1|2}$. The parameters $(y,\vartheta_A)$ would then describe pure gauge degrees of freedom.

One might be concerned that our analysis given in \S\ref{sec:doubly_susy} was incomplete, since we only focused on the generators $\{J^+,J^3,K^{\tilde{a}},S^{+AI}\}$, which have a natural action on the boundary $\C\P^1$ in the patch $\omega_- \neq 0$. We justified this by our assertion that the other generators of $\mathfrak{psu}(1,1|2)_1$ describe special (super)conformal transformations, meaning they have a simpler description in the opposite patch $\omega_+\neq 0$. Hence, we need not introduce associated coordinates, in addition to those for the patch $\omega_-\neq 0$. This assumption agrees nicely with the ``Inverse Higgs Phenomenon'' of \cite{Ivanov:1975zq}. When dilations and special conformal transformations are broken, it is shown in \cite{Ivanov:1975zq} that the associated Goldstone fields are not independent --- only one field associated to the dilations needs to be introduced. The same procedure also applies to special superconformal transformations (see e.g. \cite{Bellucci:1998mk}). This agrees with how we have excluded explicit coordinates/Goldstone modes associated to the special (super)conformal transformations of the boundary.

One implication of our conjecture is that $\mathcal{Q}$ plays the same role that $\kappa$-symmetry would in a conventional GS superstring. Therefore, in the $\mathcal{N} = 2$ formulation, $\kappa$-symmetry has been replaced by local worldsheet supersymmetry. We comment on why this is natural in twistor space in Appendix \ref{sec:kappa_sym}.

\subsection*{Bulk physics from spectrally flowed states}

The integral transform \eqref{O} provided a method for recovering the bulk-boundary propagator from the vertex operators localised at the boundary with one unit of spectral flow, $w=1$. It is natural to ask how this integral transformation should generalise to $w>1$ and what bulk objects we would uncover.

Crucially, our integral transform \eqref{O} was not a path integral over fields $\omega(z)$ and $\mu(\bar{z})$, but instead we fixed $(z_i,\bar{z}_i)$ and integrated over the values of $\omega(z_i),\mu(\bar{z}_i) \in \C\P^1$. A natural generalisation, therefore, would be to integrate over the values of $\p^n\omega(z_i)$ and $\p^n\mu(\bar{z}_i)$ for $n = 0,1, \dots , (w-1)$. The additional delta functions in the vertex operators for $w>1$ will constrain these values. The corresponding bulk object would be valued in a jet bundle for the coordinates $(\phi,\gamma,\bar{\gamma})$ viewed as sections over the worldsheet. That is, the local coordinates of the jet bundle would consist of the fields $(\phi,\gamma,\bar{\gamma})$ and their worldsheet derivatives. We intend to return to this elsewhere.

We also note that, on the solution to the classical equations of motion, the bulk incidence relations can be written in the form $\omega = h(z)h^{\dagger}(\bar{z}) \pi$ from \eqref{eq:g_factorised}. This can be rearranged to $h^{-1}\omega = h^{\dagger}\pi = \text{const.}$, since it equates holomorphic and anti-holomorphic objects. From this, one deduces that $\omega_+ - a \omega_- = c e^{-\Phi_0}$ is a purely holomorphic statement at arbitrary points in the bulk, with $a(z)$ being the boundary value of $\gamma$ and $c$ is a constant. This presumably gives an $AdS_3$ realisation of the mechanism proposed in \cite{Bhat:2021dez} for the $AdS_5$ incidence relations and covering maps.

\subsection*{Generalisations to $AdS_5/CFT_4$}

It has been proposed that a string theory dual to free $\mathcal{N} = 4$ SYM can be constructed using a free field realisation for $\mathfrak{psu}(2,2|4)_1$ \cite{Gaberdiel:2021jrv,Gaberdiel:2021qbb} analogous to the free field realisation of $\mathfrak{psu}(1,1|2)_1$ discussed here. The free fields of this proposal are inherently related to the twistor space of $AdS_5$, a quadric inside $\C\P^3 \times \C\P^3$. As such, it appears that many of the ideas discussed in this paper could also be applied to the case of $AdS_5 \times S^5$, building upon the work of \cite{Gaberdiel:2021jrv,Gaberdiel:2021qbb} as well as \cite{Adamo:2016rtr}, and these generalisations are currently under investigation \cite{McStayReidEdwards}.

In particular, our expressions for the projective ground state vertex operators of the $k=1$ string should provide a foothold for constructing the vertex operators of the proposed $\mathfrak{psu}(2,2|4)_1$ theory. By construction, our $AdS_3$ vertex operators enforce that $\gamma$ behaves as a covering map from the worldsheet to the boundary, which leads to a localisation in moduli space for the correlation functions \cite{Dei:2023ivl}. It may be possible, therefore, to deduce similar localising properties for the $AdS_5$ case from a generalisation of our vertex operators. Moreover, we have shown that the presence of PBGS in the $k=1$ string implied the existence of an $\mathcal{N} = 2$ description of the theory. A similar conjecture was made in \cite{Gaberdiel:2021jrv,Gaberdiel:2021qbb} that there exists an $\mathcal{N} = 4$ description for their proposed $\mathfrak{psu}(2,2|4)_1$ theory. It would be interesting to see whether a similar argument to the one given here could be used to support this conjecture and to study its implications for the physical constraints. Indeed, one might expect generalisations of the $\mathcal{Q}$ constraint to appear in this model.

To elucidate slightly on the generalisation to $AdS_5 \times S^5$, consider the following. Similarly to Euclidean $AdS_3$, which we studied in \S\ref{sec:euclidean_AdS3}, $AdS_5$ is a coset rather than a group,
$$AdS_5 = \frac{SO(2,4)}{SO(1,4)}=G/H.$$
Following \cite{Adamo:2016rtr}, we can define complexified $AdS_5$ as a subset of $\C\P^5$. Specifically, $AdS_5$ can be charted by the projective, anti-symmetric $g_{AB}$ where $A = 1, \dots, 4$. The locus $g^2 = 0$ defines the boundary of $AdS_5$, such that $\C\P^5\backslash M$, where $M = \{g\in\C\P^5|g^2 = 0\}$, combined with an appropriate metric describes complexified $AdS_5$. The natural incidence relation is
$$\omega_A = g_{AB}\pi^B,$$
and the anti-symmetry of $g_{AB}$ implies $\omega_A\pi^A = 0$. This constraint generates a scaling symmetry. Note that there is no canonical way to raise or lower a single index, but anti-symmetric pairs of indices may be raised or lowered via $\epsilon^{ABCD}$ and $\epsilon_{ABCD}$, respectively. An $AdS_5$ twistor,
$$Z = \begin{pmatrix}
    \omega_A\\
    \pi^B
\end{pmatrix},$$
lives in the twistor space
$$\P\T(AdS_5) = \{(\omega_A,\pi^B) \in \C\P^3\times \C\P^3| \omega\cdot\pi = 0\},$$
which is equivalent to the ambitwistor space of $S^4$.

We could now follow the procedure of \S\ref{sec:euclidean_AdS3}. For Euclidean $AdS_3$, the coset model has a current algebra generated by one copy of $SL(2,\C)$. This is in contrast to the WZW model which had two copies of $SL(2,\C)$. For $AdS_5$, the index $A$ represents a fundamental index of $SU(2,2)$, the universal cover of $SO(2,4)$, and hence the coset representative $g_{AB}$ is a bispinor of $SU(2,2)$. This means that there is one diagonal action of $SU(2,2)$,
$$g_{AB} \mapsto h_A{}^Ch_B{}^D g_{CD},$$
for $h \in SU(2,2)$. We could understand this action as being generated by $J_A{}^B = \omega_A\pi^B$, which is naturally trace-free since $\omega\cdot\pi =0$. At the boundary, we can interpret $\omega$ as a twistor of $S^4$. In terms of boundary four-dimensional spinor indices, we can write $\omega_A = (\mu_{\dot{\alpha}},\lambda^{\alpha})$. The incidence relation becomes
$$g^{AB}\omega_B = g^{AB}g_{BC}\pi^C = 0.$$
$g$ may be parameterised similarly to \eqref{eq:g_wakimoto}, leading to the boundary incidence relation
$$\lambda^{\alpha} = x^{\alpha\dot{\alpha}}\mu_{\dot{\alpha}},$$
where $x^{\alpha\dot{\alpha}}$ describe coordinates in a patch of $S^4$. We could then augment the above with Grassmann fields to generate supertwistors in the usual way.

A more detailed understanding of how the analysis performed in this work for $\M\T(AdS_3)$ can be applied to the twistor space of $AdS_5$ will be given elsewhere \cite{McStayReidEdwards}.

\subsection*{Similarities with the twistor string}

In our construction of the sigma model for superstring theory on $\M\T(AdS_3)$, we encountered an interesting conundrum in \S\ref{sec:almost}. The natural action \eqref{eq:CP1_string}, which looked like a lower-dimensional analogue of Berkovits' description of the twistor string \cite{Berkovits:2004hg}, described a non-critical theory with a central charge of $c = -28$. In \cite{Berkovits:2004hg}, this issue is resolved by coupling the theory to the current algebra of a group carrying $c = +28$. However, one of our key results was to show that this is not the correct thing to do for $AdS_3 \times S^3$. Instead, by demonstrating the existence of PBGS, we argued that \eqref{eq:CP1_string} represented the gauge fixed form of an $\mathcal{N} = 2$ formulation of the theory. Such a formulation further implies the existence of PCOs and we used these to deduce an additional physical constraint on states denoted $\mathcal{Q}$. It was the ghost content associated to $\mathcal{Q}$ that canceled the conformal anomaly of the theory.

We should ask, therefore, whether the constructions motivated here can provide a new perspective on the twistor string \cite{Berkovits:2004hg}? The constructions of \cite{Witten:2003nn} and \cite{Berkovits:2004hg} lead to theories of conformal supergravity \cite{Berkovits:2004jj} in four dimensions. By contrast, for our string theory on $\M\T(AdS_3)$, the additional structure given by the ${\cal Q}$-constraint lead to the sigma model of the hybrid formalism of \cite{Berkovits:1999im} at $k=1$. In the large $k$ limit, the hybrid formalism is known to reproduce Einstein supergravity in the bulk, so it appears that $\mathcal{Q}$ may play an important role in ensuring a coupling to non-conformal theories of gravity in the bulk. It would be interesting to see whether a better understanding of the origin of this constraint could lead to a systematic way of constructing novel Einstein gravity twistor strings.\footnote{Gauging a worldsheet constraint in a twistor string to produce a theory with Einstein gravity has been considered before \cite{Abou-Zeid:2006akj}; however, the resulting theory only described self-dual gravity \cite{Nair:2007md}.} Moreover, it was pointed out in \cite{Berkovits:2004jj} that, whilst the vertex operators in \cite{Witten:2003nn} preserved the volume form on $\C\P'^{3|4}$, a clear motivation for this constraint from the perspective of \cite{Berkovits:2004hg} was lacking. Since $\mathcal{Q}$ is the pullback of the volume form on $\C\P^{1|2}$ to the worldsheet \cite{McStay:2023thk}, it appears we have motivated the lower-dimensional analogue of this constraint.

Going the other way, twistor string theory may be able to shed some light on the physics of the $k=1$ string. It has been suggested that the $k=1$ string is in fact a topological theory and the $k=1$ string has similarities with the Berkovits twistor string \cite{Berkovits:2004hg}, which can also be understood as a deformation of the topological $B$-model \cite{Witten:2003nn}. And so a natural question is whether the $k=1$ string admits a similar alternative description, along the lines of \cite{Witten:2003nn}, which makes any connection with topological string theory more apparent. An alternative formulation may provide a clearer route to addressing many of the other questions raised here, such as generalizations to $AdS_5\times S^5$ and applications beyond $k=1$.

\subsection*{$AdS_3$ string theory beyond $k=1$}

An important final question is whether the constructions presented here admit an obvious generalization to $k>1$? The techniques we employed made use of very specific features that occur only for the shortened representation of $\mathfrak{psu}(1,1|2)_1$ that appears at $k=1$, and a generalization for $k>1$ is not at all obvious to us. In particular, the existence of a conformal embedding $\mathfrak{sl}(2;\R)_1 \oplus\mathfrak{su}(2)_1 \subset \mathfrak{psu}(1,1|2)_1$ at $k=1$ was a key reason for why our twistorial construction was possible. In \S\ref{sec:almost}, the bilinears of the bosonic fields $(\omega_{\alpha},\lambda^{\alpha})$ realised $\mathfrak{sl}(2;\R)_1$ whilst the fermions $(\psi_A,\chi^A)$ realised $\mathfrak{su}(2)_1$. Taking the cross terms of a boson and a fermion then extended the realisation to $\mathfrak{psu}(1,1|2)_1$. Since $\mathfrak{sl}(2;\R)_k \oplus \mathfrak{su}(2)_k \subset \mathfrak{psu}(1,1|2)_k$ is not a conformal embedding for $k>1$, a simple generalisation appears unlikely.

A more hopeful line of enquiry might be the modification to include Ramond flux, the original motivation for the hybrid construction. The pure NS-NS background is in some sense a singular case \cite{Seiberg:1999xz} and the more general case, with non-vanishing Ramond flux, might allow for a more natural framework in which to generalise the discussion presented here. After all the $k=1$ string, unlike the case for $k>1$, does not include a continuum of long string states, a property it shares with the Ramond flux background. Following \cite{Berkovits:1999im}, the starting point for such a generalisation would be to make the minitwistors (and hence the currents) in \eqref{eq:J} neither purely holomorphic nor purely anti-holomorphic.

\begin{center}
    \textbf{Acknowledgements}
\end{center}
The authors would like to thank Nathan Berkovits, Wei Bu, Maciej Dunajski, Rajesh Gopakumar, Sean Seet, David Skinner, Dima Sorokin and Vit Sriprachyakul for insightful conversations. We are particularly grateful to Bob Knighton for comments on an early draft of the paper.

This work has been partially supported by STFC consolidated grants ST/T000694/1 and ST/X000664/1. NM is supported by an EPSRC studentship. RR is the Thomas and Stephan K\"{o}rner Fellow at Trinity Hall and is grateful to the Avery-Wong Foundation for their continued support of this Fellowship. The authors would also like to thank the Isaac Newton Institute for Mathematical Sciences, Cambridge, for support and hospitality during the programme Twistor theory, where some work on this paper was undertaken. This work was supported by EPSRC grant EP/Z000580/1.

\appendix

\section{Twistors for complexified $\mathbf{AdS_3}$}\label{sec:twistors}

\subsection{Definitions}

\subsection*{Minitwistor space of $H_{3\C}$}

We can identify complexified $AdS_3$ with three-dimensional hyperbolic space, $H_{3\C}$. The minitwistor space for $H_{3\C}$ is the space of oriented geodesics and is given by
$$
\mathbb{MT}(H_{3\C})=\{(\omega,\pi)\in\C\P^1\times\C\P^1|\langle\omega\bar{\pi}\rangle\neq 0\}.
$$
The locus $\langle\omega\bar{\pi}\rangle = 0$ is a $\C\P^1$, since for fixed $\bar{\pi}$, this relationship projectively fixes $\omega$. We therefore often write $\M\T(H_{3\C}) = (\C\P^1\times \C\P^1)\backslash\C\P^1$. As shall be reviewed in \S\ref{sec:reality_conditions}, the minitwistor space for a real slice of $H_{3\C}$, such as Euclidean $H_3^+$, is achieved by imposing an additional reality condition on $\M\T(H_{3\C})$.

By considering the case of $H_3^+$, we can understand why $\M\T(H_{3\C}) = (\C\P^1\times \C\P^1) \backslash\C\P^1$ in the following way \cite{Bu:2023cef}. $\CP^1=S^2$ is a copy of the boundary of $H_3^+$ and thus, $\C\P^1\times\C\P^1$ parameterises two copies of the boundary. An oriented geodesic that starts and ends on the boundary can be denoted by two such points (the start and end) and this is a point in $\C\P^1\times\C\P^1$. The set on which $\langle\omega\bar{\pi}\rangle:= \omega_{\alpha}\bar{\pi}^{\alpha} =0$ are those geodesics that start and end at the same point --- a degenerate case, which we remove by omitting this locus. Indeed, we saw in \S\ref{sec:Boundary_spacetime} that $\langle\omega\bar{\pi}\rangle = 0$ specifies the boundary. The geodesics are oriented as their start and endpoints are labelled by which $\C\P^1$ they are attached to.

\subsection*{Ambitwistor space of $S^2$}

The projective ambitwistor space $\mathbb{A}$ of a Riemannian manifold $M_{\R}$ is defined as the space of complex null geodesics in the complexified manifold $M_{\C}$. For the case of our boundary $S^2$, we have that
$$
\mathbb{A}(S^2)=\{(\omega,\pi)\in\C\P^1\times\C\P^1\}.
$$
The two-dimensional case is somewhat degenerate as the twistor space is just a copy of the spacetime with the canonical complex structure.

In addition we can specify the real Euclidean ambitwistor space as
$$
\mathbb{A}_E(S^2)=\{(\omega,\pi)\in\C\P^1\times\C\P^1| \langle\omega\bar{\pi}\rangle =0\}.
$$
One can think of this as the diagonal $\C\P^1\subset \C\P^1\times\C\P^1$. Following the description above, we identify it as the boundary spacetime of $H_3^+$. By definition, the spaces are clearly related by
$$
\mathbb{MT}(H_{3\C})\cup \mathbb{A}_E(S^2)=\mathbb{A}(S^2).
$$

\subsection{Reality conditions}\label{sec:reality_conditions}

The reality conditions for $\M\T(H_{3\C})$ descend from reality conditions on $\P\T = \C\P^3\backslash\C\P^1$. As such, we shall briefly review how different reality conditions are imposed on $\P\T$ and how they relate to a choice of signature for a real slice of $\C^4$ \cite{Adamo:2017qyl}. We emphasise that our conventions for the incidence relations are $\omega_{\alpha} = X_{\alpha\dot{\alpha}}\pi^{\dot{\alpha}}$, in line with the conventions for $\M\T$ in the main text. This differs by a factor of $i$ from the conventions of \cite{Adamo:2017qyl}. For the most part, our choice of convention removes various factors of $i$ from the main text and neatly reproduces known conventions from the $k=1$ literature. However the price we pay is that the conventions of this appendix are non-standard.

\subsubsection*{Lorentzian signature}

For spinors $\omega_{\alpha} = (a,b)$ and $\pi^{\dot{\alpha}} = (c,d)$, complex conjugation is defined via
$$\bar{\omega}_{\dot{\alpha}} = (\bar{a},\bar{b}), \qquad \bar{\mu}_{\alpha} = (\bar{c},\bar{d}).$$
Note that, as well as conjugating $a,b,c,d \in \C$, the complex conjugation exchanges the spinor representations. This implies that the incidence relation $\omega_{\alpha}=X_{\alpha\dot{\alpha}}\pi^{\dot{\alpha}}$ complex conjugates to $\bar{\omega}_{\dot{\alpha}}=\bar{\pi}^{\alpha}X_{\alpha\dot{\alpha}}^{\dagger}$. Then, for a twistor $Z^I \in \P\T$, we define a quaternionic conjugation via 
$$
Z^I=(\omega_{\alpha},\pi^{\dot{\alpha}}) \mapsto \tilde{Z}_I=(\bar{\pi}^{\alpha},-\bar{\omega}_{\dot{\alpha}}).
$$
As such,
$$
Z\cdot \tilde{Z}= \langle\omega\bar{\pi} \rangle - [\bar{\omega}\pi]= \bar{\pi}^{\alpha}(X-X^{\dagger})_{\alpha\dot{\alpha}}\pi^{\dot{\alpha}}.
$$
Thus, on the locus $Z\cdot \tilde{Z}=0$, the matrix $X$ is Hermitian. Since $X_{\alpha\dot{\alpha}} = X_{\mu}\sigma^{\mu}_{\alpha\dot{\alpha}}$, the coordinates $X^{\mu}$ are real, giving a real Lorentzian spacetime inside $\C^4$. This space is called
$$
\P\N=\{Z\in\P\T|Z\cdot \tilde{Z}=0\}.
$$

\subsubsection*{Euclidean signature}

A natural quaternionic conjugation on the spinors is generated by
$$
-i\sigma^{\alpha\dot{\alpha}}_2=\left(\begin{array}{cc}
   0  & -1 \\
    1 & 0
\end{array}\right),
$$
acting as
$$
\omega_{\alpha}=(\omega_+,\omega_-)\rightarrow \hat{\omega}_{\alpha}=(-\bar{\omega}_-,\bar{\omega}_+),  \qquad  \pi^{\dot{\alpha}}=(\pi_+,\pi_-)\rightarrow \hat{\pi}^{\dot{\alpha}}=(-\bar{\pi}^-,\bar{\pi}^+),
$$
where the bar denotes complex conjugation. Similarly to the Lorentzian case described above, one can then construct a map $Z^I \mapsto \hat{Z}^I$ such that the locus of $Z\cdot \hat{Z} = 0$ selects a Euclidean subspace of $\C^4$.

\subsubsection*{Split signature}

We can also make use of the difference between Hermitian conjugation and complex conjugation of $X$. The complex conjugate $\bar{X}_{\alpha\dot{\alpha}}$ is the Hermitian conjugate without the transpose. The condition $X=\bar{X}$ picks out a split signature real space from $\C^4$. The resulting twistor space is isomorphic to $\R\P^3\subset \C\P^3$.

\subsubsection*{$AdS$ signature}

We recover complexified $AdS_3$ from the quadric
$$
AdS_{3\C}=\{X\in\C^4|X^2 = 1\},
$$
where $X^2 = X_0^2 - X_1^2 - X_2^2 - X_3^2$. We recover different signatures by taking different slices, ${\cal X}_n=\{X\in\R^{4-n,n}|X^2=1\}$. For example, Euclidean $AdS_3$, $H_3^+$, is given by the Lorentzian ambient space, $\mathcal{X}_1$, for which the $X^{\mu}$ are all real.

\section{The uniqueness of $\mathbf{\mathcal{Q}}$}\label{sec:uniqueness}

In \S\ref{sec:ground_up}, we proposed that, for a superstring theory defined on the minitwistor space of $AdS_3$, there must exist a PCO and we made the following ansatz:
$$\mathcal{P}_+ = f_{ghost} \mathcal{O}_{matter}.$$
We conjectured that $\mathcal{O}_{matter}$ should satisfy
\begin{equation}\label{eq:O}
    \left[ \mathcal{O}_{matter}(z), A_0 \right]_{\pm} = 0, \qquad \left[ \mathcal{O}_{matter}(z), S^{\alpha A -}_0 \right]_{\pm} \neq 0,
\end{equation}
for $A = J^a, K^{\tilde{a}}$ or $S^{\alpha A +}$, where we use the anticommutator, $[\cdot,\cdot]_+$, if both operators are fermionic, and the commutator, $[\cdot,\cdot]_-$ otherwise. We make no assumption yet as to whether $\mathcal{O}_{matter}$ is bosonic or fermionic. Since PCOs map between physical states, we also require that
\begin{equation}\label{eq:scaling_constraint}
    \left[ \mathcal{O}_{matter}(z), (\mathcal{C}_L)_n \right]_- = 0
\end{equation}
for all $n \geq 0$. We would now like to determine candidate operators for $\mathcal{O}_{matter}$. We shall work with the free fields $(\omega_{\alpha}, \lambda^{\alpha}, \psi_{A}, \chi^A)$ to do this. To clearly lay out our conventions, we define a free field realisation of $\mathfrak{psu}(1,1|2)_1$ as \cite{Eberhardt:2018ouy,Dei:2020zui}
\begin{equation}\label{eq:psu_bilinears}
    \begin{aligned}
        J^{\pm} &= \pm\lambda^{\pm}\omega_{\mp} , & J^3 &= \frac{1}{2} \left( \lambda^+\omega_+ - \lambda^-\omega_- \right), & S^{\pm A+} &= \pm \omega_{\mp}\chi^A, \\
    K^{\pm} &= \chi^{\pm}\psi_{\mp}, & K^3 &= \frac{1}{2}\left( \chi^+\psi_+ - \chi^- \psi_- \right), & S^{\alpha\pm -} &= \mp \lambda^{\alpha}\psi_{\mp}.
    \end{aligned}
\end{equation}
The $GL(1)$ current $\mathcal{C}_L = \frac{1}{2}( \lambda^{\alpha}\omega_{\alpha} + \chi^A \psi_A )$ will also be needed. Moreover, for simplicity of notation, we will drop the word ``matter'' from $\mathcal{O}_{matter}$.
\\

The requirement that $\mathcal{O}(z)$ (anti-)commutes with $S^{\alpha A+}_0$ but does not (anti-)commute with $S^{\alpha A-}_0$ implies that
$$\mathcal{O} = f\left( \omega_{\alpha}, \chi^A \right),$$
for some function $f$, with non-trivial dependence on each field. We consider the space of functions
$$\mathcal{O} \in \left\{ f \in \text{algebra of } \partial^m\omega_{\alpha}, \partial^n \chi^A \text{ for } m, n\in\Z_{\geq 0} \, | \, f \text{ is a conformal primary} \right\}.$$
That is to say, $\mathcal{O}$ is formed of arbitrary sums and products of the fields $(\omega_{\alpha},\chi^A)$ and their derivatives, such that the overall expression is primary with respect to the free field stress tensor
$$T= \frac{1}{2} :\left( \omega_{\alpha}\p\lambda^{\alpha} - \lambda^{\alpha} \p\omega_{\alpha} \right): - \frac{1}{2}:\left( \psi_A\partial\chi^A + \chi^A\partial\psi_A \right):.$$

It is simple to check that
\begin{align*}
    \left[ J_0^3, \partial^n \omega_{\pm} \right] &= \pm \frac{1}{2}\partial^n \omega_{\pm}, & \left[ K_0^3, \partial^n \chi^{\pm} \right] &= \pm \frac{1}{2}\partial^n \chi^{\pm},\\
    \left[ ( \mathcal{C}_L )_0, \partial^n \omega_{\alpha} \right] &= -\frac{1}{2}\partial^n \omega_{\alpha}, & \left[ ( \mathcal{C}_L )_0, \partial^n \chi^A \right] &= \frac{1}{2}\partial^n \chi^A.
\end{align*}
Hence, any solution to \eqref{eq:O} must have an equal number of each of the fields $(\omega_+,\omega_-, \chi^+, \chi^-)$ or their derivatives. We deduce that $\mathcal{O}$ is a bosonic operator built from sums and products of terms of the form
$$\partial^{m} \chi^- \partial^{n} \chi^+ \partial^{p} \omega_- \partial^{q} \omega_+,$$
for non-negative integers $m,n,p$ and $q$. We may rewrite such expressions in terms of the combinations
\begin{equation}\label{eq:Delta}
    \begin{split}
        \Delta^{\pm}_{p,q} &= \partial^p\omega_-\partial^q\omega_+ \pm \partial^q\omega_-\partial^p\omega_+,\\
        \tilde{\Delta}^{\pm}_{m,n} &= \partial^m\chi^-\partial^n\chi^+ \pm \partial^n\chi^-\partial^m\chi^+.
    \end{split}
\end{equation}
Therefore, we use $\tilde{\Delta}^{\pm}_{m,n}\Delta^{\pm}_{p,q}$ (all signs) as the building blocks for the algebra in which $\mathcal{O}$ lives. The requirement that $\mathcal{O}$ commutes with $J^{\pm}_0$ and $K^{\pm}_0$ leaves only one invariant combination, $\tilde{\Delta}^+_{m,n}\Delta^-_{p,q}$, such that
\begin{equation}\label{eq:O_expansion}
    \mathcal{O} = \sum_{k}\sum_{\{\mathbf{m},\mathbf{n},\mathbf{p}, \mathbf{q} \}} c_{\mathbf{m}, \mathbf{n}, \mathbf{p}, \mathbf{q}} \tilde{\Delta}^+_{m_1,n_1}\tilde{\Delta}^+_{m_2,n_2} \dots \tilde{\Delta}^+_{m_k,n_k} \Delta^-_{p_1,q_1} \Delta^-_{p_2,q_2} \dots \Delta^-_{p_k,q_k} ,
\end{equation}
where $(\mathbf{m}, \mathbf{n}, \mathbf{p}, \mathbf{q})$ are all $k$-dimensional vectors. For $\mathcal{O}$ of fixed weight $h$, the coefficients $c_{\{\mathbf{m}, \mathbf{n}, \mathbf{p}, \mathbf{q}\}}$ can only be non-vanishing provided
$$\sum_{i=1}^k(m_i+n_i+p_i+q_i) = h - 2k.$$
Moreover, the ordering of the terms in the product does not matter, since the components $\tilde{\Delta}^+_{m_i,n_i}$ and $\Delta^-_{p_i,q_i}$ all commute. It is clear from the definitions \eqref{eq:Delta} that $\Delta^-_{p,q} = -\Delta^-_{q,p}$ and $\tilde{\Delta}^+_{m,n} = \tilde{\Delta}^+_{n,m}$. This gives symmetry relations on the $c_{\{\mathbf{m}, \mathbf{n}, \mathbf{p}, \mathbf{q}\}}$: it is symmetric under the exchange $m_i \leftrightarrow n_i$, $(m_i,n_i) \leftrightarrow (m_j,n_j)$ or $(p_i,q_i) \leftrightarrow (p_j,q_j)$ whilst it is antisymmetric under $p_i \leftrightarrow q_i$. If we were to also impose \eqref{eq:scaling_constraint} for $n>0$, we would derive further constraints on the coefficients $c_{\{\mathbf{m}, \mathbf{n}, \mathbf{p}, \mathbf{q}\}}$.

There also exists a version of the Jacobi identity for the $\tilde{\Delta}^+_{m_i,n_i}$, given by
$$\tilde{\Delta}^+_{m_1,n_1}\tilde{\Delta}^+_{m_2,n_2} + \tilde{\Delta}^+_{m_1,m_2}\tilde{\Delta}^+_{n_2,n_1} + \tilde{\Delta}^+_{m_1,n_2}\tilde{\Delta}^+_{n_1,m_2} = 0,$$
and similarly for $\Delta^-_{p_i,q_i}$. This highlights the fact that the terms in \eqref{eq:O_expansion} are not linearly independent. This is a particularly useful identity for the bilinears in the fermions, $\tilde{\Delta}^+_{m,n}$, which is symmetric. We have that
$$\left( \tilde{\Delta}^+_{m,n} \right)^2 = -\frac{1}{2}\tilde{\Delta}^+_{m,m}\tilde{\Delta}^+_{n,n},$$
which is non-vanishing when $m\neq n$, whilst $\left( \tilde{\Delta}^+_{m,n} \right)^3 = 0$. We can therefore choose to expand $\mathcal{O}$ as in \eqref{eq:O_expansion} such that there are no repeated terms $\tilde{\Delta}^+_{m,n}$. More generally, if there exist any repeated indices among the $(m_i,n_i)$, then we will choose to write this as
\begin{equation}\label{eq:repeated_index}
    \tilde{\Delta}^+_{m_1,n}\tilde{\Delta}^+_{m_2,n} = -\frac{1}{2} \tilde{\Delta}^+_{m_1,m_2}\tilde{\Delta}^+_{n,n},
\end{equation}
and an index cannot be repeated three times.\\

We would now like to consider which $c_{\mathbf{m}, \mathbf{n}, \mathbf{p}, \mathbf{q}}$ in \eqref{eq:O_expansion} are allowed for $\mathcal{O}$ to be a primary operator. Using the stated symmetries in the indices $c_{\mathbf{m}, \mathbf{n}, \mathbf{p}, \mathbf{q}}$, we may first choose representative elements in each of the equivalence classes generated by these symmetries. In particular, we will fix $m_i \geq n_i$ and $p_i > q_i$ for all $i$ (by antisymmetry, $p_i\neq q_i$). Next, the invariance under swapping pairs $(m_i,n_i) \leftrightarrow (m_j,n_j)$ or $(p_i,q_i) \leftrightarrow (p_j,q_j)$ gives a permutation $S_k \times S_k$ symmetry. Hence, we may restrict the sum over $\{\mathbf{m}, \mathbf{n}, \mathbf{p}, \mathbf{q}\}$ to
$$\mathcal{O} = \sum_k \sum_{\substack{\{\mathbf{m}, \mathbf{n}, \mathbf{p}, \mathbf{q}\}/(S_k \times S_k) \\ \text{s.t. } m_i\geq n_i, \; p_i > q_i}} c_{\mathbf{m}, \mathbf{n}, \mathbf{p}, \mathbf{q}} \tilde{\Delta}^+_{m_1,n_1}\tilde{\Delta}^+_{m_2,n_2} \dots \tilde{\Delta}^+_{m_k,n_k} \Delta^-_{p_1,q_1} \Delta^-_{p_2,q_2} \dots \Delta^-_{p_k,q_k}.$$
As we will be considering the OPE with $T(z)$ to look for primary operators, it will be helpful to split this sum up further into linearly independent pieces. Clearly, terms can only be linearly dependent if they have the same value $k$ for the number of each type of free field in the product. Assuming this to be the case, the Jacobi identity implies that terms labelled by $(\mathbf{m}, \mathbf{n}, \mathbf{p}, \mathbf{q})$ and $(\mathbf{m}', \mathbf{n}', \mathbf{p}', \mathbf{q}')$ could potentially be linearly dependent if $(\mathbf{m}, \mathbf{n})$ and $(\mathbf{m}', \mathbf{n}')$ are related by a permutation of $S_{2k}$ and likewise for $(\mathbf{p}, \mathbf{q})$ and $(\mathbf{p}', \mathbf{q}')$. Let
$$\Omega^k_{\mathbf{m}, \mathbf{n}} = \left\{ (\mathbf{m}', \mathbf{n}')/S_{k} \; | \; m_i'\geq n_i' \text{ and } \exists \, \sigma \in S_{2k} \text{ s.t. } \sigma((\mathbf{m}', \mathbf{n}')) = (\mathbf{m}, \mathbf{n})   \right\}$$
label this space of possible linearly dependent terms for the fermionic part and similarly $\Omega^k_{\mathbf{p}, \mathbf{q}}$ for the bosonic part. We deduce that $\mathcal{O}$ is built from the linearly independent
\begin{equation}\label{eq:omega}
    \Lambda^k_{\mathbf{m}, \mathbf{n}, \mathbf{p}, \mathbf{q}} = \sum_{\substack{(\mathbf{m}',\mathbf{n}') \in \Omega^k_{\mathbf{m}, \mathbf{n}}, \\ (\mathbf{p}',\mathbf{q'}) \in \Omega^k_{\mathbf{p}, \mathbf{q}}}} c_{\mathbf{m}', \mathbf{n}', \mathbf{p}', \mathbf{q}'} \tilde{\Delta}^+_{m_1',n_1'}\tilde{\Delta}^+_{m_2',n_2'} \dots \tilde{\Delta}^+_{m_k',n_k'} \Delta^-_{p_1',q_1'} \Delta^-_{p_2',q_2'} \dots \Delta^-_{p_k',q_k'}.
\end{equation}
To impose that $\mathcal{O}$ is primary, we will need to consider the OPE of such expressions with $T(z)$. The following OPEs will be helpful in our computation. Assuming $m>n$ and $p>q$,
$$\left. T(z) \tilde{\Delta}^+_{m,n}\right|_{(z-w)^{-m-2}} = \frac{1}{2}(m+1)! \tilde{\Delta}^+_{0,n}, \quad \left. T(z) \Delta^-_{p,q}\right|_{(z-w)^{-p-2}} = \frac{1}{2}(p+1)! \Delta^-_{0,q},$$
whilst if $m=n$,
$$\left. T(z) \tilde{\Delta}^+_{m,m}\right|_{(z-w)^{-m-2}} = (m+1)! \tilde{\Delta}^+_{0,m}.$$

To simplify the derivation, let us first consider an operator built solely from the bosons,
$$\Lambda^{k,B}_{\mathbf{p}, \mathbf{q}} = \sum_{\Omega^k_{\mathbf{p}, \mathbf{q}}} c_{\mathbf{p}', \mathbf{q}'} \Delta^-_{p_1',q_1'} \Delta^-_{p_2',q_2'} \dots \Delta^-_{p_k',q_k'},$$
which to contribute non-trivially to $\mathcal{O}$ we assume is non-zero. As mentioned above, the sets $\Omega^k_{\mathbf{p},\mathbf{q}}$ label linearly independent sectors. These sectors remain linearly independent even after taking the OPE with $T(z)$, and so it is sufficient to solve for primary $\Lambda^{k,B}_{\mathbf{p},\mathbf{q}}$ and not consider sums of such fields (the same result also holds in the fermionic case below). Suppose that the largest derivative that appears in $\Lambda^{k,B}_{\mathbf{p},\mathbf{q}}$ is $N$ and that it appears $l$ times (we know that $N$ occurs the same number of times for each element of $\Omega^k_{\mathbf{p}, \mathbf{q}}$). Without loss of generality, we may take $p_i' = N$ for $1\leq i \leq l$. Expanding the corresponding $\Delta^-_{p_i',q_i'}$,
\begin{equation}\label{eq:omega_B}
    \Lambda^{k,B}_{\mathbf{p}, \mathbf{q}} = \sum_{l_1=0}^l \left( \partial^N\omega_- \right)^{l_1} (-\partial^N \omega_+)^{l-l_1} \Lambda^{k,B,l_1}_{\mathbf{p}, \mathbf{q}},
\end{equation}
where
$$\Lambda^{k,B,l_1}_{\mathbf{p}, \mathbf{q}} = \sum_{\Omega^k_{\mathbf{p}, \mathbf{q}}} c_{\mathbf{p}', \mathbf{q}'} \Delta^-_{p_{l+1}',q_{l+1}'} \dots \Delta^-_{p_k',q_k'} \sum_{\sigma \in S_l} \partial^{q_{\sigma(1)}}\omega_+ \dots \partial^{q_{\sigma(l_1)}}\omega_+\partial^{q_{\sigma(l_1 + 1)}}\omega_- \dots \partial^{q_{\sigma(l)}}\omega_-.$$
The leading order contribution to the OPE of $T(z) \Lambda^{k,B}_{\mathbf{p},\mathbf{q}}(w)$ comes at order $(z-w)^{-N-2}$, which we denote by $(N+1)!\, \tilde{\Lambda}^{k,B}_{\mathbf{p},\mathbf{q}}$, where
\begin{align*}
    \tilde{\Lambda}^{k,B}_{\mathbf{p},\mathbf{q}} &= \sum_{l_1 = 0}^l \left[ l_1 \omega_- (\partial^N\omega_-)^{l_1 - 1}(-\partial^N\omega_+)^{l-l_1} - (l-l_1)\omega_+ (\partial^N\omega_-)^{l_1} (-\partial^N\omega_+)^{l-l_1-1} \right] \Lambda^{k,B,l_1}_{\mathbf{p}, \mathbf{q}},\\
    &= \sum_{l_1 = 0}^{l-1} (\partial^N\omega_-)^{l_1}(-\partial^N\omega_+)^{l-l_1-1}\left[ (l_1+1)\omega_- \Lambda^{k,B,l_1+1}_{\mathbf{p},\mathbf{q}} - (l-l_1)\omega_+ \Lambda^{k,B,l_1}_{\mathbf{p},\mathbf{q}} \right].
\end{align*}
If we want $\Lambda^{k,B}_{\mathbf{p},\mathbf{q}}$ to be primary, then this term in the OPE must vanish for $N>0$. Suppose $\tilde{\Lambda}^{k,B}_{\mathbf{p},\mathbf{q}} = 0$, then
$$(l_1+1)\omega_- \Lambda^{k,B,l_1+1}_{\mathbf{p},\mathbf{q}} = (l-l_1)\omega_+\Lambda^{k,B,l_1}_{\mathbf{p},\mathbf{q}},$$
for $l_1 = 0, \dots, l-1$. This constraint is solved by
$$(\omega_-)^{l_1}\Lambda^{k,B,l_1}_{\mathbf{p},\mathbf{q}} = 
\begin{pmatrix}
    l\\
    l_1
\end{pmatrix}
(\omega_+)^{l_1} \Lambda^{k,B,0}_{\mathbf{p},\mathbf{q}},$$
for all $l_1 = 0, \dots, l$. Substituting this back in to \eqref{eq:omega_B}, we find
$$(\omega_-)^l\Lambda^{k,B}_{\mathbf{p},\mathbf{q}} = \sum_{l_1 = 0}^l
\begin{pmatrix}
    l\\
    l_1
\end{pmatrix}
(\omega_+\partial^N\omega_-)^{l_1} (-\omega_-\partial^N\omega_+)^{l-l_1}\Lambda^{k,B,0}_{\mathbf{p},\mathbf{q}} = \Lambda^{k,B,0}_{\mathbf{p},\mathbf{q}}(\Delta^-_{N,0})^l.$$
For this identity to hold for general $l$, there must be a factor of $(\omega_-)^l$ contained in $\Lambda^{k,B,0}_{\mathbf{p},\mathbf{q}}$ which, from the definition, implies $q_i' = 0$ for $1\leq i \leq l$. Therefore,
$$\Lambda^{k,B}_{\mathbf{p},\mathbf{q}} = \sum_{\Omega^k_{\mathbf{p}, \mathbf{q}}} c_{\mathbf{p}', \mathbf{q}'} \Delta^-_{p_{l+1}',q_{l+1}'} \dots \Delta^-_{p_k',q_k'} (\Delta^-_{N,0})^l.$$
If $N>1$, then the term of order $(z-w)^{-N-1}$ in the OPE with $T(z)$ must also vanish. There are two types of contribution at this order: if any $p_i' = N-1$, they will contribute analogously to $p_i' = N$ at order $(z-w)^{-N-2}$, whilst we also have contributions from $\Delta^-_{N,0}$,
$$\left. T(z) \Delta^-_{N,0}\right|_{(z-w)^{-N-1}} = \frac{1}{2}[(N+1)!+N!] \Delta^-_{1,0}.$$
The two types of contribution will be linearly independent of one another (as can be seen by counting the number of derivatives of order $N$), so each contribution must vanish. It is clear, therefore, that the only primary fields of the bosons are
$$\Lambda^{k,B}_{\mathbf{p},\mathbf{q}} = (\Delta_{1,0})^k,$$
for some $k \in \mathbb{N}$, up to an overall rescaling.

Now let's consider an operator built solely from the fermions,
$$\Lambda^{k,F}_{\mathbf{m},\mathbf{n}} = \sum_{\Omega^k_{\mathbf{m},\mathbf{n}}} c_{\mathbf{m}',\mathbf{n}'} \tilde{\Delta}^+_{m_1',n_1'}\tilde{\Delta}^+_{m_2',n_2'}\dots \tilde{\Delta}^+_{m_k',n_k'},$$
which again we assume to be non-zero to contribute non-trivially to $\mathcal{O}$. As mentioned earlier, an index can appear at most twice for a non-zero contribution. Suppose the largest derivative is of order $N$. Without loss of generality, we take $m_1' = N$ and if $N$ is repeated, then we also take $n_1' = N$ using \eqref{eq:repeated_index}. In either case, up to a constant of proportionality, the OPE with $T(z)$ at order $(z-w)^{-N-2}$ returns
$$\tilde{\Lambda}^{k,F}_{\mathbf{m},\mathbf{n}} = \sum_{\Omega^k_{\mathbf{m},\mathbf{n}}} c_{\mathbf{m}',\mathbf{n}'} \tilde{\Delta}^+_{0,n_1'}\tilde{\Delta}^+_{m_2',n_2'}\dots \tilde{\Delta}^+_{m_k',n_k'}.$$
For this to vanish as well as lower order poles, it must be the case that the index $0$ is repeated three times.\footnote{It is clear that repeating the $0$ index three times is the simplest solution to $\tilde{\Lambda}^{k,F}_{\mathbf{m},\mathbf{n}} = 0$. However, there are solutions when this is not the case through the Jacobi identity. For example, $\Lambda^{k,F}_{\mathbf{m},\mathbf{n}} = \tilde{\Delta}^+_{N,2} \tilde{\Delta}^+_{2,0} \tilde{\Delta}^+_{1,1} + \tilde{\Delta}^+_{N,2} \tilde{\Delta}^+_{2,1} \tilde{\Delta}^+_{1,0} + \tilde{\Delta}^+_{0,2} \tilde{\Delta}^+_{2,1} \tilde{\Delta}^+_{N,1}$ with $N>2$. The fine tuning of the coefficients for the order $(z-w)^{-N-2}$ pole to vanish whilst $\Lambda^{k,F}_{\mathbf{m},\mathbf{n}} \neq 0$ means that the poles of lower order will not vanish, as can be verified explicitly in this example. Therefore, repeating the $0$ index is the generic solution.\label{footnote:OPE}} Without loss of generality, we therefore set $m_k'=n_k'=0$. This implies that for $N=0$ or $N=1$, the unique solutions (up to rescaling) are respectively
$$\tilde{\Delta}^+_{0,0} \quad \text{or} \quad \tilde{\Delta}^+_{1,1}\tilde{\Delta}^+_{0,0}.$$
For $N>1$, we may iterate this argument. For example, at order $(z-w)^{-N-1}$ in the OPE with $T(z)$, we collect two types of terms as in the bosonic case: those from $\tilde{\Delta}^+_{m_i',n_i'}$ with $m_i' = N-1$ and those from Taylor expanding the OPE with $\tilde{\Delta}^+_{N,n_1'}$. Since, $m_k' = n_k' = 0$, the first type give a vanishing contribution, whilst the second type lead to terms proportional to $\tilde{\Delta}^+_{1,n_1'}$. To vanish, the index $1$ must now be repeated three times in the OPE and without loss of generality, we deduce $m_{k-1}'=n_{k-1}' = 1$. Iterating the argument for all poles of order greater than two, the unique solution is
$$\Lambda^{k,F}_{\mathbf{m},\mathbf{n}} = \tilde{\Delta}^+_{0,0} \tilde{\Delta}^+_{1,1}\dots \tilde{\Delta}^+_{k-1,k-1}.$$

We can now return to the problem of interest, which is to determine the primary operators $\mathcal{O}$ that are built from the linearly independent pieces $\Lambda^k_{\mathbf{m},\mathbf{n},\mathbf{p},\mathbf{q}}$ as in \eqref{eq:omega}. In the cases of only bosons and only fermions, a key result that we used was that the linearly independent sectors ($\Omega^k_{\mathbf{p},\mathbf{q}}$ for the bosons and $\Omega^k_{\mathbf{m},\mathbf{n}}$ for the fermions) remain linearly independent after taking the OPE with $T(z)$. This is not quite true in the mixed case for $N>1$, where the linearly independent sectors are labelled by $\Omega^k_{\mathbf{m},\mathbf{n}}\times \Omega^k_{\mathbf{p}, \mathbf{q}}$. As an example, consider
$$\tilde{\Delta}^+_{2,1}\Delta^-_{1,0} + \tilde{\Delta}^+_{1,0}\Delta^-_{2,1}$$
for which $N=2$ and $k=1$. The two terms are linearly independent and yet, the order $(z-w)^{-N-2} = (z-w)^{-4}$ term in the OPE with $T(z)$ vanishes through a cancellation between the bosonic and fermionic contributions to the OPE. However, the subleading terms in the OPE will not cancel in such cases, rendering such examples non-primary (see also footnote \ref{footnote:OPE}). We therefore deduce that the primary operators \eqref{eq:O_expansion} are built from primary operators in each sector $\Omega^k_{\mathbf{m},\mathbf{n}}\times \Omega^k_{\mathbf{p}, \mathbf{q}}$. These are necessarily given by a product of the bosonic and fermionic operators we found earlier, $\Lambda^k_{\mathbf{m},\mathbf{n},\mathbf{p},\mathbf{q}} = \Lambda^{k,F}_{\mathbf{m},\mathbf{n}} \Lambda^{k,B}_{\mathbf{p},\mathbf{q}}$. Finally, the sum over such terms in \eqref{eq:O_expansion} is constrained to have definite conformal weight. For fixed $h$, this gives a unique primary operator $\mathcal{O}$ satisfying \eqref{eq:O} (up to rescaling)
$$\mathcal{O} = \tilde{\Delta}^+_{0,0}\tilde{\Delta}^+_{1,1}\dots \tilde{\Delta}^+_{k-1,k-1}(\Delta^-_{1,0})^k,$$
where $h = k^2 + 2k$.

What's interesting is that these operators can be rewritten in terms of $\mathcal{Q}$. For $k=1$ this is clear, as $\mathcal{Q} = \half \tilde{\Delta}^+_{0,0}\Delta^-_{0,1}$. The operator $\mathcal{Q}\partial^2 \mathcal{Q}$ is non-zero when one of the derivatives acts on each of the fermions in $\mathcal{Q}$. As such, $\mathcal{Q}\partial^2\mathcal{Q} = \frac{1}{2} \tilde{\Delta}^+_{0,0} \tilde{\Delta}^+_{1,1}(\Delta^-_{0,1})^2$. Working inductively,
$$\mathcal{Q}\partial^2 \mathcal{Q} \dots \partial^{2(k-1)}\mathcal{Q} = \nu_k \tilde{\Delta}^+_{0,0}\tilde{\Delta}^+_{1,1}\dots \tilde{\Delta}^+_{k-1,k-1}(\Delta^-_{0,1})^k,$$
where
$$\nu_k = \frac{1}{2^k}\prod_{m=1}^{k-1}
\begin{pmatrix}
    2m\\
    m
\end{pmatrix}.$$
Hence, the set of primary operators that solve \eqref{eq:O} are given by
\begin{equation}\label{eq:O^k}
    \mathcal{O}^{(k)} = \mathcal{Q}\partial^2\mathcal{Q}\dots \partial^{2(k-1)}\mathcal{Q},
\end{equation}
for $k\in\mathbb{N}$.

\section{$\kappa$-symmetry and twistor space}\label{sec:kappa_sym}

In our discussion of PBGS in \S\ref{sec:PBGS}, we noted that it was natural to interpret the non-critical theory \eqref{eq:CP1_string} as the $\kappa$-gauge fixed phase of a fully covariant target space description, and as such it is missing compensating ghost terms. It appears, therefore, that we have recast the gauge constraints of $\kappa$-symmetry into a redundant worldsheet supersymmetry \eqref{eq:superRS}. This highlights the simplifications that can occur when a system with $\kappa$-symmetry is written in twistorial variables. Indeed, comparing to the flat space GS superstring \cite{Green:1983wt}, we appear to have circumvented two of the conventional problems with $\kappa$-symmetry: firstly, the vanishing of the Noether charges associated to $\kappa$-symmetry means a generator cannot be written down; and secondly, the infinite reducibility (see \cite{Batalin:1983ggl}) of $\kappa$-symmetry makes covariant quanitization challenging.

On the second point, recall the usual setting of a type II GS superstring on flat space in ten dimensions. This is given by embedding fields $(X^{\mu},\theta^{ai})$ where $\mu = 0,\dots , 9$ are spacetime vector indices, $a = 1, \dots , 32$ are spacetime spinor indices and $i = 1, 2$ label the two supersymmetries. The $\theta^{ai}$ describe Majorana-Weyl fermions, so describe 16 real degrees of freedom for each $i$. Then, $\kappa$-symmetry is a local symmetry which reduces the number of degrees of freedom in $\theta^{ai}$ appearing in physical states and acts as \cite{Green:1987sp}
\begin{equation}\label{eq:kappa_sym}
    \delta \theta^{ai} = 2i \left(\Gamma^{\mu} 
\right)^a{}_{b} \Pi_{\mu}\kappa^{bi}, \qquad \delta X^{\mu} = i\bar{\theta}^{ai} \left( \Gamma^{\mu} \right)_{ab} \delta \theta^{bi},
\end{equation}
where $\Pi^\mu$ is the holomorphic component of the (pullback of) supersymmetric invariant 1-forms,
$$\Pi^{\mu} = \p X^{\mu} - i\bar{\theta}^{ai} \left( \Gamma^{\mu} \right)_{ab}\p\theta^{bi},$$
and $\Gamma^{\mu}$ are 10-dimensional gamma matrices. Gamma matrix identities can be used to show that the matrix $\Gamma \cdot \Pi$ is half rank on-shell, meaning that half of the degrees of freedom in $\kappa^{ai}$ are projected out. This is the sense in which $\kappa$-symmetry is infinitely reducible: if we introduce ghost fields for the gauge-redundancy in $\kappa^{ai}$, then we must further introduce ghosts for ghosts for the degrees of freedom that were projected out from $\kappa^{ai}$ in \eqref{eq:kappa_sym}, and this process repeats ad infinitum.

So why should this simplify in twistor space? For our interests in the space $\C\P^{1|2} \times \C\P^{1|2}$ which has two complex bosonic dimensions, it is simplest to draw an analogy with the two-complex-dimensional flat space GS superstring with extended supersymmetry. The $\kappa$-symmetry transformations take an analogous form to \eqref{eq:kappa_sym}, except we now have dotted and undotted spinor indices, with $\alpha,\dot{\alpha}=\pm$. Since $\Gamma\cdot\Pi$ is half-rank, we can explicitly write $(\Gamma \cdot \Pi)_{\alpha\dot{\alpha}} = \lambda_{\alpha} \tilde{\lambda}_{\dot{\alpha}}$ for bosonic spinors $\lambda_{\alpha}$ and $\tilde{\lambda}_{\dot{\alpha}}$. Then,
$$\delta \theta^{\alpha i} = 2i\epsilon^i(\tau,\sigma) \lambda^{\alpha},$$
for $\epsilon^i(\tau,\sigma) = \tilde{\lambda}_{\dot{\alpha}}\kappa^{\dot{\alpha}i}$, such that the $\kappa$-symmetry transformation has been recast into a local supersymmetry transformation relating bosonic and fermionic spinors.\footnote{This is closely related to the idea in \cite{Witten:1985nt}, where $\kappa$-symmetry in ambitwistor space was found to take on an especially simple form, generating an orbit of the string in half of the fermionic directions. See also \cite{Carabine:2018kdg} for a similar phenomenon in a different context.} This property has been well-studied in the super-embedding approach of \cite{Sorokin:1988nj,Sorokin:1988jor,Bandos:1995zw,Sorokin:1999jx} (see also \cite{Volkov:1988vf,Volkov:1989ct,Berkovits:1989zq,Tonin:1991ii}), which by construction contains both worldsheet and spacetime supersymmetry. For this reason, it has sometimes been dubbed as a ``doubly supersymmetric'' membrane theory, where the $\kappa$-symmetry is replaced by worldsheet (or worldvolume) supersymmetry of the twistor variables. We expect that the superembedding formalism should provide a natural language for constructing the $\mathcal{N}=2$ description of the $k=1$ string as outlined in \S\ref{sec:PBGS}.

\bibliographystyle{JHEP}
\bibliography{AdS3}

\end{document}